%% file: main.tex
\documentclass[journal,10pt,twocolumn,final,]{IEEEtran}
\pdfoutput=1
\input{Paper/packages.tex}

\input{Paper/tikz.tex}

\input{Paper/commands.tex}

\begin{document}

\title{Robust M-Estimation Based Bayesian Cluster Enumeration for Real Elliptically Symmetric Distributions}

\author{Christian~A.~Schroth and Michael Muma,~\IEEEmembership{Member,~IEEE}
\thanks{C. A. Schroth and M. Muma are with the Signal Processing Group at Technische Universit\"at Darmstadt, Germany. mail: \{schroth, muma\}@spg.tu-darmstadt.de}
\thanks{Manuscript submitted, May 6, 2020.}}

\markboth{Submitted to IEEE Transactions on Signal Processing (accepted)}%
{Schroth \MakeLowercase{\textit{et al.}}: Robust M-Estimation Based Bayesian Cluster Enumeration for Real Elliptically Symmetric Distributions}

\maketitle

\begin{abstract}
	Robustly determining the optimal number of clusters in a data set is an essential factor in a wide range of applications. Cluster enumeration becomes challenging when the true underlying structure in the observed data is corrupted by heavy-tailed noise and outliers. Recently, Bayesian cluster enumeration criteria have been derived by formulating the cluster enumeration problem as a maximization of the posterior probability of candidate models. This article generalizes robust Bayesian cluster enumeration so that it can be used with any arbitrary Real Elliptically Symmetric (RES) distributed mixture model. Our framework also covers the case of M-estimators. These robust estimators allow for mixture models, which are decoupled from a specific probability distribution. Examples of Huber's and Tukey's M-estimators are discussed. We derive a robust criterion for data sets with finite sample size, and also provide an asymptotic approximation to reduce the computational cost at large sample sizes. The algorithms are applied to simulated and real-world data sets, including radar-based person identification and remote sensing, and they show a significant robustness improvement in comparison to existing methods.
\end{abstract}


\IEEEpeerreviewmaketitle

\input{Paper/intro.tex}
\input{Paper/res_dist.tex}
\input{Paper/gen_clu_enum.tex}
\input{Paper/res_clu_enum.tex}

\input{Paper/clu_enum_alg.tex}
\input{Paper/sim.tex}

\section{Conclusion}
\label{sec:conclusion}
We have presented a general Robust Bayesian cluster enumeration framework. This was done by deriving an EM algorithm for arbitrary RES distributions and adapting the generic BIC from \cite{Teklehaymanot.2018} to the class of RES distributions and to the class of M-estimators. Some robust M-estimators, such as Huber's, or the t M-estimator, may correspond to ML estimators for a specific RES distribution. Our framework, however, also allows for non-ML loss functions such as Tukey's loss function. The performance was evaluated on simulated and real world examples, which show a superior robustness against outliers, compared to existing work. Further research may be done to derive alternatives for the EM algorithm or to include further challenging problems, like skewed data distributions \cite{Schroth.2020c} or high-dimensionalty \cite{Ollila.2014, Liu.2019, Roizman.2019}.

\section*{Acknowledgment}
We thank Ann-Kathrin Seifert for her support in providing us with the Radar Data set and Freweyni K. Teklehaymanot for the inspiring scientific discussions. Christian A. Schroth is supported by the DFG Project Number 431431951. The work of Michael Muma has been funded by the LOEWE initiative (Hesse, Germany) within the emergenCITY centre and is supported by the ‘Athene Young Investigator Programme’ of Technische Universität Darmstadt, Hesse, Germany.

\appendices
\input{Paper/ap_FIM_RES_mumu.tex}
\input{Paper/ap_ML_RES_short.tex}

\renewcommand*{\bibfont}{\footnotesize}
\printbibliography

\end{document}


\title{Supplementary Information: Robust M-Estimation Based Bayesian Cluster Enumeration for Real Elliptically Symmetric Distributions}

\author{Christian~A.~Schroth and Michael Muma,~\IEEEmembership{Member,~IEEE}
\thanks{C. A. Schroth and M. Muma are with the Signal Processing Group at Technische Universit\"at Darmstadt, Germany. mail: \{schroth, muma\}@spg.tu-darmstadt.de}
\thanks{Manuscript submitted, May 6., 2020.}}

\markboth{Submitted to IEEE Transactions on Signal Processing (accepted)}%
{Schroth \MakeLowercase{\textit{et al.}}: Supplementary Information: Robust M-Estimation Based Bayesian Cluster Enumeration for Real Elliptically Symmetric Distributions}

\maketitle

\IEEEpeerreviewmaketitle

\section{Structure}

This Supplementary Information  for the paper 'Robust M-Estimation Based Bayesian Cluster Enumeration for Real Elliptically Symmetric Distributions' is organized as follows: In Appendix~\ref{ch:appDerivative} a detailed step-by-step solution of the second derivatives of the log-likelihood function for the FIM is given. Afterwards the ML estimates for $\bShat_{m}$ and $\bmuhat_{m}$ based on the first derivatives are calculated and some used identities are shown. Finally we provide a comprehensive summary of the used matrix calculus in Appendix~\ref{ch:matcalc}.

\appendices
\input{Appendix/ap_FIM_RES.tex}

\input{Appendix/ap_ML_RES.tex}
\input{Appendix/ap_matcalc.tex}

\ifCLASSOPTIONcaptionsoff
  \newpage
\fi

\printbibliography

%% file: Paper/packages.tex
\usepackage[backend=biber, style=ieee, isbn=false, doi=false]{biblatex}   
\bibliography{references.bib}
\usepackage{tabu}
\usepackage{multirow} 
\usepackage{rotating}
\usepackage{booktabs}

\usepackage[ruled,vlined]{algorithm2e}
\DontPrintSemicolon

\usepackage{mathtools} 
\usepackage{amssymb}   
\usepackage{subdepth} 

\usepackage{amsthm}
\theoremstyle{plain}

\newtheorem{theorem}{Theorem}

\usepackage{empheq}

\usepackage{tikz}

\usepackage{pgfplots}
\usepackage{varwidth}

\allowdisplaybreaks

\usepackage[caption=false,font=footnotesize]{subfig}

\usepackage{url}

\makeatletter
\newcommand{\removelatexerror}{\let\@latex@error\@gobble}
\makeatother

\usepackage[most]{tcolorbox}
\newtcolorbox{empheqboxed}{colback=white, colframe=black, sharpish corners, top=-2mm, bottom=0pt, left=0mm}

%% file: Paper/tikz.tex
\newlength{\figurewidth}
\newlength{\figureheight}

\usetikzlibrary{arrows}

\definecolor{matlabblue}{rgb}{0 0.4470 0.741}
\definecolor{matlaborange}{rgb}{0.8500 0.3250 0.0980}
\definecolor{matlabyellow}{rgb}{0.9290 0.6940 0.1250}
\definecolor{matlabpurple}{rgb}{0.4940 0.1840 0.5560}
\definecolor{matlabgreen}{rgb}{0.4660 0.6740 0.1880}
\definecolor{matlablightblue}{rgb}{0.3010 0.7450 0.9330}
\definecolor{matlabred}{rgb}{0.6350 0.0780 0.1840}

\pgfplotsset{compat=newest}

\pgfplotscreateplotcyclelist{matlabcolor}{%
	matlabblue, mark=diamond*\\%
	matlaborange, mark=square*\\%
	matlabyellow, mark=*\\%
	matlabpurple, mark=triangle*\\%
	matlabgreen, mark=star\\%
	matlablightblue\\%
	matlabred\\%
}

\pgfplotscreateplotcyclelist{matlabcolorNOmark}{%
	matlabblue\\%
	matlaborange\\%
	matlabyellow\\%
	matlabpurple\\%
	matlabgreen\\%
	matlablightblue\\%
	matlabred\\%
}

\pgfplotsset{
	colormap={parula}{
		rgb255=(53,42,135)
		rgb255=(15,92,221)
		rgb255=(18,125,216)
		rgb255=(7,156,207)
		rgb255=(21,177,180)
		rgb255=(89,189,140)
		rgb255=(165,190,107)
		rgb255=(225,185,82)
		rgb255=(252,206,46)
		rgb255=(249,251,14)
	},
}

\newcommand{\plotfile}[1]{
	\pgfplotstableread[col sep=tab]{#1}{\table}
	\pgfplotstablegetcolsof{\table}
	\pgfmathtruncatemacro\numberofcols{\pgfplotsretval-1}
	\pgfplotsinvokeforeach{1,...,\numberofcols}{
		\pgfplotstablegetcolumnnamebyindex{##1}\of{\table}\to{\colname}
		\addplot+[thick,line width=1.5pt] table [y index=##1, col sep=tab] {#1}; 
		\addlegendentryexpanded{\colname}
	}
}

\newcommand{\plotfileNOlegend}[1]{
	\pgfplotstableread[col sep=tab]{#1}{\table}
	\pgfplotstablegetcolsof{\table}
	\pgfmathtruncatemacro\numberofcols{\pgfplotsretval-1}
	\pgfplotsinvokeforeach{1,...,\numberofcols}{
		\pgfplotstablegetcolumnnamebyindex{##1}\of{\table}\to{\colname}
		\addplot+[thick,line width=1.5pt] table [y index=##1, col sep=tab] {#1}; 
	}
}

%% file: Paper/commands.tex
\newcommand{\xn}{\boldsymbol{x}_{n}}

\newcommand{\xinX}{\boldsymbol{x}_{n} \in \mathcal{X}_{m}}
\newcommand{\that}{\skew{1}\hat{t}}

\newcommand{\ba}{\boldsymbol{a}}

\newcommand{\bx}{\boldsymbol{x}}
\newcommand{\bA}{\boldsymbol{A}}

\newcommand{\bD}{\boldsymbol{D}}
\newcommand{\bF}{\boldsymbol{F}}

\newcommand{\bI}{\boldsymbol{I}}

\newcommand{\bK}{\boldsymbol{K}}
\newcommand{\bS}{\boldsymbol{S}}

\newcommand{\bSigma}{\boldsymbol{\Sigma}}
\newcommand{\btheta}{\boldsymbol{\theta}}
\newcommand{\bTheta}{\boldsymbol{\Theta}}
\newcommand{\bmu}{\boldsymbol{\mu}}
\newcommand{\bPhi}{\boldsymbol{\Phi}}

\newcommand{\bgamma}{\boldsymbol{\gamma}}
\newcommand{\bPsi}{\boldsymbol{\Psi}}

\newcommand{\bxhat}{\hat{\boldsymbol{x}}}
\newcommand{\bxt}{\tilde{\boldsymbol{x}}}

\newcommand{\bFhat}{\skew{3}\hat{\boldsymbol{F}}}
\newcommand{\bShat}{\skew{3}\hat{\boldsymbol{S}}}
\newcommand{\bJhat}{\skew{5}\hat{\boldsymbol{J}}}
\newcommand{\bthetahat}{\skew{3}\hat{\btheta}}
\newcommand{\bmuhat}{\skew{1}\hat{\boldsymbol{\mu}}}

\newcommand{\bPhihat}{\hat{\boldsymbol{\Phi}}}

\DeclareMathOperator*{\argmax}{arg\,max}

\DeclareMathOperator{\Tr}{Tr}

\newcommand{\vecop}{\text{vec}}
\newcommand{\vechop}{\text{vech}}

%% file: Paper/intro.tex
\section{Introduction}
\IEEEPARstart{C}{luster} enumeration refers to the task of answering the question: How many subgroups of similar points are there in a given data set? Robustly determining the optimal number of clusters, $K$, is an essential factor in a wide range of applications. Providing a universal and objective answer, however, is challenging. It depends on the users' understanding of what constitutes a cluster and how to deal with outliers and uncertainty about the data. Popular clustering algorithms \cite{Dempster.1977, Lloyd.1982, Jain.1988, Arthur.2007, Xu.2015, Batool.2019} rely on small distances (or other measures of similarity) between cluster members, dense areas of the data space, or mixture models of particular statistical distributions.

The focus of this work lies on robust statistical model-based cluster analysis. The algorithms should provide reliable results, even if the cluster distribution is heavy-tailed or if the data set contains outliers. These are untypical data points that may not belong to any of the clusters. The methods should also work for cases where the data size is not huge, such that, clusters may have a relatively small number of associated data samples. Compared to purely data-driven unsupervised approaches, model-based methods allow for incorporating prior knowledge and assumptions. Statistically robust methods \cite{Huber.2011, Zoubir.2018, Maronna.2018} such as M-estimators \cite{Huber.2011} can deal with uncertainty by accounting for the fact that the prior knowledge is inexact and the assumptions are only approximately fulfilled. 

M-estimators are a generalization of Maximum-Likelihood-Estimators (MLE) where the negative log-likelihood function may be replaced by a robustness inducing objective function. For example, M-estimators can be designed based on the likelihood function of a Real Elliptically Symmetric (RES) distribution.  This wide family of distributions is useful in statistically modeling the non-Gaussian behavior of noisy data in many practical applications \cite{Fortunati.2016, Fortunati.2019b, Fortunati.2020, Ollila.2012}. RES distributions include, for example, Gaussian, the Generalized Gaussian \cite{Pascal.2013}, the t-distribution, the Compound Gaussian \cite{Pascal.2008}, and Huber's distribution, as special cases. Some M-estimators are not a MLE. For example, Tukey's estimator is designed to completely reject outlying observations by giving them zero-weight. This behavior is beneficial when outliers are generated by a contaminating distribution that strongly differs from the assumed distribution (often the Gaussian). 

A popular strategy in robust cluster enumeration is to use model selection criteria such as the Bayesian Information Criterion (BIC) derived by Schwarz \cite{Schwarz.1978, Cavanaugh.1999} in combination with robust clustering algorithms. Existing approaches include outlier detection and removal \cite{Wang.2018, Neykov.2007, Gallegos.2009, Gallegos.2010}, modeling noise or outliers
using an additional component in a mixture model \cite{Fraley.1998, Dasgupta.1998}, or modeling the data as a mixture of heavy-tailed distributions \cite{Andrews.2012, McNicholas.2012}. A ``robustified likelihood'' is complemented by a general penalty term to establish a trade-off between robust data-fit and model complexity. However, Schwarz' BIC is generic and it does not take the specific clustering problem into account. The penalty term only depends on the number of model parameters and on the number of data points. Therefore, it penalizes two structurally different models the same way if they have the same number of unknown parameters \cite{Djuric.1998, Stoica.2004}.

Recently, a BIC for cluster analysis has been derived by formulating the cluster enumeration problem as a maximization of the posterior probability of candidate models \cite{Teklehaymanot.2018, Teklehaymanot.2018d}. For these approaches, the penalty term incorporates more information about the clustering problem. It depends on the number of model parameters, the assumed data distribution, the number of data points per cluster, and the estimated parameters. A first attempt at robust Bayesian cluster enumeration has been recently derived by formulating the cluster enumeration problem as a maximization of the posterior probability of multivariate t-distributed candidate models \cite{Teklehaymanot.2018b}. Although this heavy-tailed model provided a significant increase in robustness compared to using Gaussian candidate models, it still relied on a specific distributional model. Our main contribution is to generalize robust Bayesian cluster enumeration so that it can be used with any arbitrary RES distributed mixture model, and even M-estimators. These robust estimators allow for mixture models that are decoupled from a specific probability distribution.

The paper is organized as follows. Section~\ref{sec:res_dis} gives a brief introduction to RES distributions and possible loss functions, including a more detailed discussion of the Huber distribution and Tukey's loss function. Section~\ref{sec:bic_g} introduces the BIC for general distributions, followed by Section~\ref{sec:bic_res} which is dedicated to the derivation of the proposed cluster enumeration criterion. Section~\ref{sec:algo} details the proposed robust cluster enumeration algorithm. Simulations and real-world examples are provided in Section~\ref{sec:experiments}. Finally, conclusions are drawn in Section~\ref{sec:conclusion}.  The appendices include derivatives for the Fisher Information Matrix (FIM) as well as ML estimators for RES distributions. Further details on the derivation of the FIM can be found in the online supplementary material.

\textbf{Notation:} Normal-font letters ($a, A$) denote a scalar, bold lowercase ($\ba$) a vector and bold uppercase ($\bA$) a matrix; calligraphic letters ($\mathcal{A}$) denote a set, with the exception of $\mathcal{L}$, which is reserved for the likelihood function; $\mathbb{R}$ denotes the set of real numbers and $\mathbb{R}^{r \times 1}$, $\mathbb{R}^{r \times r}$ the set of column vectors of size $r \times 1$, matrices of size $r \times r$, respectively; $\bA^{-1}$ is the matrix inverse; $\bA^{\top}$ is the matrix transpose; $|a|$ is the absolute value of a scalar; $|\bA|$ is the determinant of a matrix; $\otimes$ represents the Kronecker product; $\vecop(\cdot)$ is the vectorization operator, $\bD$ is the duplication matrix and $\vechop(\cdot)$ is the vector half operator as defined in \cite{Magnus.2007, Abadir.2005}.

%% file: Paper/res_dist.tex
\section{RES Distributions \& Loss Functions}
This section briefly revisits RES distributions and introduces some important loss functions.
\label{sec:res_dis}
\subsection{RES Distributions}
Assuming that the observed data $\bx \in \mathbb{R}^{r \times 1}$ follows a RES distribution, let $\bmu  \in \mathbb{R}^{r \times 1}$ be the centroid and let $\bS \in \mathbb{R}^{r \times r}$ be the positive definite symmetric scatter matrix of a distribution with a pdf, see \cite[p.~109]{Zoubir.2018} and \cite{Sahu.2003}:
\begin{equation}
	f(\bx| \bmu, \bS, g) = \left| \bS \right|^{-\frac{1}{2}} g\left( \left(\bx - \bmu\right)^{\top} \bS^{-1} \left(\bx - \bmu\right)\right),
	\label{eqn:res}
\end{equation}
where the squared Mahalanobis distance is denoted by \mbox{$t = \left(\bx - \bmu\right)^{\top} \bS^{-1} \left(\bx - \bmu\right)$}. The function $g$, often referred to as the density generator, is a function defined by
\begin{align}
	g(t) = \frac{\Gamma\left(\frac{r}{2}\right)}{\pi^{r/2}} \left(\int_{0}^{\infty} u^{r/2-1} h(u; r) \text{d}u\right)^{-1} h(t; r),
\end{align}
where $ h(t; r)$ is a function such that
\begin{align}
	\int_{0}^{\infty} u^{r/2-1} h(u; r) \text{d}u < \infty
\end{align}
holds. Note that $h(t; r)$ can be a function of multiple parameters, not only of $r$. 

\subsection{Loss Functions}
Assuming an observation of $N$ independent and identically distributed (iid) random variables denoted by \mbox{$\mathcal{X} = \{\bx_{1}, \dots, \bx_{N}\}$}, the likelihood function is given by
\begin{equation}
	\mathcal{L}(\bmu, \bS|\mathcal{X}) = \prod_{n = 1}^{N}\left| \bS^{-1} \right|^{\frac{1}{2}} g\left( t_{n}\right)
\end{equation}
with $t_{n} = \left(\bx_{n} - \bmu\right)^{\top} \bS^{-1} \left(\bx_{n} - \bmu\right)$. The MLE minimizes the negative log-likelihood function
\begin{align}
	-\ln\left(\mathcal{L}(\bmu, \bS|\mathcal{X}) \right) =& -\ln\left( \prod_{n = 1}^{N}\left| \bS^{-1} \right|^{\frac{1}{2}} g\left( t_{n}\right)\right)\notag\\
	=& \sum_{n = 1}^{N} -\ln\left(g(t_{n})\right) - \frac{N}{2} \ln\left(\left| \bS^{-1} \right|\right) \notag\\
	=& \sum_{n = 1}^{N} \rho_{\text{ML}}(t_{n}) + \frac{N}{2} \ln\left(\left| \bS \right|\right)
\end{align}	
with the associated ML loss function \cite[p.~109]{Zoubir.2018}
\begin{equation}
	\label{eq:ml-loss}
	\rho_{\text{ML}}(t_{n}) = -\ln\left(g(t_{n})\right).
\end{equation}
The corresponding first and second derivatives are denoted, respectively, by
\begin{equation}
	\psi_{\text{ML}}(t_{n}) = \frac{\partial \rho_{\text{ML}}(t_{n})}{\partial t_{n}}, \quad \eta_{\text{ML}}(t_{n}) = \frac{\partial \psi_{\text{ML}}(t_{n})}{\partial t_{n}}.
	\label{eqn:psi}
\end{equation}
The basic idea of M-estimation \cite{Huber.2011} is to replace the ML loss function $\rho_{\text{ML}}(t_{n})$ in Eq.~\eqref{eq:ml-loss} with a more general loss function $\rho(t_{n})$ that may not correspond to a ML estimator. A Non-ML loss function is not based on a specific distribution, but is designed to downweight outlying data points according to desired characteristics.

\begin{figure}
	\centering
	\resizebox{.6\columnwidth}{!}{\input{figures/review/psi_gaus_t_huber.tex}}
	\caption{Visualization of different weight functions $\psi(t)$ for the scalar case with mean $\mu = 0$, variance $S = 1$ and different values for the tuning parameters, as defined in Section~\ref{sec:tuning}. For $q_{H} \rightarrow 1$, the Huber distribution converges to the Gaussian distribution, while smaller values of $q_{H}$ result in heavier tails and a stronger downweighting of outliers. Tukey's is the only weighting function which completely downweights samples to zero for finite values of~$t$.}
	\label{fig:psi_gaus_t_huber}
\end{figure}
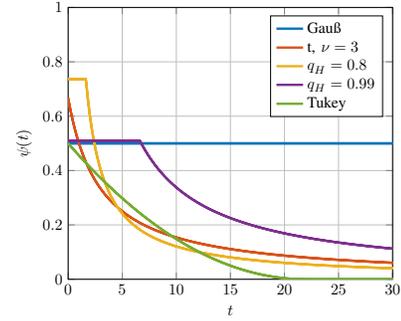

\subsection{Examples for RES Distributions and Loss Functions}

An overview of some exemplary loss functions and their derivatives can be found in Tables~ \ref{tb:g_rho} and \ref{tb:psi_eta}, a visualization of the weight associated weight functions is given in Figure~\ref{fig:psi_gaus_t_huber}, where we illustrate how robustness is obtained by downweighting data points based on their squared Mahalanobis distance. Since the Gaussian and t distribution are well-known they will not be further discussed, but a brief discussion is provided for the Huber distribution and Tukey's loss function.

\subsubsection{Huber Distribution}

As \cite[p.~115]{Zoubir.2018} and \cite[p.~8]{Ollila.2016b} point out, Huber's M-estimator can be viewed as a ML estimator for a RES distribution, which we will call Huber distribution. It is defined by
\begin{align}
	h(t; r, c) =& \exp \left(-\frac{1}{2} \rho_{\text{H}}(t;c)\right)
\end{align}
with
\begin{equation}
	\rho_{\text{H}}(t;c) = 
	\begin{dcases}
	\frac{t}{b} &, t \leq c^{2}\\
	\frac{c^{2}}{b}\left(\ln\left(\frac{t}{c^{2}}\right)+1\right) &, t > c^{2}
	\end{dcases}
\end{equation}
and to obtain Fisher consistency 
\begin{equation}
	b = F_{\chi_{r+2}^{2}}\left(c^{2}\right) + \frac{c^{2}}{r}\left(1 - F_{\chi_{r}^{2}}\left(c^{2}\right)\right),
\end{equation}
where $F_{\chi_{r}^{2}}\left(\cdot\right)$ is the Chi-square cumulative distribution function with degree of freedom $r$. To obtain a valid pdf the normalization factor, according to \cite{Sahu.2003}, has to be calculated as
\begin{align}
	\int_{0}^{\infty}& u^{r/2-1} h(u; r, c)  \text{d}u \notag \\
	=& \int_{0}^{c^{2}} u^{r/2-1} e^{-\frac{u}{2b}} \text{d}u + \int_{c^{2}}^{\infty} u^{r/2-1} \left(\frac{e u}{c^{2}} \right)^{-\frac{c^{2}}{2b}} \text{d}u \notag \\
	=& (2b)^{r/2}\left(\Gamma\left(\frac{r}{2}\right) - \Gamma\left(\frac{r}{2}, \frac{c^{2}}{2b}\right)\right) + \frac{2 b c^{r} \exp \left(-\frac{c^{2}}{2b} \right)}{c^2 - br},
\end{align}
with the gamma function $\Gamma(\cdot)$ and the upper incomplete gamma function $\Gamma(\cdot, \cdot)$.
We can now write the density generator of a Huber distribution as
\begin{align}
	g(t) =& 
	\begin{dcases}
		A_{\text{H}} \exp \left(-\frac{t}{2b}\right) &, t \leq c^{2}\\
		A_{\text{H}} \left(\frac{t}{c^{2}} \right)^{-\frac{c^{2}}{2b}} \exp \left(-\frac{c^{2}}{2b} \right) &, t > c^{2}
	\end{dcases}
\end{align}
with
\begin{equation}
	\resizebox{\columnwidth}{!}{$
	A_{\text{H}} = \frac{\Gamma\left(\frac{r}{2}\right)}{\pi^{r/2}} \left((2b)^{\frac{r}{2}}\left(\Gamma\left(\frac{r}{2}\right) - \Gamma\left(\frac{r}{2}, \frac{c^{2}}{2b}\right)\right)+\frac{2 b c^{r} e^{-\frac{c^{2}}{2b} }}{c^2 - br}\right)^{-1}$}.
\end{equation}

\subsubsection{Tukey's Loss Function}

One of the most commonly used Non-ML loss functions is Tukey's. It is a redescending loss function because it redescends to zero, i.e., it gives zero weight to values larger than $c$. In \cite[11]{Zoubir.2018}, Tukey's loss function, for the univariate case, is given as
\begin{equation}
	\rho(x) = 
	\begin{dcases}
		\frac{x^{6}}{6 c^{4}} - \frac{x^{4}}{2 c^{2}} + \frac{x^{2}}{2}&, |x| \leq c\\
		\frac{c^{2}}{6} &, |x| > c,
	\end{dcases}
\end{equation}
which can be generalized to the multivariate case with $x^{2} = t_{n}$ and $|x| = \sqrt{t_{n}}$. By adding the constant $\frac{r}{2} \ln\left(2 \pi\right)$ it can be ensured that for $c \rightarrow \infty$, Tukey's loss function is equal to the Gaussian loss function. The resulting expression for Tukey's $\rho(t_{n})$ is given in Table~\ref{tb:g_rho}, while  $\psi(t_{n})$ and $\eta(t_{n})$ can be found in Table~\ref{tb:psi_eta}.

\begin{table*}[t]
	\renewcommand{\arraystretch}{1.3}
	\caption{Overview of $g(t_{n})$ and $\rho(t_{n})$ functions}
	\label{tb:g_rho}
	\centering
	{\tabulinesep=1.7mm
	\begin{tabu}{ccc}
		 & $g(t_{n})$ & $\rho(t_{n})$  \\\hline
		Gaussian
		& $(2 \pi)^{-\frac{r}{2}} \exp \left(-\frac{1}{2} t_{n}\right)$ 
		& $\frac{1}{2} t_{n} + \frac{r}{2} \ln\left(2 \pi\right)$ \\
		t
		& $\frac{\Gamma\left((\nu+r)/2\right)}{\Gamma\left(\nu/2 \right)(\pi \nu)^{r/2}} \left(1 + \frac{t_{n}}{\nu}\right)^{-(\nu + r)/2}$ 
		& $-\ln\left(\frac{\Gamma\left((\nu+r)/2\right)}{\Gamma\left(\nu/2 \right)(\pi \nu)^{r/2}}\right) + \frac{\nu + r}{2} \ln\left(1 + \frac{t_{n}}{\nu}\right)$ \\
		Huber
		& $\begin{dcases}
		A_{\text{H}} \exp \left(-\frac{t_{n}}{2b}\right) &, t_{n} \leq c^{2}\\
		A_{\text{H}} \left(\frac{t_{n}}{c^{2}} \right)^{-\frac{c^{2}}{2b}} \exp \left(-\frac{c^{2}}{2b} \right) &, t_{n} > c^{2}
		\end{dcases}$ 
		& $	\begin{dcases}
		-\ln\left(A_{\text{H}}\right) + \frac{t_{n}}{2b} &, t_{n} \leq c^{2}\\
		-\ln\left(A_{\text{H}}\right) + \frac{c^{2}}{2b} \left(\ln\left(\frac{t_{n}} {c^{2}}\right)+1\right) &, t_{n} > c^{2}
		\end{dcases}$ \\
		Tukey
		& n.a. 
		& $\begin{dcases}
		\frac{t_{n}^{3}}{6c^{4}} - \frac{t_{n}^{2}}{2c^{2}} + \frac{t_{n}}{2} + \frac{r}{2} \ln\left(2 \pi\right)&, t_{n} \leq c^{2}\\
		\frac{c^{2}}{6} + \frac{r}{2} \ln\left(2 \pi\right) &, t_{n} > c^{2}
		\end{dcases}$  \\\hline
	\end{tabu}}
\end{table*}

\begin{table}[t]
	\caption{Overview of $\psi(t_{n})$ and $\eta(t_{n})$ functions}
	\label{tb:psi_eta}
	\centering
	\resizebox{\columnwidth}{!}{
	{\tabulinesep=1.7mm
	\begin{tabu}{ccc}
		&  $\psi(t_{n})$ & $\eta(t_{n})$ \\\hline
		Gaussian
		& $\frac{1}{2}$ 
		& $0$ \\
		t
		& $\frac{1}{2} \cdot\frac{\nu + r}{\nu + t_{n}}$
		& $- \frac{1}{2} \cdot \frac{\nu + r}{(\nu + t_{n})^{2}}$
		 \\
		Huber
		& $	\begin{dcases}
		\frac{1}{2b} &, t_{n} \leq c^{2}\\
		\frac{c^{2}}{2bt_{n}} &, t_{n} > c^{2}
		\end{dcases}$ 
		& $	\begin{dcases}
		0 &, t_{n} \leq c^{2}\\
		-\frac{c^{2}}{2bt_{n}^{2}} &, t_{n} > c^{2}
		\end{dcases}$  \\
		Tukey
		& $\begin{dcases}
		\frac{t_{n}^{2}}{2c^{4}} - \frac{t_{n}}{c^{2}} + \frac{1}{2}&, t_{n} \leq c^{2}\\
		0 &, t_{n} > c^{2}
		\end{dcases}$ 
		& $\begin{dcases}
		\frac{t_{n}}{c^{4}} - \frac{1}{c^{2}} &, t_{n} \leq c^{2}\\
		0 &, t_{n} > c^{2}
		\end{dcases}$ \\\hline
	\end{tabu}}}
\end{table}

%% file: figures/review/psi_gaus_t_huber.tex
\begin{tikzpicture}
\begin{axis}[
	scaled ticks=false,
	tick label style={/pgf/number format/.cd},
	width = \figurewidth,
	height = \figureheight,
	xmin = 0,
	xmax = 30,
	ymax = 1,
	ymin = 0,
	grid,
	xlabel= $t$,
	ylabel= $\psi(t)$,
	legend entries={{Gauß}\\
		{t, $\nu = 3$}\\
		{$q_{H} = 0.8$}\\
		{$q_{H} = 0.99$}\\
		{Tukey}\\},
	legend pos=north east,
	legend cell align={left},
	]
	\addplot+[thick,matlabblue,no marks,line width=1.5pt] table[x index=0,y index=1,col sep=tab]  {figures/review/psi_gaus_t_huber.csv};
	\addplot+[thick,matlaborange,no marks,line width=1.5pt] table[x index=0,y index=2,col sep=tab] {figures/review/psi_gaus_t_huber.csv};
	\addplot+[thick,matlabyellow,no marks,line width=1.5pt] table[x index=0,y index=4,col sep=tab] {figures/review/psi_gaus_t_huber.csv};
	\addplot+[thick,matlabpurple,no marks,line width=1.5pt] table[x index=0,y index=5,col sep=tab] {figures/review/psi_gaus_t_huber.csv};
	\addplot+[thick,matlabgreen,no marks,line width=1.5pt] table[x index=0,y index=6,col sep=tab] {figures/review/psi_gaus_t_huber_tukey.csv};
	%
\end{axis}
\end{tikzpicture}

%% file: Paper/gen_clu_enum.tex
\section{Bayesian Cluster Enumeration for a General Distribution}
\label{sec:bic_g}
This section briefly revisits the BIC for cluster analysis, which formulates the cluster enumeration problem as a maximization of the posterior probability of candidate models \cite{Teklehaymanot.2018}. The general definition in \cite{Teklehaymanot.2018} provides the conceptual basis to derive the specific robust criteria that are presented in Section~\ref{sec:bic_res}. Following the definition and notation in \cite{Teklehaymanot.2018, Teklehaymanot.2019}, $\mathcal{X} = \{\bx_{1},\dots,\bx_{N}\}$ is the observed data set of length $N$. It can be partitioned into $K$ mutually exclusive subsets (clusters) $\{\mathcal{X}_{1},\dots,\mathcal{X}_{K}\}$, each cluster $\mathcal{X}_{k} \subseteq \mathcal{X}$, $k \in \mathcal{K} = \{1,\dots,K\}$ containing $N_{k} > 0$ observations of iid random variables $\bx_{k}\in \mathbb{R}^{r \times 1}$. 
The set of candidate models is defined as $\mathcal{M} = \{M_{L_{\min}},\dots,M_{L_{\max}}\}$, where each $M_{l} \in \mathcal{M}$ represents the partitioning of $\mathcal{X}$ into $l \in \{L_{\min},\dots,L_{\max}\}$, $l \in \mathbb{Z}^{+}$ subsets  $\mathcal{X}_{m}$, $m = 1,\dots,l$. The true number of subsets $K$ is assumed to lie within $L_{\min} \leq K \leq L_{\max} $. For each $M_{l}$, the parameters are stored in $\bTheta_{l} = [\btheta_{1},\dots,\btheta_{l}] \in \mathbb{R}^{q \times l}$, with $q$ being the number of parameters per cluster. Now \cite[p.~18]{Teklehaymanot.2019} derives a Bayesian criterion specifically for the cluster enumeration problem as
\begin{align}
\text{BIC}&_{\text{G}}(M_{l}) \triangleq \ln\left(p(M_{l}|\mathcal{X})\right) \notag \\
\begin{split}
\approx& \ln\left(p(M_{l})\right) + \ln\left(f\left(\hat{\bTheta}_{l}|M_{l}\right)\right) + \ln\left(\mathcal{L}\left(\hat{\bTheta}_{l}|\mathcal{X}\right)\right) 
\\&+ \frac{lq}{2}\ln(2\pi)  - \frac{1}{2} \sum_{m=1}^{l} \ln\left(\left|\bJhat_{m}\right|\right) - \ln(f(\mathcal{X})),
\end{split}
\end{align}
where $p(M_{l})$ is the discrete prior on the model $M_{l} \in \mathcal{M}$, 
\begin{equation}
f\left(\hat{\bTheta}_{l}|M_{l}\right) = \prod_{m=1}^{l} f\left(\bthetahat_{m}|M_{l}\right)
\end{equation}
is a prior on the parameter vectors in $\hat{\bTheta}_{l}$ given $M_{l}$, 
\begin{equation}
\mathcal{L}\left(\hat{\bTheta}_{l}|\mathcal{X}\right) = \prod_{m=1}^{l} \mathcal{L}\left(\bthetahat_{m}|\mathcal{X}_{m}\right)
\end{equation}
is the likelihood function,
\begin{equation}
\bJhat_{m} = - \frac{d^{2}\ln\left(\mathcal{L}\left(\bthetahat_{m}|\mathcal{X}_{m}\right)\right) }{d\bthetahat_{m} d \bthetahat_{m}^{\top}} \in \mathbb{R}^{q \times q}
\end{equation}
is the FIM, and $f(\mathcal{X})$ is the pdf of $\mathcal{X}$. We can further simplify the $\text{BIC}_{\text{G}}$ by assuming an equal prior and noting that $f(\mathcal{X})$ is model independent, hence we can remove both terms. Lastly, we can assume that each parameter vector is equally probable as follows
\begin{equation}
f\left(\hat{\bTheta}_{l}|M_{l}\right) = \prod_{m=1}^{l} f\left(\bthetahat_{m}|M_{l}\right) = \prod_{m=1}^{l} \frac{1}{l} = l^{-l}
\end{equation}
and finally
\begin{align}
\text{BIC}_{\text{G}}(M_{l}) \approx& \sum_{m=1}^{l} \ln\left( \mathcal{L}\left(\bthetahat_{m}|\mathcal{X}_{m}\right)\right) - l \ln\left(l\right) \notag \\*
&+ \frac{ql}{2}\ln(2\pi) - \frac{1}{2} \sum_{m=1}^{l} \ln\left(\left|\bJhat_{m}\right|\right).
\label{eqn:BICG}
\end{align}
The number of clusters can be estimated by evaluating
\begin{equation}
\hat{K} = \argmax_{l = L_{\min},\dots,L_{\max}}  \text{BIC}_{\text{G}}(M_{l}).
\label{eqn:K_hat}
\end{equation}

%% file: Paper/res_clu_enum.tex
\section{Proposed Bayesian Cluster Enumeration for RES Distributions and M-Estimation}
\label{sec:bic_res}
\subsection{Proposed Finite Sample Criterion}
Our first main result is stated in Theorem \ref{theo:1}. Based on Eq. \eqref{eqn:BICG}, we derive a BIC that can be used for any RES distribution and even for Non-ML loss functions such as Tukey's M-estimator. Firstly, the parameter vector is defined as $\bthetahat_{m} = \left[\bmuhat_{m}^{\top}, \vechop(\bShat_{m})^{\top}\right]^{\top} \in \mathbb{R}^{q \times 1}$, $q = \frac{r}{2}(r+3)$. Because $\bShat_{m}$ is symmetric, it has only $\frac{r}{2}(r+1)$ unique elements, therefore, $\vechop(\bShat_{m})$ has to be used \mbox{{\cite[p.~367]{Abadir.2005}}}. The $\vechop$ (vector half) operator takes a symmetric  $r \times r$ matrix and stacks the lower triangular half into a single vector of length $\frac{r}{2}(r+1)$.
\begin{theorem}
The posterior probability of $M_{l}$ given $\mathcal{X}$, based on any ML or Non-ML loss function $\rho(t)$, can be calculated by
\label{theo:1}
\end{theorem}
\begin{empheqboxed}
\begin{align}
	\begin{split}
		\text{BIC}&_{\text{F}}(M_{l}) 
		\\ \approx &-\sum_{m=1}^{l} \left(\sum_{\xinX} \rho(\that_{nm})\right) + \sum_{m=1}^{l} N_{m} \ln \left(N_{m}\right) - l \ln\left(l\right) 
		\\&- \sum_{m=1}^{l}\frac{N_{m}}{2} \ln\left(\left| \bShat_{m} \right|\right) + \frac{ql}{2}\ln(2\pi) - \frac{1}{2} \sum_{m=1}^{l} \ln\left(\left|\bJhat_{m}\right|\right)
	\end{split}
	\label{eqn:bic_RES}
\end{align}
\end{empheqboxed}
\noindent\textit{with $\left|\bJhat_{m}\right|$ given in Eq.~\eqref{eqn:detFIM}, using Eqs.~\eqref{eqn:FIM_mat}-\eqref{eqn:dmuShat}.}

Theorem~\ref{theo:1} is derived from Eq.~\eqref{eqn:BICG} by ignoring model independent terms in the log-likelihood function for an arbitrary RES distribution
\begin{align}
\ln(\mathcal{L}(\bthetahat_{m}|\mathcal{X}_{m} )) =& \ln\left(\prod_{\xinX} p(\xinX) f(\xn| \bthetahat_{m}) \right)\notag\\
=& \sum_{\xinX} \ln\left(\frac{N_{m}}{N} \left| \bShat_{m}^{-1} \right|^{\frac{1}{2}} g\left(\that_{nm}\right)\right)\notag\\
\begin{split}
=& -\sum_{\xinX} \rho(\that_{nm}) + N_{m} \ln \left(N_{m}\right) 
\\&- N_{m}\ln \left(N\right) - \frac{N_{m}}{2} \ln\left(\left| \bShat_{m}\right|\right),
\end{split}
\label{eqn:loglikelihood}
&\end{align}
and computing the FIM
\begin{equation}
\bJhat_{m} = \begin{bmatrix}
-\bFhat_{\bmu\bmu} & -\bFhat_{\bmu\bS}\\
-\bFhat_{\bS\bmu} & -\bFhat_{\bS\bS}
\end{bmatrix} \in \mathbb{R}^{q \times q}.
\label{eqn:FIM_mat}	
\end{equation}
All derivatives are evaluated with the ML estimates of $\bS_{m}$ and $\bmu_{m}$, respectively, $\bShat_{m}$ and $\bmuhat_{m}$. The proof of Theorem~\ref{theo:1} is provided in Appendix~\ref{ch:appDerivative}. Due to limited space, some detailed explanations are left out. A complete and comprehensive step-by-step derivation for all elements of the FIM in Eq.~\eqref{eqn:FIM_mat} is given in the online supplementary material. The final resulting expressions are as follows:
\begin{equation}
\begin{split}
\bFhat_{\bmu\bmu} =& - 4 \bShat_{m}^{-1} \left(\sum_{\xinX} \eta(\that_{nm})\bxhat_{n} \bxhat_{n}^{\top}\right) \bShat_{m}^{-1} 
\\&- 2 \bShat_{m}^{-1} \sum_{\xinX}\psi(\that_{nm}) \in \mathbb{R}^{r \times r},
\end{split}
\label{eqn:dmumuhat}
\end{equation}
\begin{align}
\begin{split}
\bFhat_{\bmu\bS} = & \bFhat_{\bS\bmu}^{\top} = -2 \sum_{\xinX} \eta(\that_{nm}) \times 
\\ &\left( \bShat_{m}^{-1} \bxhat_{n} \bxhat_{n}^{\top} \bShat_{m}^{-1} \otimes \bxhat_{n}^{\top}\bShat_{m}^{-1} \right)\bD_{r} \in \mathbb{R}^{r \times \frac{r}{2}(r+1)},
\end{split}
\label{eqn:dmuShat}
\end{align}
\begin{align}
\begin{split}
\bFhat&_{\bS\bS} = - \bD_{r}^{\top} \left( \bShat_{m}^{-1} \otimes \bShat_{m}^{-1} \right) \times
\\&\left(\sum_{\xinX} \eta(\that_{nm})\left( \bxhat_{n} \bxhat_{n}^{\top} \otimes \bxhat_{n} \bxhat_{n}^{\top}\right)\right) \left( \bShat_{m}^{-1} \otimes \bShat_{m}^{-1} \right)\bD_{r}
\\&- \frac{N_{m}}{2}\bD_{r}^{\top}\left( \bShat_{m}^{-1} \otimes \bShat_{m}^{-1} \right)\bD_{r} \in \mathbb{R}^{\frac{r}{2}(r+1) \times \frac{r}{2}(r+1)}.
\end{split}
\label{eqn:dSShat}
\end{align}
Here, $\bD_{r} \in \mathbb{R}^{r^{2} \times \frac{r}{2}(r+1)}$ is the duplication matrix, and $\bxhat_{n} \triangleq \xn - \bmuhat_{m}$. The FIM is a partitioned matrix \mbox{\cite[p.~114]{Abadir.2005}} and the determinant follows as
\begin{equation}
\left|\bJhat_{m}\right| = \left|-\bFhat_{\bmu\bmu}\right| \cdot \left|-\bFhat_{\bS\bS} + \bFhat_{\bS\bmu}\bFhat_{\bmu\bmu}^{-1} \bFhat_{\bmu\bS}\right|.
\label{eqn:detFIM}
\end{equation}

Based on \eqref{eqn:bic_RES}, the number of clusters can be estimated by evaluating
\begin{equation}
\hat{K} = \argmax_{l = L_{\min},\dots,L_{\max}}  \text{BIC}_{\text{F}}(M_{l}).
\label{eqn:K_hat_res}
\end{equation}

\subsection{Asymptotic Sample Penalty Term}
Our second main result is stated in Theorem~\ref{theo:2}. Because it can be numerically expensive to calculate the FIM, especially for large sample sizes as discussed in Section~\ref{sec:complex}, it may be advantageous to asymptotically approximate the FIM. 
\begin{theorem}
	Ignoring terms in Eq.~\eqref{eqn:detFIM} that do not grow as $N \rightarrow \infty$, the posterior probability of $M_{l}$ given $\mathcal{X}$ becomes
	\label{theo:2}
\end{theorem}
\begin{empheqboxed}
\begin{align}
	\begin{split}
	\text{BIC}&_{\text{A}}(M_{l}) 
	\\ \approx& -\sum_{m=1}^{l} \left(\sum_{\xinX} \rho(\that_{nm})\right) + \sum_{m=1}^{l} N_{m} \ln \left(N_{m}\right) 
	\\& - \sum_{m=1}^{l}\frac{N_{m}}{2} \ln\left(\left| \bShat_{m} \right|\right) - \frac{q}{2} \sum_{m=1}^{l} \ln\left(\varepsilon_{m}\right)
	\end{split}
	\label{eqn:bic_aRES}
\end{align}
\end{empheqboxed}
\noindent\textit{with $\varepsilon_{m}$ given in Eq.~\eqref{eqn:eps_max}.}

The scalar variable $\varepsilon_{m}$ is computed, such that
\begin{equation}
\left|\frac{1}{\varepsilon_{m}} \bJhat_{m}\right| = \text{const},
\label{eqn:constJ}
\end{equation}
leads to a term that does not grow as $N \rightarrow \infty$. From Eqs.~\eqref{eqn:dmumuhat}, \eqref{eqn:dmuShat} and \eqref{eqn:dSShat} we can extract three normalization factors to fulfill Eq.~\eqref{eqn:constJ} the maximum must be taken, which yields
\begin{equation}
\varepsilon_{m} = \max \left(\left|\sum_{\xinX} \psi(\that_{nm})\right|, \left|\sum_{\xinX} \eta(\that_{nm})\right|, N_{m}\right).
\label{eqn:eps_max}
\end{equation}
Based on \eqref{eqn:bic_aRES}, the number of clusters can be estimated by evaluating
\begin{equation}
\hat{K} = \argmax_{l = L_{\min},\dots,L_{\max}}  \text{BIC}_{\text{A}}(M_{l}).
\label{eqn:K_hat_ares}
\end{equation}

\subsection{Discussion of the Theoretical Results}
The derived BIC are composed of two parts, the likelihood term (or more generally the measure of data fit term) and the model complexity penalty term. The function $\rho(t)$ plays a key role to introduce robustness into the measure of data fit by downweighting outlying data points, as illustrated in Figure~\ref{fig:psi_gaus_t_huber}. In contrast to simply using robust estimates in the popular Schwarz BIC, the proposed criteria also introduce robustness into the penalty term, either via the FIM in the case of the $\text{BIC}_{\text{F}}$, or via the asymptotic approximation $\varepsilon_{m}$ for the $\text{BIC}_{\text{A}}$. This is because the criteria are derived specifically for cluster enumeration based on the theoretical framework of \cite{Teklehaymanot.2018} and, therefore, the penalty term carries structural information about the clustering problem. Compared to existing criteria that have been derived for the Gaussian \cite{Teklehaymanot.2018, Teklehaymanot.2018d} and the t distribution \cite{Teklehaymanot.2018b}, the two derived BIC generalize Bayesian cluster enumeration by decoupling the loss-function $\rho(t)$ in Eq.~\eqref{eqn:loglikelihood} from a specific distribution. Based on Theorems~\ref{theo:1} and \ref{theo:2}, it is possible to compute robust \mbox{M-estimation} type cluster enumeration, and apply a broad set of available loss-functions that have been designed based on robustness considerations rather than optimality under a specific distributional model. This includes non-ML loss functions such as Tukey's, which is a popular member of the redescending class of M-estimators. A further advantage compared to \cite{Teklehaymanot.2018, Teklehaymanot.2018d, Teklehaymanot.2018b} is that the decoupling allows us to easily change the weight function in the BIC, without having to recalculate the likelihood function, FIM or penalty term. We will later show in the simulation section that these generalizations are beneficial in terms of introducing robustness against outliers into the calculation of the BIC. We will also evaluate the advantages and disadvantages of the two criteria based on performance, robustness and computational complexity.

%% file: Paper/clu_enum_alg.tex
\section{Proposed Robust Cluster Enumeration Algorithm}
\label{sec:algo}
To evaluate the BIC, our approach requires a robust clustering algorithm to partition the data according to the number of clusters specified by each candidate model and to compute the associated parameter estimates. Accordingly, we will derive an expectation maximization (EM) algorithm for RES distributions in Section~\ref{sec:em}. The resulting two-step approach is summarized in Algorithm \ref{alg:ce}, where we provide a unified framework for the robust estimation of the number of clusters and cluster memberships.

{
	\centering
	\removelatexerror
	\begin{algorithm}[!htb]
		\KwIn{$\mathcal{X}$, $L_{\min}$, $L_{\max}$}
		\KwOut{$\hat{K}$}
		\For{$l = L_{\min},\dots,L_{\max}$}
		{%
			Compute parameter estimates using Algorithm \ref{alg:em}.\;
			
			Hard clustering:\;
			\For{$m = 1,\dots,l$}
			{
				\For{$n = 1,\dots,N$}
				{
					$\gamma_{nm} = \begin{dcases}
						1 &, m = \argmax_{j = 1,\dots,l} \hat{v}_{nj}^{(i)}\\
						0 &, \text{else}
					\end{dcases}$
				}
			}
			\For{$m = 1,\dots,l$}
			{
				$N_{m} = \sum_{n = 1}^{N} \gamma_{nm}$
			}
			Calculate $\text{BIC}(M_{l})$ according to \eqref{eqn:bic_RES} or \eqref{eqn:bic_aRES}.\;
		}
		Estimate $\hat{K}$ with Eq.~\eqref{eqn:K_hat_res} or \eqref{eqn:K_hat_ares}. \;
		\caption{Proposed enumeration algorithm.}
		\label{alg:ce}
	\end{algorithm}
}

\subsection{Expectation Maximization (EM) Algorithm for a Mixture of RES Distributions}
\label{sec:em}
This section describes the EM algorithm that is used to find ML estimates of the RES mixture model parameters \cite{Dempster.1977, Teklehaymanot.2019, Bishop.2009}, and the cluster memberships of the data vectors $\xn$, which are latent variables. For a mixture of $l$ RES distributions,  the log-likelihood function is given by 
\begin{align}
\ln\left(\mathcal{L}(\bPhi_{l}|\mathcal{X} )\right) =& \sum_{n = 1}^{N} \ln \left(\sum_{m=1}^{l} \gamma_{m} \left| \bS_{m} \right|^{-\frac{1}{2}} g\left( t_{nm}\right)\right)
\label{eqn:emml}
\end{align}
with $\gamma_{m}$ being the mixing coefficient, $\bS_{m}$ the scatter matrix, $g\left( t_{nm}\right)$ the density generator, and  $\bPhi_{l} = [\bgamma_{l}, \bTheta_{l}^{\top}]$ with $ \bgamma_{l} = [\gamma_{1},\dots, \gamma_{l}]^{\top}$. Using the matrix calculus rules from \cite{Magnus.2007, Abadir.2005}, we define $\bF$ as a $1 \times 1$ scalar function of the $r \times 1$ vector $\bmu_{m}$. Hence, the resulting Jacobian matrix is of size $1 \times r$. Setting $\bF$ equal to \eqref{eqn:emml}
\begin{align}
\bF(\bmu_{m}) = \sum_{n = 1}^{N} \ln \left(\sum_{m=1}^{l} \gamma_{m} \left| \bS_{m} \right|^{-\frac{1}{2}} g\left( t_{nm}\right)\right)
\end{align}
and applying the differential
\begin{align}
\text{d}\bF&(\bmu_{m}) \notag \\
=& \sum_{n = 1}^{N} \text{d}\ln \left(\sum_{m=1}^{l} \gamma_{m} \left| \bS_{m}^{-1} \right|^{\frac{1}{2}} g\left( t_{nm}\right)\right)\notag \\
=& -\sum_{n = 1}^{N} \frac{\gamma_{m} \left| \bS_{m}^{-1} \right|^{\frac{1}{2}} g'\left( t_{nm}\right)  2 \left(\xn - \bmu_{m}\right)^{\top} \bS_{m}^{-1}\text{d}\bmu_{m} }{ \sum_{j=1}^{l} \gamma_{j} \left| \bS_{j}^{-1} \right|^{\frac{1}{2}} g\left( t_{nj}\right)} 
\end{align}
the Jacobain matrix follows as
\begin{align}
\text{D}\bF(\bmu_{m}) =& \sum_{n = 1}^{N} v_{nm} \psi(t_{nm}) 2 \left(\xn - \bmu_{m}\right)^{\top} \bS_{m}^{-1} 
\label{eqn:emdmu}
\end{align}
with
\begin{equation}
g'\left( t_{nm}\right) = - \psi\left(t_{nm}\right) g\left( t_{nm}\right)
\label{eqn:gprime}
\end{equation}
and
\begin{equation}
v_{nm} = \frac{\gamma_{m} \left| \bS_{m}^{-1} \right|^{\frac{1}{2}} g\left( t_{nm}\right)}{\sum_{j=1}^{l} \gamma_{j} \left| \bS_{j}^{-1} \right|^{\frac{1}{2}} g\left( t_{nj}\right)}.
\end{equation}
The ML estimate can be calculated by setting \eqref{eqn:emdmu} equal to zero which yields
\begin{align}
& \bmuhat_{m} = \frac{\sum_{n = 1}^{N} v_{nm} \psi\left(\that_{nm}\right) \xn}{\sum_{n = 1}^{N} v_{nm} \psi\left(\that_{nm}\right)}.
\end{align}

Now $\bF$ is defined as a $1 \times 1$ scalar function of the $r \times r$ matrix $\bS_{m}$. Hence, the resulting Jacobian matrix is of size $1 \times r^2$. Setting $\bF$ equal to \eqref{eqn:emml} and applying the differential
\begin{align}
\text{d}\bF&(\bS_{m}) \notag \\*
=&  \sum_{n = 1}^{N} \text{d}\ln \left(\sum_{m=1}^{l} \gamma_{m} \left| \bS_{m}^{-1} \right|^{\frac{1}{2}} g\left( t_{nm}\right)\right) \notag \\
\begin{split}
=&  \sum_{n = 1}^{N} \frac{\gamma_{m}}{\sum_{j=1}^{l} \gamma_{j} \left| \bS_{j}^{-1} \right|^{\frac{1}{2}} g\left( t_{nj}\right)} \Bigl[\text{d}\left(\left| \bS_{m} \right|^{-\frac{1}{2}}\right) g\left( t_{nm}\right) 
\\&+ \left| \bS_{m}^{-1} \right|^{\frac{1}{2}} \text{d}g\left( t_{nm}\right)\Bigr] 
\end{split}\notag \\
\begin{split}
=&  \sum_{n = 1}^{N} \frac{\gamma_{m}}{\sum_{j=1}^{l} \gamma_{j} \left| \bS_{j}^{-1} \right|^{\frac{1}{2}} g\left( t_{nj}\right)}\times   
\\& \Biggl[-\frac{1}{2} \left| \bS_{m} \right|^{-\frac{1}{2} - 1}  \left| \bS_{m} \right| \Tr\left( \bS_{m}^{-1} \text{d} \bS_{m} \right) g\left( t_{nm}\right)
\\& - \left| \bS_{m}^{-1} \right|^{\frac{1}{2}} g'\left( t_{nm}\right) \bxt_{n}^{\top} \bS_{m}^{-1} \text{d} \bS_{m} \bS_{m}^{-1} \bxt_{n} \Biggr]
\end{split} \notag \\
\begin{split}
=&  \sum_{n = 1}^{N} \Bigl[-\frac{v_{nm}}{2} \Tr\left( \bS_{m}^{-1} \text{d} \bS_{m} \right)  
\\&+ v_{nm} \psi\left(t_{nm}\right) \bxt_{n}^{\top} \bS_{m}^{-1} \text{d} \bS_{m} \bS_{m}^{-1} \bxt_{n} \Bigr]
\end{split}
\end{align}
with $\bxt_{n} = \bx_{n} - \bmu_{m}$, followed by the vectorization
\begin{align}
\text{d}\vecop&\left(\bF(\bS_{m})\right)  \notag \\
\begin{split}
=& \sum_{n = 1}^{N} \Bigl[-\frac{v_{nm}}{2} \Tr\left( \bS_{m}^{-1} \text{d} \bS_{m} \right) 
\\& + v_{nm} \psi\left(t_{nm}\right) \vecop\left(\bxt_{n}^{\top} \bS_{m}^{-1} \text{d} \bS_{m} \bS_{m}^{-1} \bxt_{n}\right) \Bigr] 
\end{split} \notag \\
\begin{split}
=& \sum_{n = 1}^{N} \Bigl[-\frac{v_{nm}}{2}  \vecop\left(\bS_{m}^{-1}\right)^{\top}   
\\&+ v_{nm} \psi\left(t_{nm}\right) \left(\bxt_{n}^{\top} \bS_{m}^{-1} \otimes \bxt_{n}^{\top} \bS_{m}^{-1} \right)  \Bigr] \text{d}\vecop\left(\bS_{m}\right)
\end{split} 
\end{align}
leads to the Jacobian matrix
\begin{align}
\begin{split}
\text{D}\bF(\bS_{m}) =& \sum_{n = 1}^{N} \Bigl[v_{nm} \psi\left(t_{nm}\right) \left(\bxt_{n}^{\top} \bS_{m}^{-1} \otimes \bxt_{n}^{\top} \bS_{m}^{-1} \right) 
\\&- \frac{v_{nm}}{2}  \vecop \left(\bS_{m}^{-1} \right)^{\top} \Bigr].
\end{split}
\label{eqn:emdS}
\end{align}
The ML estimate can be calculated by setting \eqref{eqn:emdS} equal to zero which yields
\begin{align}
\begin{split}
\Rightarrow &\sum_{n = 1}^{N} v_{nm} \psi\left(t_{nm}\right) \left(\bxt_{n}^{\top} \otimes \bxt_{n}^{\top} \right) \left(\bS_{m}^{-1} \otimes \bS_{m}^{-1} \right) 
\\&= \sum_{n = 1}^{N} \frac{v_{nm}}{2}  \vecop \left(\bS_{m}^{-1} \right)^{\top}
\end{split} \notag \\
\begin{split}
\Rightarrow &\sum_{n = 1}^{N} \frac{v_{nm}}{2} \vecop \left(\bS_{m}^{-1} \right)^{\top} \left(\bS_{m} \otimes \bS_{m} \right) 
\\&= \sum_{n = 1}^{N} v_{nm} \psi\left(t_{nm}\right) \left(\bxt_{n}^{\top} \otimes \bxt_{n}^{\top} \right) 
\end{split} \notag \\
\Rightarrow & \vecop \left(\bShat_{m} \right) = \frac{2\sum_{n = 1}^{N} v_{nm} \psi\left(t_{nm}\right) \left(\bxhat_{n}\otimes \bxhat_{n}\right)}{\sum_{n = 1}^{N} v_{nm}} \notag \\
\Rightarrow & \bShat_{m} = \frac{2\sum_{n = 1}^{N} v_{nm} \psi\left(\that_{nm}\right) \left(\xn - \bmuhat_{m}\right) \left(\xn - \bmuhat_{m}\right)^{\top} }{\sum_{n = 1}^{N} v_{nm}}.
\end{align}
Finally, we have to maximize with regard to the mixing coefficients $\gamma_{m}$. A Lagrange multiplier is used to fulfill the constraint
\begin{equation}
\sum_{m=1}^{l}\gamma_{m} = 1,
\end{equation}
which yields
\begin{align}
\begin{split}
\text{d}\bF(\gamma_{m}) =& \sum_{n = 1}^{N} \text{d}\ln \left(\sum_{m=1}^{l} \gamma_{m} \left| \bS_{m}^{-1} \right|^{\frac{1}{2}} g\left( t_{nm}\right)\right) 
\\&+ \lambda \text{d}\left(\sum_{m=1}^{l}\gamma_{m} - 1\right) 
\end{split} \notag \\
=& \sum_{n = 1}^{N} \frac{\left| \bS_{m}^{-1} \right|^{\frac{1}{2}} g\left( t_{nm}\right)}{\sum_{j=1}^{l} \gamma_{j} \left| \bS_{j}^{-1} \right|^{\frac{1}{2}} g\left( t_{nj}\right)} + \lambda.
\end{align}
First, we solve for $\lambda$, which leads to 
\begin{align}
\Rightarrow 0 =& \sum_{n = 1}^{N} \sum_{m=1}^{l} \frac{\gamma_{m} \left| \bS_{m}^{-1} \right|^{\frac{1}{2}} g\left( t_{nm}\right)}{\sum_{j=1}^{l} \gamma_{j} \left| \bS_{j}^{-1} \right|^{\frac{1}{2}} g\left( t_{nj}\right)} + \lambda  \sum_{m=1}^{l}\gamma_{m} \notag \\
\Rightarrow \lambda =& -N
\end{align}
and after the elimination of $\lambda$ we obtain
\begin{align}
\hat{\gamma}_{m} =& \frac{1}{N} \sum_{n = 1}^{N} v_{nm}.
\end{align}
The resulting iterative EM algorithm to compute these parameters is summarized in Algorithm~\ref{alg:em}.

\subsection{Computational Complexity}
\label{sec:complex}
For one candidate model $M_{l}$, the computational complexity of the first step, namely the EM algorithm, is given by $\mathcal{O}(Nr^{3}li_{\max})$, with $i_{\max}$ denoting the maximum number of iterations of the EM algorithm. In the complexity analysis, we can neglect the initialization of the EM algorithm with a K-means/K-medians, which is only performed for a few iterations. The asymptotic and Schwarz BIC are of the same complexity as the EM algorithm, since they use the Mahalanobis distance that has already been computed in the EM, and the complexity of evaluating the penalty term can be neglected, in comparison. Hence, the overall complexity of our algorithm with the asymptotic penalty and that of the Schwarz BIC is $\mathcal{O}(Nr^{3} (L_{\min} + \dots + L_{\max})i_{\max})$. By contrast, using the finite BIC significantly increases the runtime of the algorithm, because it requires the computation of multiple Kronecker products and the determinant of the FIM. The complexity of the calculation of the determinant of the FIM can be reduced to $\mathcal{O}(r^{3})$, by exploiting the block matrix structure of the FIM. Hence, it can be neglected compared to the Kronecker products in $\bFhat_{\bS\bS}$, which have a complexity of $\mathcal{O}(N r^{4})$ and, therefore, dominate the runtime, especially for increasing values of $r$.  A detailed experimental runtime analysis that confirms this theoretical analysis of the computational complexity can be found in Sections~\ref{sec:run_N} and \ref{sec:run_r}.

%% file: Paper/sim.tex
\section{Experimental Results}
\label{sec:experiments}
\begin{figure}[t]
	\centering
	\subfloat[single replacement outlier]{\resizebox{0.48\columnwidth}{!}{\input{figures/data_31_Nk_250_eps_00013333.tex}}%
		\label{fig:data_31_1_out}}
	\hfil
	\subfloat[$\epsilon = 10\%$ replacement outliers]{\resizebox{0.48\columnwidth}{!}{\input{figures/data_31_Nk_250_eps_01.tex}}%
		\label{fig:data_31_0.1_out}}
	\caption{Two exemplary realizations of the data set.}
	\label{fig:data_31}
\end{figure}
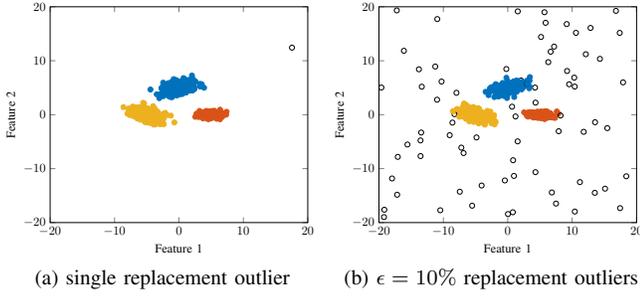

The proposed cluster enumeration framework allows for a variety of possible algorithms which include the recently proposed cluster enumeration criteria for the Gaussian distribution \cite{Teklehaymanot.2018}, \cite{Teklehaymanot.2018d}  and for the t-distribution \cite{Teklehaymanot.2018b}, as special cases. Further, as a benchmark comparison, Schwarz penalty can be combined with the robust data fit, as provided by the EM algorithm. The code that implements our proposed two-step algorithm for robust Bayesian cluster enumeration is available at: \url{https://github.com/schrchr/Robust-Cluster-Enumeration}

We use the same simulated data as in \cite{Teklehaymanot.2018, Teklehaymanot.2018b, Teklehaymanot.2018d}, to be able to compare the results. Our results can therefore be compared to the Robust Trimmed BIC \cite{Neykov.2007} and the Robust Gravitational Clustering Method \cite{Binder.2018}. The simulated data set is defined by $\bx_{k} \sim \mathcal{N}\left(\bmu_{k}, \bSigma_{k}\right)$, $k = 1,2,3$, with the cluster centroids $\bmu_{1} = [0, 5]^{\top}$, $\bmu_{2} = [5, 0]^{\top}$ and $\bmu_{3} = [-5, 0]^{\top}$ and the covariance matrices
\begin{equation}
\resizebox{\columnwidth}{!}{
$\bSigma_{1} = \begin{bmatrix}
2 & 0.5\\
0.5 & 0.5\end{bmatrix}, 
\bSigma_{2} = \begin{bmatrix}
1 & 0\\
0 & 0.1\end{bmatrix},
\bSigma_{3} = \begin{bmatrix}
2 & -0.5\\
-0.5 & 0.5\end{bmatrix}.$}\notag
\end{equation}
Every cluster has $N_{k}$ data points and the outliers are replacement outliers where $\epsilon$ is the percentage of replaced data points. These replacements are uniformly distributed in the range of $[-20, 20]$ in each dimension. Two exemplary realizations with different values of $\epsilon$ are shown in Figure~\ref{fig:data_31}. 

\subsection{Choice of Tuning Parameters}
\label{sec:tuning}
For the Huber distribution, \cite[p.~116]{Zoubir.2018} suggest to choose $c^2$ as the $q_{H}^{\text{th}}$ upper quantile of a $\chi_{r}^{2}$ distribution, i.e.
\begin{equation}
	c^{2} = F_{\chi_{r}^{2}}^{-1}\left(q_{H}\right), \quad 0 < q_{H} < 1.
	\label{eqn:qh}
\end{equation}
In \cite[121]{Zoubir.2018}, a value of $q_{H} = 0.8$ is used, which leads to $c = 1.282$. From \cite[23]{Zoubir.2018}, we have the value $c = 1.345$, which will achieve an asymptotic relative efficiency (ARE) of $95\%$. Since both values are quite similar, there should not be a large performance difference and we choose to use $q_{H} = 0.8$ in all simulations. For Tukey's loss function we will use $c = 4.685$, according to \cite[23]{Zoubir.2018}. 

\subsection{Results of Numerical Simulations}
\subsubsection{Analysis of the sensitivity of the cluster enumeration algorithms to a single outlier}
\begin{figure}[t]
	\centering
	\subfloat[EM: Gauß, $\text{BIC}_{\text{F}}$: Gauß]
	{\resizebox{0.48\columnwidth}{!}{\input{figures/paper_sens_EM_Gaus_BIC_Gaus_RES_Nk_50_step_2_MC_500.tex}}%
		\label{fig:sens_Gaus_RES_10}}
	\hfil
	\subfloat[EM: Gauß, $\text{BIC}_{\text{A}}$: Gauß]
	{\resizebox{0.48\columnwidth}{!}{\input{figures/paper_sens_EM_Gaus_BIC_Gaus_aRES_Nk_50_step_2_MC_500.tex}}%
		\label{fig:sens_Gaus_aRES_10}} \\
	\subfloat[EM: Huber, $\text{BIC}_{\text{F}}$: Huber]
	{\resizebox{0.48\columnwidth}{!}{\input{figures/paper_sens_EM_Huber_BIC_Huber_RES_Nk_50_step_2_MC_500.tex}}%
		\label{fig:sens_Huber_RES_10}}
	\hfil
	\subfloat[EM: Huber, $\text{BIC}_{\text{A}}$: Huber]
	{\resizebox{0.48\columnwidth}{!}{\input{figures/paper_sens_EM_Huber_BIC_Huber_aRES_Nk_50_step_2_MC_500.tex}}%
		\label{fig:sens_Huber_aRES_10}} \\
	\subfloat[EM: Huber, $\text{BIC}_{\text{F}}$: Tukey]
	{\resizebox{0.48\columnwidth}{!}{\input{figures/paper_sens_EM_Huber_BIC_Tukey_RES_Nk_50_step_2_MC_500.tex}}%
		\label{fig:sens_Huber_tuk_RES_10}} 
	\hfil
	\subfloat[EM: Huber, $\text{BIC}_{\text{A}}$: Tukey]
	{\resizebox{0.48\columnwidth}{!}{\input{figures/paper_sens_EM_Huber_BIC_Tukey_aRES_Nk_50_step_2_MC_500.tex}}%
		\label{fig:sens_Huber_tuk_aRES_10}}
	\caption{Sensitivity curves for $N_{k} = 50$ that show six exemplary results for the empirical probability of correctly deciding for $K=3$ clusters as a function of the single replacement outlier position. }
	\label{fig:sensitivity}
\end{figure}
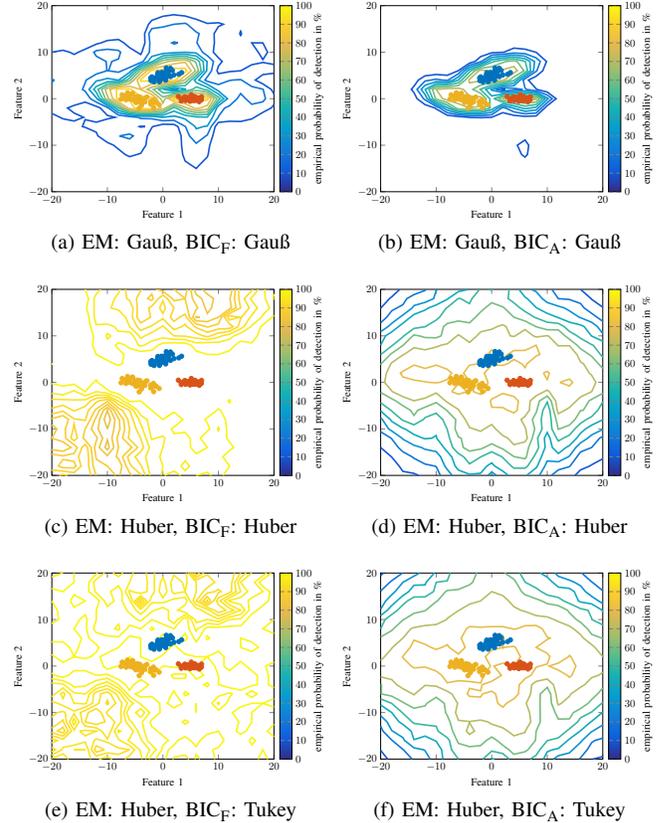

To evaluate the sensitivity of the proposed cluster enumeration algorithm to the position of a single replacement outlier, we computed the sensitivity curves, which display the average performance over 500 Monte Carlo iterations with $N_{k} = 50$. Here, we replaced a randomly selected data point with an outlier that takes values over the range $[-20; 20]$ on each variate at each iteration. In Figure~\ref{fig:sensitivity}, six exemplary results for the resulting empirical probability of correctly deciding for $K=3$ clusters are shown as a function of the outlier position. The first row is based on the $\text{BIC}_{\text{F}}$ and the second row on the $\text{BIC}_{\text{A}}$. Due to the relatively small sample size, $\text{BIC}_{\text{F}}$ clearly performs better than $\text{BIC}_{\text{A}}$ for all shown loss functions. As expected, the Gaussian loss function is not robust against outliers and it only has a very small area with a high probability of detection. A Huber and Tukey based loss function increases the probability of detection significantly. The difference between the two robust loss functions is less prominent, but when comparing Figures~\ref{fig:sens_Huber_RES_10} and \ref{fig:sens_Huber_aRES_10} with Figures~\ref{fig:sens_Huber_tuk_RES_10} and \ref{fig:sens_Huber_tuk_aRES_10} one can observe a higher probability of detection for Tukey's loss function because it completely rejects large outliers.

\subsubsection{Breakdown point analysis}
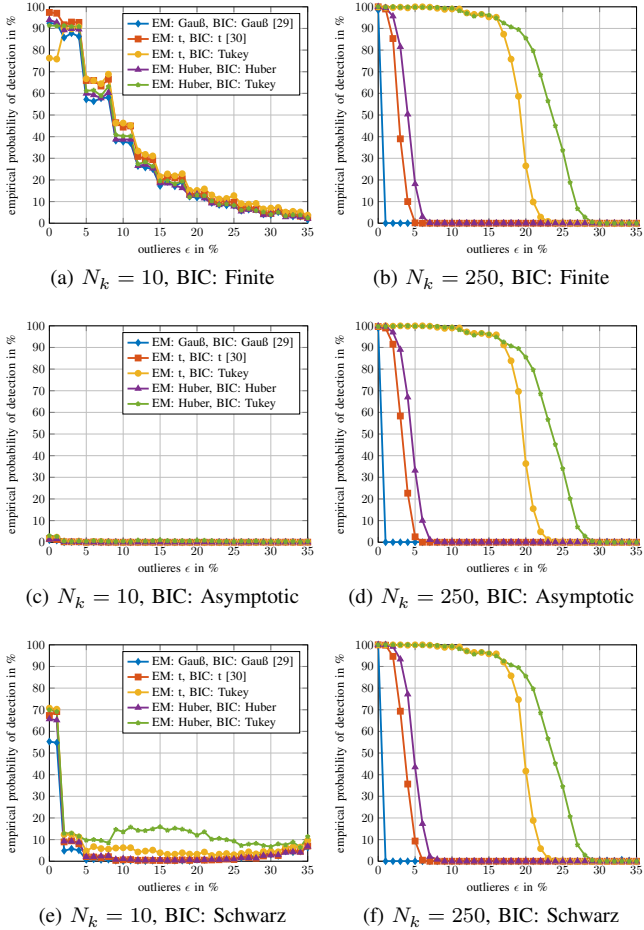
\begin{figure}[t]
	\centering
	\subfloat[$N_{k} = 10$, BIC: Finite] {\resizebox{0.48\columnwidth}{!}{\input{figures/outliers_BIC-RES_MC_2000_Nk_10.tex}}%
		\label{fig:breakdown_RES_10}}
	\hfil
	\subfloat[$N_{k} = 250$, BIC: Finite] {\resizebox{0.48\columnwidth}{!}{\input{figures/outliers_BIC-RES_MC_600_Nk_250.tex}}%
		\label{fig:breakdown_RES_250}} \\
	\subfloat[$N_{k} = 10$, BIC: Asymptotic] {\resizebox{0.48\columnwidth}{!}{\input{figures/outliers_BIC-aRES_MC_2000_Nk_10.tex}}%
	\label{fig:breakdown_aRES_10}}
	\hfil
	\subfloat[$N_{k} = 250$, BIC: Asymptotic] {\resizebox{0.48\columnwidth}{!}{\input{figures/outliers_BIC-aRES_MC_600_Nk_250.tex}}%
		\label{fig:breakdown_aRES_250}} \\
	\subfloat[$N_{k} = 10$, BIC: Schwarz] {\resizebox{0.48\columnwidth}{!}{\input{figures/outliers_BIC-Schwarz_MC_2000_Nk_10.tex}}%
	\label{fig:breakdown_Schwarz_10}}
	\hfil
	\subfloat[$N_{k} = 250$, BIC: Schwarz] {\resizebox{0.48\columnwidth}{!}{\input{figures/outliers_BIC-Schwarz_MC_600_Nk_250.tex}}%
	\label{fig:breakdown_Schwarz_250}}
	\caption{Breakdown point simulation for two different samples per cluster $N_{k}$.}
	\label{fig:breakdown}
\end{figure}

Figure~\ref{fig:breakdown} shows the robustness against a fraction of replacement outliers, where the contaminating distribution is a uniform distribution in the interval $[20,20]$ for each outlier variate in each Monte Carlo iteration. The uniform distribution is chosen so that the outliers do not form a cluster, which would lead to an ambiguity in the cluster enumeration results for larger amounts of outliers. The first row of plots in Figure~\ref{fig:breakdown} represents the results for a cluster size of $N_k = 10$ while $N_k = 250$ for the second row. We can observe two different behaviors based on the number of samples. Firstly, for $N_k = 10$, the results are similar for the same penalty term. Hence, in Figure~\ref{fig:breakdown_RES_10} the finite based BIC is able to perform quite well for all applied distributions. In contrast, Figure~\ref{fig:breakdown_aRES_10} shows that the asymptotic BIC is not able to detect anything and the Schwarz based BIC in Figure \ref{fig:breakdown_Schwarz_10} also does not perform well. In the second row, a different observation can be made. In Figures~\ref{fig:breakdown_RES_250}, \ref{fig:breakdown_aRES_250} and \ref{fig:breakdown_Schwarz_250} the best performing combination is always observed for a similar loss function combination. The EM with a Huber distribution and Tukey BIC, followed by an EM with t distribution and Tukey BIC always have the best performance. This effect can be explained by the actual values of the likelihood and the penalty term of the BIC. For $N_k = 10$ the values of the likelihood and penalty term are in the same magnitude, whereas for $N_k = 250$ the values of the likelihood and penalty term are one to two magnitudes apart. Hence, for low sample sizes, the penalty term has a large influence and for large sample sizes, the penalty term has almost no influence.

\subsubsection{Analysis of the behavior of the different BICs for increasing sample numbers}
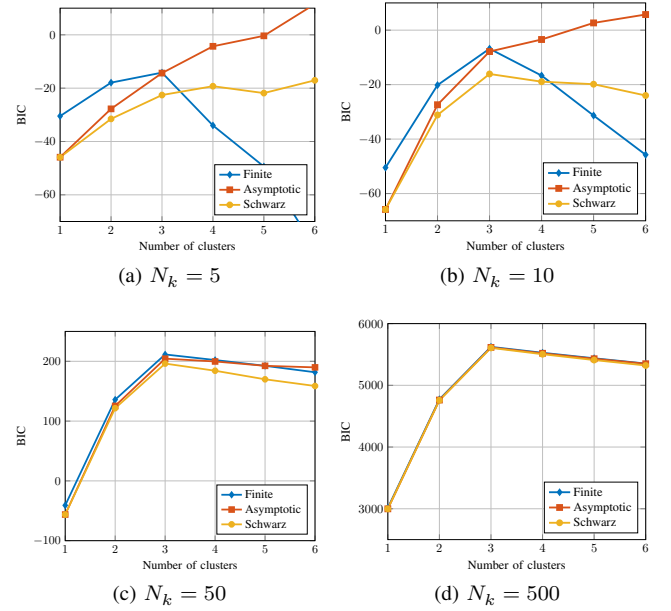
\begin{figure}[t]
	\centering
	\subfloat[$N_{k} = 5$] {\resizebox{0.47\columnwidth}{!}{\input{figures/review/bic_Nk_5_eps_0_EM_Huber_BIC_Huber.tex}}%
		\label{fig:bic_N_5}}
	\hfil
	\subfloat[$N_{k} = 10$] {\resizebox{0.48\columnwidth}{!}{\input{figures/review/bic_Nk_10_eps_0_EM_Huber_BIC_Huber.tex}}%
		\label{fig:bic_N_10}} \\
	\subfloat[$N_{k} = 50$] {\resizebox{0.47\columnwidth}{!}{\input{figures/review/bic_Nk_50_eps_0_EM_Huber_BIC_Huber.tex}}%
		\label{fig:bic_N_50}}
	\hfil
	\subfloat[$N_{k} = 500$] {\resizebox{0.48\columnwidth}{!}{\input{figures/review/bic_Nk_500_eps_0_EM_Huber_BIC_Huber.tex}}%
		\label{fig:bic_N_500}} 
	\caption{EM: Huber, BIC: Huber with $\epsilon = 0\%$ over different amount of samples per cluster. For increasing number of samples per cluster all three BICs converge to the same values.}
	\label{fig:data_bic_N}
\end{figure}

In Figure~\ref{fig:data_bic_N}, we show the behavior of the three BICs as a function of the sample size, exemplarily using EM: Huber, BIC: Huber, for a two-dimensional feature space with three clusters. It can be noted that all BICs converge to the same values if the sample size becomes sufficiently large. In this example, $N_{k} = 500$ is sufficient and, at this point, the Schwarz BIC, or the asymptotic BIC are preferred over the finite BIC, due to their lower computational complexity. While it generally holds that the difference between the asymptotic and the finite criterion will disappear asymptotically (and be insignificant beyond some sample size), the exact value for which the asymptotic criterion will begin to work depends on the specific example. More precisely, the minimum sample size required for the asymptotic BIC will depend on a combination of the following factors: the number of clusters, outliers, distance between clusters, cluster overlap, cluster shape, cluster imbalance and dimensionality. For example, a high-dimensional feature space will be less densely populated for the same value of $N_{k}$. Therefore, a larger sample size is required for the asymptotic BIC to work, given that all other parameters remain unchanged.

\subsubsection{Analysis of the behavior of the different BICs for cluster imbalance}
\begin{figure}[t]
	\centering
	\subfloat[$N_{1} = 5$, $N_{2} = 25$, $N_{3} = 100$] {\resizebox{0.47\columnwidth}{!}{\input{figures/review/bic_imb_5_eps_0_EM_Huber_BIC_Huber.tex}}%
		\label{fig:bic_N_5_imb}}
	\hfil
	\subfloat[$N_{1} = 50$, $N_{2} = 250$, $N_{3}~=~1000$] {\resizebox{0.47\columnwidth}{!}{\input{figures/review/bic_imb_50_eps_0_EM_Huber_BIC_Huber.tex}}%
		\label{fig:bic_N_50_imb}}
	\hfil
	\caption{EM: Huber, BIC: Huber with $\epsilon = 0\%$ over different amount of samples per cluster and heavily imbalanced clusters. For increasing number of samples per cluster all three BICs  converge to the same values.}
	\label{fig:data_bic_N_imb}
\end{figure}
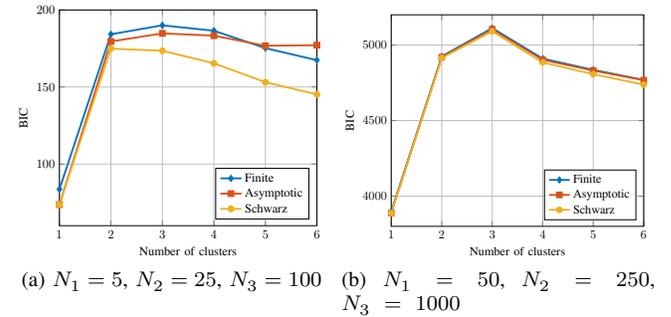

Compared to the popular Schwarz BIC, the derived asymptotic BIC is advantageous in some situations, for example, when we have cluster imbalance, i.e., when $N_{k}$ is very different for differing values of $k$. An example is given in  Figure~\ref{fig:data_bic_N_imb}, where we show a simulation with heavily imbalanced clusters. Here, the asymptotic BIC outperforms the Schwarz BIC, because it is able to incorporate information about the number of feature vectors per cluster into its penalty term. For asymptotic (or very large sample sizes) the BICs, again, converge to the same values.

\subsubsection{Runtime analysis of the proposed cluster enumeration algorithms in terms of the sample number}
\label{sec:run_N}
\begin{figure}[t]
	\centering
	\subfloat[50 to 15000 data points] {\resizebox{0.49\columnwidth}{!}{\input{figures/review/runtime_samples_eps_0_EM_Huber_BIC_Huber.tex}}%
		\label{fig:runtime_over_samples}}
	\hfil
	\subfloat[2 to 24 features] {\resizebox{0.49\columnwidth}{!}{\input{figures/review/runtime_r_EM_Huber_BIC_Huber.tex}}%
		\label{fig:runtime_over_r}}
	\caption{Runtime of EM: Huber, BIC: Huber over number of samples and number of features}
	\label{fig:runtime}
\end{figure}
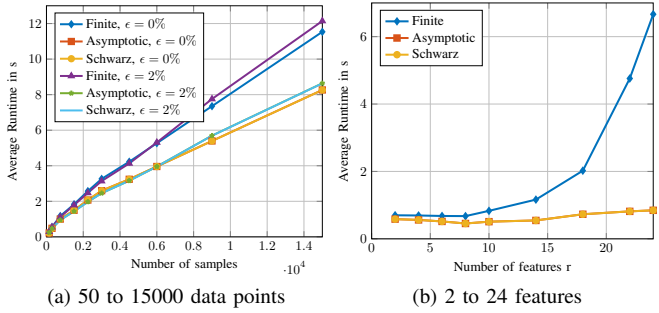

Figure~\ref{fig:runtime_over_samples} shows the average runtime for different numbers of data points. It can be noted that increasing the percentage of outliers leads to a small increase of the runtime. This behavior is expected, because with more outliers, the EM algorithm will take longer to converge. More noteworthy, the dependence of the runtime on the number of samples is linear, as we expected from the theoretical analysis in Section~\ref{sec:complex}. Hence, increasing the sample size will not lead to an exponentially longer runtime.

\subsubsection{Probability of detection analysis of the proposed cluster enumeration algorithms over number of samples per cluster}
\begin{figure}[t]
	\centering
	\subfloat[Finite BIC] {\resizebox{0.49\columnwidth}{!}{\input{figures/review2/pdet_over_N_finite.tex}}%
		\label{fig:pdet_over_N_finite}}
	\hfil
	\subfloat[Asymptotic BIC] {\resizebox{0.49\columnwidth}{!}{\input{figures/review2/pdet_over_N_asymp.tex}}%
		\label{fig:pdet_over_N_asymp}}
	\caption{Empirical probability of detection over different amount of samples per cluster.}
	\label{fig:pdet_over_N}
\end{figure}
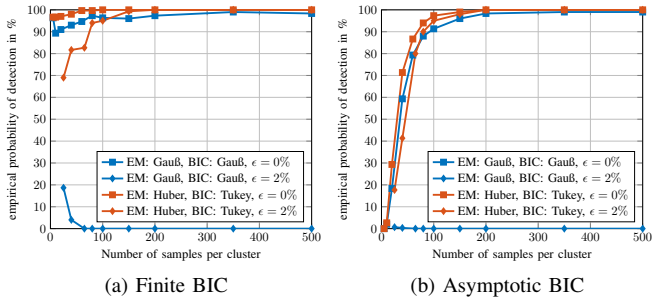

In Figure~\ref{fig:pdet_over_N}, we analyze the influence of the amount of samples on the empirical probability of detection. We compare the non-robust Gaussian and the robust EM: Huber, BIC: Tukey based estimator. Generally speaking, the EM: Huber, BIC: Tukey estimator is able to obtain a high probability of detection with and without outliers, whereas the non-robust Gaussian estimator completely breaks down when outliers are present. When comparing the finite and the asymptotic case, the finite estimators are able to maintain a high probability of detection even for very low sample sizes, in comparison to the asymptotic BIC, which requires a certain minimal amount of samples (in this example $N_{k} > 100$) to work properly.

\subsubsection{Runtime analysis of the proposed cluster enumeration algorithms over number of features}
\label{sec:run_r}
For this simulation, the following data set is used: $x_{k} \sim t_{3}(\bmu_{k}, \bPsi_{k})$ for $k = 1,2$, the cluster centroids and scatter matrices are given as $\bmu_{k} = c \, \boldsymbol{1}_{r \times 1}$ and $\Psi_{k} = \bI_{r}$, with $c \in \{0, 15\}$ and $N_{k} = 500$. Since we generate the data based on a $t_{3}$ distribution, the data is heavy-tailed. Figure~\ref{fig:runtime_over_r} shows the average runtime for an increasing number of features. It can be noted that the finite BIC has a  polynomial complexity in $r$, whereas the asymptotic approximation and the Schwarz BIC seem to be, at most, of cubic complexity. The slight decrease in runtime up until $r = 8$, probably can be attributed to other variables dominating the runtime over the number of features. The experimental findings confirm our  expectations based on the theoretical complexity analysis from Section~\ref{sec:complex}.

\subsection{Real Data Simulations}

\subsubsection{Radar Data}
\begin{figure}
	\centering
	\subfloat[Proposed criteria]
	{\resizebox{0.48\columnwidth}{!}{\input{figures/bic_radar.tex}}%
		\label{fig:bic_radar}}
	\hfil
	\subfloat[Schwarz criterion] {\resizebox{0.48\columnwidth}{!}{\input{figures/bic_radar_Schwarz.tex}}%
		\label{fig:bic_radar_schwarz}}
	\caption{Cluster enumeration results of the radar-based human gait data.}
\end{figure}
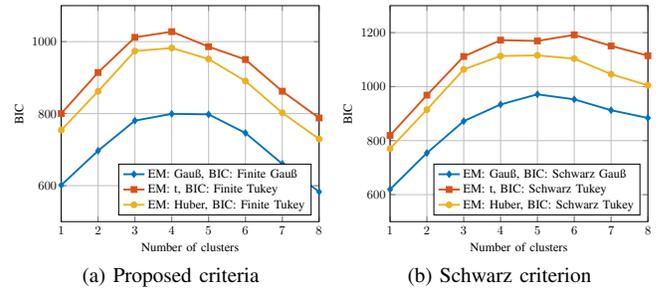

The data set is composed of four walking persons. Their gait signatures, measured by a 24GHz radar system, were processed to calculate the spectrogram followed by a feature extraction step \cite{Teklehaymanot.2018c}. To reduce the dimensionality from $r = 12800$, a PCA was applied and the first five components were selected to form the final data set with $N = 187$ and $r = 5$. A subset of the first three components is shown in Figure~\ref{fig:data_radar}. The correct number of different persons is estimated by a $\text{BIC}_{\text{F}}$ with EM: Gaussian, BIC: Gaussian (also used by \cite{Teklehaymanot.2018c}), EM: t, BIC: Tukey and EM: Huber, BIC: Tukey, as shown in Figure~\ref{fig:bic_radar}. In comparison to the method used by \cite{Teklehaymanot.2018c}, one can note that the peaks in the newly proposed methods are more prominent, hence, they lead to a more stable result. Additionally, in Figure~\ref{fig:bic_radar_schwarz}, we show the results based on a Schwarz penalty term that clearly overestimates the number of clusters due to the small sample size. 
\begin{figure}
	\centering
	\subfloat[First three PCA features of the radar-based human gait data.]
	{\resizebox{0.51\columnwidth}{!}{\input{figures/data_radar.tex}}%
		\label{fig:data_radar}}
	\hfil
	\subfloat[First three features of the remote sensing data set.] {\resizebox{0.45\columnwidth}{!}{\input{figures/review/data_forest_non_vs_hinoki_N_255.tex}}%
		\label{fig:data_forest}}
	\caption{Exemplary visualizations of the used data sets.}
\end{figure}
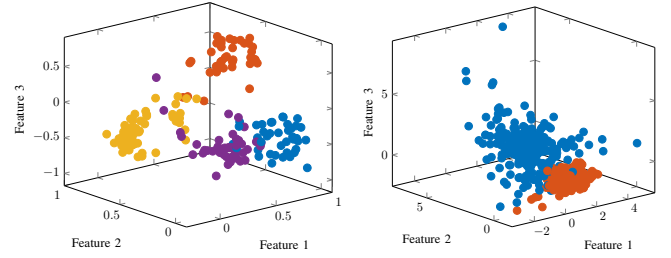

\subsubsection{Remote Sensing Data}
The remote sensing data used in this paper is a subset of the data set studied in \cite{Johnson.2012} and it can be downloaded from \url{https://archive.ics.uci.edu/ml/datasets/Forest+type+mapping}. More specifically, for our experiment, we selected the two classes of 'Hinoki forest' and 'non forest'. The data set contains different spectral characteristics at visible-to-near infrared wavelengths and maps them to different types of forest. As it can be observed in Figure~\ref{fig:data_forest}, the data set contains outlying data points and also the two clusters partially overlap. To keep the cluster imbalance as in the original data set while downsampling, we sample each cluster separately, but proportional to the complete data set.  The results of our numerical experiments are reported in Table~\ref{tb:forest_data}. For $N = 32$, the best results are generally obtained by the finite BIC, with the overall best result from a EM: Huber, BIC: finite Tukey. For the medium sample size, the EM: Huber Schwarz based BICs have the best empirical probability of detection. And finally, for $N = 255$, almost all robust variations have a very high empirical probability of detection. As discussed in the previous section, the point at which the criteria provide similar results (i.e. what can be considered as a regime where the asymptotic approximation holds well) depends heavily on the data set and it can vary significantly between data sets. Generally speaking, we can also conclude, that our newly proposed cluster enumeration algorithms, outperform the non-robust variations for this real data set.
\begin{table}
	\centering
	\hfill
	\caption{Empirical probability of detection of the different cluster enumeration algorithms for the remote sensing data set.}
	\label{tb:forest_data}
	\begin{tabu}{cc cccc}
		\toprule
		EM                     & \multicolumn{2}{c}{BIC}       & $N = 32$ & $N = 128$ &$N = 255$ \\ \midrule
		\multirow{3}{*}{Gauß}  & \multirow{3}{*}{Gauß}     & F \cite{Teklehaymanot.2018d} & 67 & 51 & 42  \\
		                       &                           & A \cite{Teklehaymanot.2018d} & 8 & 46 & 44 \\
		                       &                       	   & S \cite{Teklehaymanot.2018d} & 46 & 77 & 89  \\\cmidrule(l){1-6}
		\multirow{6}{*}{t}     & \multirow{3}{*}{t}        & F \cite{Teklehaymanot.2018b} & 81 & 94 & 100  \\
							   &                           & A \cite{Teklehaymanot.2018b} & 14 & 67 & 94  \\
							   &                           & S \cite{Teklehaymanot.2018b} & 72 & 94 & 100  \\\cmidrule(l){2-6}
						       & \multirow{3}{*}{Tukey}    & F & 80 & 95 & 100 \\
							   &                           & A & 19 & 74 & 95 \\
							   &                           & S & 78 & 93 & 100 \\\cmidrule(l){1-6}
		\multirow{6}{*}{Huber} & \multirow{3}{*}{Huber}    & F & 81 & 90 & 100 \\
							   &                           & A & 14 & 76 & 100 \\
						       &                           & S & 74 & \textbf{98} & 100 \\\cmidrule(l){2-6}
							   & \multirow{3}{*}{Tukey}    & F & \textbf{84} & 95 & 100 \\
							   &                           & A & 25 & 86 & 100 \\
							   &                           & S & 77 & 97 & 100 \\\bottomrule & 
	\end{tabu}
	\hfill
\end{table}

\subsubsection{Wine Data}
The wine data used in this paper is a subset of the data set from \url{https://archive.ics.uci.edu/ml/datasets/Wine}. Its goal is to distinguish between three different wine cultivars by using the chemical composition of the wine. In our experiments we reduced the size of the features from 13 down to five, namely 'Total phenols', 'Flavanoids', 'Color intensity', 'Hue' and 'OD280/OD315 of diluted wines'. Additionally, we conducted our experiments for different sample sizes of 158, 168 and 178 samples. The smaller sample sizes were obtained by randomly removing samples in each Monte Carlo run. Table~\ref{tb:wine_data} summarizes the results of our simulations. It can be seen that a slight decrease in the sample size can have a drastic effect on the empirical probability of detection. Except for the Gaussian based BIC, all BICs work almost perfectly for $N = 178$, but with lower sample sizes all BIC have a significantly decreased empirical probability of detection. In this specific example, the asymptotic BIC is able to outperform the finite BIC for $N=168$ and $N=158$, even if the sample size may be considered small for $r = 5$. This experiment illustrates, that it is very difficult to predict the influence of one specific parameter on the result, and that the best performing BIC may vary depending on the data set. A more general observation is that this example, again, displays the superiority of the proposed robust cluster enumeration framework over the standard Gaussian based framework in the presence of heavy-tailed clusters, rendering it a useful addition to the existing cluster enumeration frameworks.
\begin{table}
	\centering
	\hfill
	\caption{Empirical probability of detection of the different cluster enumeration algorithms for the wine data set.}
	\label{tb:wine_data}
	\begin{tabu}{cc cccc}
		\toprule
		EM                     & \multicolumn{2}{c}{BIC}       & $N = 158$ & $N = 168$ & $N = 178$ \\ \midrule
		\multirow{3}{*}{Gauß}  & \multirow{3}{*}{Gauß}         & F \cite{Teklehaymanot.2018d} & 3 & 1 & 0  \\
		&                      & A \cite{Teklehaymanot.2018d}  & 36 & 44 & 66 \\
		&                      & S \cite{Teklehaymanot.2018d}  & 3 & 1 & 0  \\\cmidrule(l){1-6}
		\multirow{6}{*}{t}     & \multirow{3}{*}{t}            & F \cite{Teklehaymanot.2018b} & 66 & 71 & 100  \\
		&                      & A \cite{Teklehaymanot.2018b}  & 74 & 87 & 100  \\
		&                      & S \cite{Teklehaymanot.2018b}  & 66 & 71 & 100  \\\cmidrule(l){2-6}
		& \multirow{3}{*}{Tukey}    & F & 47 & 64 & 100 \\
		&                           & A & 68 & 83 & 100 \\
		&                           & S & 49 & 65 & 100 \\\cmidrule(l){1-6}
		\multirow{6}{*}{Huber} & \multirow{3}{*}{Huber}    & F & 63 & 81 & 100 \\
		&                           & A & \textbf{79} & \textbf{89} & 99 \\
		&                           & S & 61 & 79 & 100 \\\cmidrule(l){2-6}
		& \multirow{3}{*}{Tukey}    & F & 43 & 64 & 100 \\
		&                           & A & 66 & 81 & 100 \\
		&                           & S & 32 & 49 & 100 \\\bottomrule & 
	\end{tabu}
	\hfill
\end{table}

%% file: figures/data_31_Nk_250_eps_00013333.tex
\begin{tikzpicture}
	\begin{axis}[
	/pgf/number format/.cd,
	fixed,
	1000 sep={},
	width = \figurewidth,
	height = \figureheight,
	xmin = -20,
	xmax = 20,
	ymin = -20,
	ymax = 20,
	xlabel= Feature 1,
	ylabel= Feature 2,
	]
	\addplot[scatter/classes={1={matlabblue},2={matlaborange},3={matlabyellow}, 4={black, mark = o}}, scatter, only marks, scatter src=explicit symbolic] table[x=x,y=y,meta=label] {figures/data_31_Nk_250_eps_00013333.csv};   
	\end{axis}
\end{tikzpicture}

%% file: figures/data_31_Nk_250_eps_01.tex
\begin{tikzpicture}
	\begin{axis}[
	/pgf/number format/.cd,
	fixed,
	1000 sep={},
	width = \figurewidth,
	height = \figureheight,
	xmin = -20,
	xmax = 20,
	ymin = -20,
	ymax = 20,
	xlabel= Feature 1,
	ylabel= Feature 2,
	]
	\addplot[scatter/classes={1={matlabblue},2={matlaborange},3={matlabyellow}, 4={black, mark = o}}, scatter, only marks, scatter src=explicit symbolic] table[x=x,y=y,meta=label] {figures/data_31_Nk_250_eps_01.csv};   
	\end{axis}
\end{tikzpicture}

%% file: figures/paper_sens_EM_Gaus_BIC_Gaus_RES_Nk_50_step_2_MC_500.tex
\begin{tikzpicture}
	\begin{axis}[
	/pgf/number format/.cd,
	fixed,
	1000 sep={},
	width = \figurewidth,
	height = \figureheight,
	xmin = -20,
	xmax = 20,
	ymin = -20,
	ymax = 20,
	xlabel= Feature 1,
	ylabel= Feature 2,
	colorbar, 
	colorbar/width=4mm,
	point meta min=0,point meta max=1, 
	colorbar style={
		xshift = -4mm,
		ylabel = empirical probability of detection in \%,
		ytick = {0,0.1,0.2,0.3,0.4,0.5,0.6,0.7,0.8,0.9,1},
		yticklabels = {$0$,$10$,$20$,$30$,$40$,$50$,$60$,$70$,$80$,$90$,$100$},
		yticklabel style={
			text width=1em,
		}
	}
	]
	\addplot[contour prepared = {labels=false,},contour prepared format=matlab,line width=1.5pt] table[header=true] {figures/paper_sens_EM_Gaus_BIC_Gaus_RES_Nk_50_step_2_MC_500_1.csv};
	\addplot[scatter/classes={1={matlabblue},2={matlaborange},3={matlabyellow}}, scatter, only marks, scatter src=explicit symbolic] table[x=x,y=y,meta=label] {figures/paper_sens_EM_Gaus_BIC_Gaus_RES_Nk_50_step_2_MC_500_2.csv};   
	\end{axis}
\end{tikzpicture}

%% file: figures/paper_sens_EM_Gaus_BIC_Gaus_aRES_Nk_50_step_2_MC_500.tex
\begin{tikzpicture}
	\begin{axis}[
	/pgf/number format/.cd,
	fixed,
	1000 sep={},
	width = \figurewidth,
	height = \figureheight,
	xmin = -20,
	xmax = 20,
	ymin = -20,
	ymax = 20,
	xlabel= Feature 1,
	ylabel= Feature 2,
	colorbar, 
	colorbar/width=4mm,
	point meta min=0,point meta max=1, 
	colorbar style={
		xshift = -4mm,
		ylabel = empirical probability of detection in \%,
		ytick = {0,0.1,0.2,0.3,0.4,0.5,0.6,0.7,0.8,0.9,1},
		yticklabels = {$0$,$10$,$20$,$30$,$40$,$50$,$60$,$70$,$80$,$90$,$100$},
		yticklabel style={
			text width=1em,
		}
	}
	]
	\addplot[contour prepared = {labels=false,},contour prepared format=matlab,line width=1.5pt] table[header=true] {figures/paper_sens_EM_Gaus_BIC_Gaus_aRES_Nk_50_step_2_MC_500_1.csv};
	\addplot[scatter/classes={1={matlabblue},2={matlaborange},3={matlabyellow}}, scatter, only marks, scatter src=explicit symbolic] table[x=x,y=y,meta=label] {figures/paper_sens_EM_Gaus_BIC_Gaus_aRES_Nk_50_step_2_MC_500_2.csv};   
	\end{axis}
\end{tikzpicture}

%% file: figures/paper_sens_EM_Huber_BIC_Huber_RES_Nk_50_step_2_MC_500.tex
\begin{tikzpicture}
	\begin{axis}[
	/pgf/number format/.cd,
	fixed,
	1000 sep={},
	width = \figurewidth,
	height = \figureheight,
	xmin = -20,
	xmax = 20,
	ymin = -20,
	ymax = 20,
	xlabel= Feature 1,
	ylabel= Feature 2,
	colorbar, 
	colorbar/width=4mm,
	point meta min=0,point meta max=1, 
	colorbar style={
		xshift = -4mm,
		ylabel = empirical probability of detection in \%,
		ytick = {0,0.1,0.2,0.3,0.4,0.5,0.6,0.7,0.8,0.9,1},
		yticklabels = {$0$,$10$,$20$,$30$,$40$,$50$,$60$,$70$,$80$,$90$,$100$},
		yticklabel style={
			text width=1em,
		}
	}
	]
	\addplot[contour prepared = {labels=false,},contour prepared format=matlab,line width=1.5pt] table[header=true] {figures/paper_sens_EM_Huber_BIC_Huber_RES_Nk_50_step_2_MC_500_1.csv};
	\addplot[scatter/classes={1={matlabblue},2={matlaborange},3={matlabyellow}}, scatter, only marks, scatter src=explicit symbolic] table[x=x,y=y,meta=label] {figures/paper_sens_EM_Huber_BIC_Huber_RES_Nk_50_step_2_MC_500_2.csv};   
	\end{axis}
\end{tikzpicture}

%% file: figures/paper_sens_EM_Huber_BIC_Huber_aRES_Nk_50_step_2_MC_500.tex
\begin{tikzpicture}
	\begin{axis}[
	/pgf/number format/.cd,
	fixed,
	1000 sep={},
	width = \figurewidth,
	height = \figureheight,
	xmin = -20,
	xmax = 20,
	ymin = -20,
	ymax = 20,
	xlabel= Feature 1,
	ylabel= Feature 2,
	colorbar, 
	colorbar/width=4mm,
	point meta min=0,point meta max=1, 
	colorbar style={
		xshift = -4mm,
		ylabel = empirical probability of detection in \%,
		ytick = {0,0.1,0.2,0.3,0.4,0.5,0.6,0.7,0.8,0.9,1},
		yticklabels = {$0$,$10$,$20$,$30$,$40$,$50$,$60$,$70$,$80$,$90$,$100$},
		yticklabel style={
			text width=1em,
		}
	}
	]
	\addplot[contour prepared = {labels=false,},contour prepared format=matlab,line width=1.5pt] table[header=true] {figures/paper_sens_EM_Huber_BIC_Huber_aRES_Nk_50_step_2_MC_500_1.csv};
	\addplot[scatter/classes={1={matlabblue},2={matlaborange},3={matlabyellow}}, scatter, only marks, scatter src=explicit symbolic] table[x=x,y=y,meta=label] {figures/paper_sens_EM_Huber_BIC_Huber_aRES_Nk_50_step_2_MC_500_2.csv};   
	\end{axis}
\end{tikzpicture}

%% file: figures/paper_sens_EM_Huber_BIC_Tukey_RES_Nk_50_step_2_MC_500.tex
\begin{tikzpicture}
	\begin{axis}[
	/pgf/number format/.cd,
	fixed,
	1000 sep={},
	width = \figurewidth,
	height = \figureheight,
	xmin = -20,
	xmax = 20,
	ymin = -20,
	ymax = 20,
	xlabel= Feature 1,
	ylabel= Feature 2,
	colorbar, 
	colorbar/width=4mm,
	point meta min=0,point meta max=1, 
	colorbar style={
		xshift = -4mm,
		ylabel = empirical probability of detection in \%,
		ytick = {0,0.1,0.2,0.3,0.4,0.5,0.6,0.7,0.8,0.9,1},
		yticklabels = {$0$,$10$,$20$,$30$,$40$,$50$,$60$,$70$,$80$,$90$,$100$},
		yticklabel style={
			text width=1em,
		}
	}
	]
	\addplot[contour prepared = {labels=false,},contour prepared format=matlab,line width=1.5pt] table[header=true] {figures/paper_sens_EM_Huber_BIC_Tukey_RES_Nk_50_step_2_MC_500_1.csv};
	\addplot[scatter/classes={1={matlabblue},2={matlaborange},3={matlabyellow}}, scatter, only marks, scatter src=explicit symbolic] table[x=x,y=y,meta=label] {figures/paper_sens_EM_Huber_BIC_Tukey_RES_Nk_50_step_2_MC_500_2.csv};   
	\end{axis}
\end{tikzpicture}

%% file: figures/paper_sens_EM_Huber_BIC_Tukey_aRES_Nk_50_step_2_MC_500.tex
\begin{tikzpicture}
	\begin{axis}[
	/pgf/number format/.cd,
	fixed,
	1000 sep={},
	width = \figurewidth,
	height = \figureheight,
	xmin = -20,
	xmax = 20,
	ymin = -20,
	ymax = 20,
	xlabel= Feature 1,
	ylabel= Feature 2,
	colorbar, 
	colorbar/width=4mm,
	point meta min=0,point meta max=1, 
	colorbar style={
		xshift = -4mm,
		ylabel = empirical probability of detection in \%,
		ytick = {0,0.1,0.2,0.3,0.4,0.5,0.6,0.7,0.8,0.9,1},
		yticklabels = {$0$,$10$,$20$,$30$,$40$,$50$,$60$,$70$,$80$,$90$,$100$},
		yticklabel style={
			text width=1em,
		}
	}
	]
	\addplot[contour prepared = {labels=false,},contour prepared format=matlab,line width=1.5pt] table[header=true] {figures/paper_sens_EM_Huber_BIC_Tukey_aRES_Nk_50_step_2_MC_500_1.csv};
	\addplot[scatter/classes={1={matlabblue},2={matlaborange},3={matlabyellow}}, scatter, only marks, scatter src=explicit symbolic] table[x=x,y=y,meta=label] {figures/paper_sens_EM_Huber_BIC_Tukey_aRES_Nk_50_step_2_MC_500_2.csv};   
	\end{axis}
\end{tikzpicture}

%% file: figures/outliers_BIC-RES_MC_2000_Nk_10.tex
\begin{tikzpicture}
	\begin{axis}[
	/pgf/number format/.cd,
	fixed,
	1000 sep={},
	width = \figurewidth,
	height = \figureheight,
	xmin = 0,
	xmax = 0.35,
	ymax = 1,
	ymin = 0,
	ytick={0,0.1,0.2,0.3,0.4,0.5,0.6,0.7,0.8,0.9,1},
	yticklabels={$0$,$10$,$20$,$30$,$40$,$50$,$60$,$70$,$80$,$90$,$100$},
	xtick={0,0.05,0.1,0.15,0.2,0.25,0.3,0.35},
	xticklabels={$0$,$5$,$10$,$15$,$20$,$25$,$30$,$35$},
	grid,
	xlabel= outlieres $\epsilon$ in \%,
	ylabel= empirical probability of detection in \%,
    legend entries={{EM: Gauß, BIC: Gauß \cite{Teklehaymanot.2018d}}\\
					{EM: t, BIC: t \cite{Teklehaymanot.2018b}}\\
					{EM: t, BIC: Tukey}\\
					{EM: Huber, BIC: Huber}\\
					{EM: Huber, BIC: Tukey}\\},
	legend pos=north east,
	legend cell align={left},
	cycle list name=matlabcolor
	]
	\plotfileNOlegend{figures/outliers_BIC-RES_MC_2000_Nk_10.csv}
	\end{axis}
\end{tikzpicture}

%% file: figures/outliers_BIC-RES_MC_600_Nk_250.tex
\begin{tikzpicture}
	\begin{axis}[
	/pgf/number format/.cd,
	fixed,
	1000 sep={},
	width = \figurewidth,
	height = \figureheight,
	xmin = 0,
	xmax = 0.35,
	ymax = 1,
	ymin = 0,
	ytick={0,0.1,0.2,0.3,0.4,0.5,0.6,0.7,0.8,0.9,1},
	yticklabels={$0$,$10$,$20$,$30$,$40$,$50$,$60$,$70$,$80$,$90$,$100$},
	xtick={0,0.05,0.1,0.15,0.2,0.25,0.3,0.35},
	xticklabels={$0$,$5$,$10$,$15$,$20$,$25$,$30$,$35$},
	grid,
	xlabel= outlieres $\epsilon$ in \%,
	ylabel= empirical probability of detection in \%,
	cycle list name=matlabcolor
	]
	\plotfileNOlegend{figures/outliers_BIC-RES_MC_600_Nk_250.csv}
	\end{axis}
\end{tikzpicture}

%% file: figures/outliers_BIC-aRES_MC_2000_Nk_10.tex
\begin{tikzpicture}
	\begin{axis}[
	/pgf/number format/.cd,
	fixed,
	1000 sep={},
	width = \figurewidth,
	height = \figureheight,
	xmin = 0,
	xmax = 0.35,
	ymax = 1,
	ymin = 0,
	ytick={0,0.1,0.2,0.3,0.4,0.5,0.6,0.7,0.8,0.9,1},
	yticklabels={$0$,$10$,$20$,$30$,$40$,$50$,$60$,$70$,$80$,$90$,$100$},
	xtick={0,0.05,0.1,0.15,0.2,0.25,0.3,0.35},
	xticklabels={$0$,$5$,$10$,$15$,$20$,$25$,$30$,$35$},
	grid,
	xlabel= outlieres $\epsilon$ in \%,
	ylabel= empirical probability of detection in \%,
    legend entries={{EM: Gauß, BIC: Gauß \cite{Teklehaymanot.2018d}}\\
					{EM: t, BIC: t \cite{Teklehaymanot.2018b}}\\
					{EM: t, BIC: Tukey}\\
					{EM: Huber, BIC: Huber}\\
					{EM: Huber, BIC: Tukey}\\},
	legend pos=north east,
	legend cell align={left},
	cycle list name=matlabcolor
	]
	\plotfileNOlegend{figures/outliers_BIC-aRES_MC_2000_Nk_10.csv}
	\end{axis}
\end{tikzpicture}

%% file: figures/outliers_BIC-aRES_MC_600_Nk_250.tex
\begin{tikzpicture}
	\begin{axis}[
	/pgf/number format/.cd,
	fixed,
	1000 sep={},
	width = \figurewidth,
	height = \figureheight,
	xmin = 0,
	xmax = 0.35,
	ymax = 1,
	ymin = 0,
	ytick={0,0.1,0.2,0.3,0.4,0.5,0.6,0.7,0.8,0.9,1},
	yticklabels={$0$,$10$,$20$,$30$,$40$,$50$,$60$,$70$,$80$,$90$,$100$},
	xtick={0,0.05,0.1,0.15,0.2,0.25,0.3,0.35},
	xticklabels={$0$,$5$,$10$,$15$,$20$,$25$,$30$,$35$},
	grid,
	xlabel= outlieres $\epsilon$ in \%,
	ylabel= empirical probability of detection in \%,
	cycle list name=matlabcolor
	]
	\plotfileNOlegend{figures/outliers_BIC-aRES_MC_600_Nk_250.csv}
	\end{axis}
\end{tikzpicture}

%% file: figures/outliers_BIC-Schwarz_MC_2000_Nk_10.tex
\begin{tikzpicture}
	\begin{axis}[
	/pgf/number format/.cd,
	fixed,
	1000 sep={},
	width = \figurewidth,
	height = \figureheight,
	xmin = 0,
	xmax = 0.35,
	ymax = 1,
	ymin = 0,
	ytick={0,0.1,0.2,0.3,0.4,0.5,0.6,0.7,0.8,0.9,1},
	yticklabels={$0$,$10$,$20$,$30$,$40$,$50$,$60$,$70$,$80$,$90$,$100$},
	xtick={0,0.05,0.1,0.15,0.2,0.25,0.3,0.35},
	xticklabels={$0$,$5$,$10$,$15$,$20$,$25$,$30$,$35$},
	grid,
	xlabel= outlieres $\epsilon$ in \%,
	ylabel= empirical probability of detection in \%,
    legend entries={{EM: Gauß, BIC: Gauß \cite{Teklehaymanot.2018d}}\\
					{EM: t, BIC: t \cite{Teklehaymanot.2018b}}\\
					{EM: t, BIC: Tukey}\\
					{EM: Huber, BIC: Huber}\\
					{EM: Huber, BIC: Tukey}\\},
	legend pos=north east,
	legend cell align={left},
	cycle list name=matlabcolor
	]
	\plotfileNOlegend{figures/outliers_BIC-Schwarz_MC_2000_Nk_10.csv}
	\end{axis}
\end{tikzpicture}

%% file: figures/outliers_BIC-Schwarz_MC_600_Nk_250.tex
\begin{tikzpicture}
	\begin{axis}[
	/pgf/number format/.cd,
	fixed,
	1000 sep={},
	width = \figurewidth,
	height = \figureheight,
	xmin = 0,
	xmax = 0.35,
	ymax = 1,
	ymin = 0,
	ytick={0,0.1,0.2,0.3,0.4,0.5,0.6,0.7,0.8,0.9,1},
	yticklabels={$0$,$10$,$20$,$30$,$40$,$50$,$60$,$70$,$80$,$90$,$100$},
	xtick={0,0.05,0.1,0.15,0.2,0.25,0.3,0.35},
	xticklabels={$0$,$5$,$10$,$15$,$20$,$25$,$30$,$35$},
	grid,
	xlabel= outlieres $\epsilon$ in \%,
	ylabel= empirical probability of detection in \%,
	cycle list name=matlabcolor
	]
	\plotfileNOlegend{figures/outliers_BIC-Schwarz_MC_600_Nk_250.csv}
	\end{axis}
\end{tikzpicture}

%% file: figures/review/bic_Nk_5_eps_0_EM_Huber_BIC_Huber.tex
\begin{tikzpicture}
	\begin{axis}[
	/pgf/number format/.cd,
	fixed,
	1000 sep={},
	width = \figurewidth,
	height = \figureheight,
	xmin = 1,
	xmax = 6,
	ymax = 10,
	ymin = -70,
	grid,
	xlabel= Number of clusters,
	ylabel= BIC,
    legend entries={{EM: Gauß, BIC: Finite Gauß}\\
					{EM: t, BIC: Finite Tukey}\\
					{EM: Huber, BIC: Finite Tukey}\\},
	legend pos=south east,
	legend cell align={left},
	cycle list name=matlabcolor
	]
	\plotfile{figures/review/bic_Nk_5_eps_0_EM_Huber_BIC_Huber.csv}
	\end{axis}
\end{tikzpicture}

%% file: figures/review/bic_Nk_10_eps_0_EM_Huber_BIC_Huber.tex
\begin{tikzpicture}
	\begin{axis}[
	/pgf/number format/.cd,
	fixed,
	1000 sep={},
	width = \figurewidth,
	height = \figureheight,
	xmin = 1,
	xmax = 6,
	ymax = 10,
	ymin = -70,
	grid,
	xlabel= Number of clusters,
	ylabel= BIC,
    legend entries={{EM: Gauß, BIC: Finite Gauß}\\
					{EM: t, BIC: Finite Tukey}\\
					{EM: Huber, BIC: Finite Tukey}\\},
	legend pos=south east,
	legend cell align={left},
	cycle list name=matlabcolor
	]
	\plotfile{figures/review/bic_Nk_10_eps_0_EM_Huber_BIC_Huber.csv}
	\end{axis}
\end{tikzpicture}

%% file: figures/review/bic_Nk_50_eps_0_EM_Huber_BIC_Huber.tex
\begin{tikzpicture}
	\begin{axis}[
	/pgf/number format/.cd,
	fixed,
	1000 sep={},
	width = \figurewidth,
	height = \figureheight,
	xmin = 1,
	xmax = 6,
	ymax = 250,
	ymin = -100,
	grid,
	xlabel= Number of clusters,
	ylabel= BIC,
    legend entries={{EM: Gauß, BIC: Finite Gauß}\\
					{EM: t, BIC: Finite Tukey}\\
					{EM: Huber, BIC: Finite Tukey}\\},
	legend pos=south east,
	legend cell align={left},
	cycle list name=matlabcolor
	]
	\plotfile{figures/review/bic_Nk_50_eps_0_EM_Huber_BIC_Huber.csv}
	\end{axis}
\end{tikzpicture}

%% file: figures/review/bic_Nk_500_eps_0_EM_Huber_BIC_Huber.tex
\begin{tikzpicture}
	\begin{axis}[
	/pgf/number format/.cd,
	fixed,
	1000 sep={},
	width = \figurewidth,
	height = \figureheight,
	xmin = 1,
	xmax = 6,
	ymax = 6000,
	ymin = 2500,
	grid,
	xlabel= Number of clusters,
	ylabel= BIC,
    legend entries={{EM: Gauß, BIC: Finite Gauß}\\
					{EM: t, BIC: Finite Tukey}\\
					{EM: Huber, BIC: Finite Tukey}\\},
	legend pos=south east,
	legend cell align={left},
	cycle list name=matlabcolor
	]
	\plotfile{figures/review/bic_Nk_500_eps_0_EM_Huber_BIC_Huber.csv}
	\end{axis}
\end{tikzpicture}

%% file: figures/review/bic_imb_5_eps_0_EM_Huber_BIC_Huber.tex
\begin{tikzpicture}
	\begin{axis}[
	/pgf/number format/.cd,
	fixed,
	1000 sep={},
	width = \figurewidth,
	height = \figureheight,
	xmin = 1,
	xmax = 6,
	ymax = 200,
	ymin = 60,
	grid,
	xlabel= Number of clusters,
	ylabel= BIC,
    legend entries={{EM: Gauß, BIC: Finite Gauß}\\
					{EM: t, BIC: Finite Tukey}\\
					{EM: Huber, BIC: Finite Tukey}\\},
	legend pos=south east,
	legend cell align={left},
	cycle list name=matlabcolor
	]
	\plotfile{figures/review/bic_imb_5_eps_0_EM_Huber_BIC_Huber.csv}
	\end{axis}
\end{tikzpicture}

%% file: figures/review/bic_imb_50_eps_0_EM_Huber_BIC_Huber.tex
\begin{tikzpicture}
	\begin{axis}[
	/pgf/number format/.cd,
	fixed,
	1000 sep={},
	width = \figurewidth,
	height = \figureheight,
	xmin = 1,
	xmax = 6,
	ymax = 5200,
	ymin = 3800,
	grid,
	xlabel= Number of clusters,
	ylabel= BIC,
    legend entries={{EM: Gauß, BIC: Finite Gauß}\\
					{EM: t, BIC: Finite Tukey}\\
					{EM: Huber, BIC: Finite Tukey}\\},
	legend pos=south east,
	legend cell align={left},
	cycle list name=matlabcolor
	]
	\plotfile{figures/review/bic_imb_50_eps_0_EM_Huber_BIC_Huber.csv}
	\end{axis}
\end{tikzpicture}

%% file: figures/review/runtime_samples_eps_0_EM_Huber_BIC_Huber.tex
\begin{tikzpicture}
	\begin{axis}[
	/pgf/number format/.cd,
	fixed,
	1000 sep={},
	width = \figurewidth,
	height = \figureheight,
	xmin = 0,
	xmax = 15000,
	ymax = 13,
	ymin = 0,
	grid,
	xlabel= Number of samples,
	ylabel= Average Runtime in s,
    legend entries={{Finite, $\epsilon = 0\%$}\\
					{Asymptotic, $\epsilon = 0\%$}\\
					{Schwarz, $\epsilon = 0\%$}\\
					{Finite, $\epsilon = 2\%$}\\
					{Asymptotic, $\epsilon = 2\%$}\\
					{Schwarz, $\epsilon = 2\%$}\\},
	legend pos=north west,
	legend cell align={left},
	cycle list name=matlabcolor
	]
	\plotfileNOlegend{figures/review/runtime_samples_eps_0_EM_Huber_BIC_Huber.csv}
	\plotfileNOlegend{figures/review/runtime_samples_eps_2_EM_Huber_BIC_Huber.csv}
	\end{axis}
\end{tikzpicture}

%% file: figures/review/runtime_r_EM_Huber_BIC_Huber.tex
\begin{tikzpicture}
	\begin{axis}[
	/pgf/number format/.cd,
	fixed,
	1000 sep={},
	width = \figurewidth,
	height = \figureheight,
	xmin = 0,
	xmax = 24,
	ymax = 7,
	ymin = 0,
	grid,
	xlabel= Number of features r,
	ylabel= Average Runtime in s,
    legend entries={{Finite}\\
					{Asymptotic}\\
					{Schwarz}\\},
	legend pos=north west,
	legend cell align={left},
	cycle list name=matlabcolor
	]
	\plotfile{figures/review/runtime_r_EM_Huber_BIC_Huber.csv}
	\end{axis}
\end{tikzpicture}

%% file: figures/review2/pdet_over_N_finite.tex
\begin{tikzpicture}
	\begin{axis}[
		/pgf/number format/.cd,
		fixed,
		1000 sep={},
		width = \figurewidth,
		height = \figureheight,
		xmin = 0,
		xmax = 500,
		ymax = 1,
		ymin = 0,
		ytick={0,0.1,0.2,0.3,0.4,0.5,0.6,0.7,0.8,0.9,1},
		yticklabels={$0$,$10$,$20$,$30$,$40$,$50$,$60$,$70$,$80$,$90$,$100$},
		grid,
		xlabel= Number of samples per cluster,
		ylabel= empirical probability of detection in \%,
		legend entries={{EM: Gauß, BIC: Gauß, $\epsilon = 0\%$}\\
			{EM: Gauß, BIC: Gauß, $\epsilon = 2\%$}\\
			{EM: Huber, BIC: Tukey, $\epsilon = 0\%$}\\
			{EM: Huber, BIC: Tukey, $\epsilon = 2\%$}\\
			{Asymptotic, $\epsilon = 2\%$}\\
			{Schwarz, $\epsilon = 2\%$}\\},
		legend pos=south east,
		legend cell align={left},
		cycle list name=matlabcolor
		]
		\addplot+[thick,matlabblue,mark=square*,line width=1.5pt] table[x index=0,y index=1,col sep=tab] {figures/review2/pdet_over_N_BIC_Finite_MC_300_eps_0.csv};
		\addplot+[thick,matlabblue,mark=diamond*,line width=1.5pt] table[x index=0,y index=1,col sep=tab] {figures/review2/pdet_over_N_BIC_Finite_MC_300_eps_0.02.csv};
		\addplot+[thick,matlaborange,mark=square*,line width=1.5pt] table[x index=0,y index=5,col sep=tab] {figures/review2/pdet_over_N_BIC_Finite_MC_300_eps_0.csv};
		\addplot+[thick,matlaborange,mark=diamond*,line width=1.5pt] table[x index=0,y index=5,col sep=tab] {figures/review2/pdet_over_N_BIC_Finite_MC_300_eps_0.02.csv};
		%
	\end{axis}
\end{tikzpicture}

%% file: figures/review2/pdet_over_N_asymp.tex
\begin{tikzpicture}
	\begin{axis}[
		/pgf/number format/.cd,
		fixed,
		1000 sep={},
		width = \figurewidth,
		height = \figureheight,
		xmin = 0,
		xmax = 500,
		ymax = 1,
		ymin = 0,
		ytick={0,0.1,0.2,0.3,0.4,0.5,0.6,0.7,0.8,0.9,1},
		yticklabels={$0$,$10$,$20$,$30$,$40$,$50$,$60$,$70$,$80$,$90$,$100$},
		grid,
		xlabel= Number of samples per cluster,
		ylabel= empirical probability of detection in \%,
		legend entries={{EM: Gauß, BIC: Gauß, $\epsilon = 0\%$}\\
			{EM: Gauß, BIC: Gauß, $\epsilon = 2\%$}\\
			{EM: Huber, BIC: Tukey, $\epsilon = 0\%$}\\
			{EM: Huber, BIC: Tukey, $\epsilon = 2\%$}\\
			{Asymptotic, $\epsilon = 2\%$}\\
			{Schwarz, $\epsilon = 2\%$}\\},
		legend pos=south east,
		legend cell align={left},
		cycle list name=matlabcolor
		]
		\addplot+[thick,matlabblue,mark=square*,line width=1.5pt] table[x index=0,y index=1,col sep=tab] {figures/review2/pdet_over_N_BIC_Asymptotic_MC_300_eps_0.csv};
		\addplot+[thick,matlabblue,mark=diamond*,line width=1.5pt] table[x index=0,y index=1,col sep=tab] {figures/review2/pdet_over_N_BIC_Asymptotic_MC_300_eps_0.02.csv};
		\addplot+[thick,matlaborange,mark=square*,line width=1.5pt] table[x index=0,y index=5,col sep=tab] {figures/review2/pdet_over_N_BIC_Asymptotic_MC_300_eps_0.csv};
		\addplot+[thick,matlaborange,mark=diamond*,line width=1.5pt] table[x index=0,y index=5,col sep=tab] {figures/review2/pdet_over_N_BIC_Asymptotic_MC_300_eps_0.02.csv};
		%
	\end{axis}
\end{tikzpicture}

%% file: figures/data_radar.tex
\begin{tikzpicture}
	\begin{axis}[
	/pgf/number format/.cd,
	fixed,
	1000 sep={},
	width = \figurewidth,
	height = \figureheight,
	xlabel= Feature 1,
	ylabel= Feature 2,
	zlabel= Feature 3,
	view={-40}{20}, 
	]
	\addplot3[scatter/classes={1={matlabblue,mark size=3pt},2={matlaborange,mark size=3pt},3={matlabyellow,mark size=3pt}, 4={matlabpurple,mark size=3pt}}, scatter, only marks, scatter src=explicit symbolic] table[x=x,y=y,z=z,meta=label] {figures/data_radar.csv};   
	\end{axis}
\end{tikzpicture}

%% file: figures/review/data_forest_non_vs_hinoki_N_255.tex
\begin{tikzpicture}
	\begin{axis}[
	/pgf/number format/.cd,
	fixed,
	1000 sep={},
	width = \figurewidth,
	height = \figureheight,
	xlabel= Feature 1,
	ylabel= Feature 2,
	zlabel= Feature 3,
	view={-40}{20}, 
	]
	\addplot3[scatter/classes={1={matlabblue,mark size=3pt},2={matlaborange,mark size=3pt},3={matlabyellow,mark size=3pt}, 4={matlabpurple,mark size=3pt}}, scatter, only marks, scatter src=explicit symbolic] table[x=x,y=y,z=z,meta=label] {figures/review/data_forest_non_vs_hinoki_N_500.csv};   
	\end{axis}
\end{tikzpicture}

%% file: figures/bic_radar.tex
\begin{tikzpicture}
	\begin{axis}[
	/pgf/number format/.cd,
	fixed,
	1000 sep={},
	width = \figurewidth,
	height = \figureheight,
	xmin = 1,
	xmax = 8,
	ymax = 1100,
	ymin = 500,
	grid,
	xlabel= Number of clusters,
	ylabel= BIC,
    legend entries={{EM: Gauß, BIC: Finite Gauß}\\
					{EM: t, BIC: Finite Tukey}\\
					{EM: Huber, BIC: Finite Tukey}\\},
	legend pos=south east,
	legend cell align={left},
	cycle list name=matlabcolor
	]
	\addplot+[thick,line width=1.5pt] table[x index=0,y index=1,col sep=tab] {figures/bic_radar_EM_Gaus_BIC_Gaus.csv};
	\addplot+[thick,line width=1.5pt] table[x index=0,y index=1,col sep=tab] {figures/bic_radar_EM_t_BIC_Tukey.csv};
	\addplot+[thick,line width=1.5pt] table[x index=0,y index=1,col sep=tab] {figures/bic_radar_EM_Huber_BIC_Tukey.csv};
	%
	\end{axis}
\end{tikzpicture}

%% file: figures/bic_radar_Schwarz.tex
\begin{tikzpicture}
	\begin{axis}[
	/pgf/number format/.cd,
	fixed,
	1000 sep={},
	width = \figurewidth,
	height = \figureheight,
	xmin = 1,
	xmax = 8,
	ymax = 1300,
	ymin = 500,
	grid,
	xlabel= Number of clusters,
	ylabel= BIC,
    legend entries={{EM: Gauß, BIC: Schwarz Gauß}\\
					{EM: t, BIC: Schwarz Tukey}\\
					{EM: Huber, BIC: Schwarz Tukey}\\},
	legend pos=south east,
	legend cell align={left},
	cycle list name=matlabcolor
	]
	\addplot+[thick,line width=1.5pt] table[x index=0,y index=3,col sep=tab] {figures/bic_radar_EM_Gaus_BIC_Gaus.csv};
	\addplot+[thick,line width=1.5pt] table[x index=0,y index=3,col sep=tab] {figures/bic_radar_EM_t_BIC_Tukey.csv};
	\addplot+[thick,line width=1.5pt] table[x index=0,y index=3,col sep=tab] {figures/bic_radar_EM_Huber_BIC_Tukey.csv};
	%
	\end{axis}
\end{tikzpicture}

%% file: Paper/ap_FIM_RES_mumu.tex
\section{Derivatives for the FIM of the RES Distribution}
\label{ch:appDerivative}

\subsection{First derivative with respect to the mean}
First, we define $\bF$ as a $1 \times 1$ scalar function of the $r \times 1$ vector $\bmu_{m}$. Hence, the resulting Jacobian matrix is of size $1 \times r$. Setting $\bF(\bmu_{m})$ equal to the log-likelihood \eqref{eqn:loglikelihood} function we obtain
\begin{equation}
\bF(\bmu_{m}) = -\sum_{\xinX} \rho(t_{nm}) + N_{m} \ln \left(\frac{N_{m}}{N}\right) + \frac{N_{m}}{2} \ln\left(\left| \bS_{m}^{-1} \right|\right)
\end{equation}
and then apply the differential 
\begin{align}
\text{d}\bF&(\bmu_{m}) \notag \\
=& -\sum_{\xinX} \text{d}\rho(t_{nm}) \notag\\
=& -\sum_{\xinX} \psi(t_{nm}) \text{d}\left(\left(\xn - \bmu_{m}\right)^{\top} \bS_{m}^{-1} \left(\xn - \bmu_{m}\right)\right)\notag\\
\begin{split}
=& -\sum_{\xinX} \psi(t_{nm}) \Bigl( \left(- \text{d}\bmu_{m}\right)^{\top} \bS_{m}^{-1} \left(\xn - \bmu_{m}\right) 
\\&+  \left(\xn - \bmu_{m}\right)^{\top} \bS_{m}^{-1} \left(- \text{d}\bmu_{m}\right)\Bigr) 
\end{split} \notag\\
=& \sum_{\xinX} 2 \psi(t_{nm}) \left(\xn - \bmu_{m}\right)^{\top} \bS_{m}^{-1} \text{d}\bmu_{m}.
\label{eqn:dt}
\end{align}
Finally, the Jacobian matrix of $\bF(\bmu_{m})$, which we will denote as $\bF_{\bmu}$, becomes
\begin{equation}
D\bF(\bmu_{m}) = \bF_{\bmu} = 2 \sum_{\xinX}  \psi(t_{nm}) \left(\xn - \bmu_{m}\right)^{\top} \bS_{m}^{-1}.
\label{eqn:dmu}
\end{equation}

For the second derivative, $\bF_{\bmu}$ is a $1 \times r$ vector function of the $r \times 1$ vector $\bmu_{m}$. Hence, the resulting Jacobian matrix is of size $r \times r$. Introducing $\bxt_{n} \triangleq \xn - \bmu_{m}$ and starting with the differential of \eqref{eqn:dmu}
\begin{align}
\text{d}\bF&_{\bmu}(\bmu_{m}) \notag \\
\begin{split}
 =& 2 \sum_{\xinX} \Bigl[\text{d}\psi(t_{nm}) \bxt_{n}^{\top} \bS_{m}^{-1} + \psi(t_{nm}) \text{d}\left(\bxt_{n}^{\top} \bS_{m}^{-1} \right) \Bigr] 
\end{split}\notag\\
\begin{split}
=& -\sum_{\xinX} \Bigl[4 \eta(t_{nm}) \bxt_{n}^{\top} \bS_{m}^{-1} \text{d}\bmu_{m}\bxt_{n}^{\top} \bS_{m}^{-1}
\\& + 2 \psi(t_{nm}) \left(\text{d}\bmu_{m}\right)^{\top} \bS_{m}^{-1}\Bigr]
\end{split}\notag
\end{align}
and applying the $\vecop$ operator
\begin{align}
\begin{split}
\text{d}\vecop& (\bF_{\bmu}(\bmu_{m})) \notag \\
=& - \sum_{\xinX} \left[4\eta(t_{nm}) \vecop\left(\bxt_{n}^{\top} \bS_{m}^{-1} \text{d}\bmu_{m} \bxt_{n}^{\top} \bS_{m}^{-1}\right) \right.\\ &\left.+ 2\psi(t_{nm}) \vecop\left( \left(\text{d}\bmu_{m}\right)^{\top} \bS_{m}^{-1}\right)\right]
\end{split}\notag\\
\begin{split}
=& - \sum_{\xinX} \Bigl[4\eta(t_{nm}) \left(\bS_{m}^{-1}\bxt_{n} \bxt_{n}^{\top} \bS_{m}^{-1}\right) 
\\&+  2\psi(t_{nm}) \bS_{m}^{-1} \Bigr]\text{d}\bmu_{m}
\end{split}\notag
\end{align}
yields the Jacobian matrix $\bF_{\bmu\bmu}$ as
\begin{align}
\begin{split}
D\bF_{\bmu}(\bmu_{m}) = \bF_{\bmu\bmu} =& - \sum_{\xinX} \Bigl[4\eta(t_{nm}) \bS_{m}^{-1}\bxt_{n} \bxt_{n}^{\top} \bS_{m}^{-1} 
\\&+  2\psi(t_{nm}) \bS_{m}^{-1} \Bigr].
\end{split}
\label{eqn:dmumu}
\end{align}
Evaluating $\bF_{\bmu\bmu}$ at $\bShat_{m}$ and $\bmuhat_{m}$ with $\bxhat_{n} \triangleq \xn - \bmuhat_{m}$ from Appendix \ref{ch:ml} leads to
\begin{align}
\begin{split}
\bFhat_{\bmu\bmu} =& - 4 \bShat_{m}^{-1} \left(\sum_{\xinX} \eta(\that_{nm})\bxhat_{n} \bxhat_{n}^{\top}\right) \bShat_{m}^{-1} 
\\&- 2 \bShat_{m}^{-1} \sum_{\xinX}\psi(\that_{nm}).
\end{split}
\end{align}

For the other second derivative, $\bF_{\bmu}$ is a $1 \times r$ vector function of the $r \times r$ matrix $\bS_{m}$. Hence the resulting Jacobian matrix should be of size $r \times r^{2}$. However because $\bS_{m}$ is a symmetric matrix and only the unique elements are needed, we use the duplication matrix $\bD_{r}$ to only keep the unique elements of $\bS_{m}$. Therefore, the resulting matrix only is of size $r \times \frac{1}{2}r(r+1)$. Starting with the differential of Eq.~\eqref{eqn:dmu}
\begin{align}
\text{d}\bF&_{\bmu}(\bS_{m}) \notag\\
=& 2 \sum_{\xinX} \left[\text{d}\psi(t_{nm}) \bxt_{n}^{\top} \bS_{m}^{-1} + \psi(t_{nm}) \text{d}\left(\bxt_{n}^{\top} \bS_{m}^{-1} \right) \right] \notag\\
\begin{split}
=& -2 \sum_{\xinX} \Bigl[\eta(t_{nm}) \bxt_{n}^{\top} \bS_{m}^{-1} \text{d}\bS_{m} \bS_{m}^{-1} \bxt_{n} \bxt_{n}^{\top} \bS_{m}^{-1} 
\\&+ \psi(t_{nm}) \bxt_{n}^{\top} \bS_{m}^{-1} \text{d}\bS_{m} \bS_{m}^{-1} \Bigr].
\end{split}
\end{align}
Application of the $\vecop$ operator leads to
\begin{align}
\text{d}\vecop& (\bF_{\bmu}(\bS_{m})) \notag \\
\begin{split}
=& - 2 \sum_{\xinX} \Bigl[\eta(t_{nm}) \left(\left(\bS_{m}^{-1} \bxt_{n} \bxt_{n}^{\top} \bS_{m}^{-1}\right)^{\top} \otimes \bxt_{n}^{\top} \bS_{m}^{-1}\right)\bD_{r} 
\\ & + \psi(t_{nm}) \left( \bS_{m}^{-1} \otimes \bxt_{n}^{\top} \bS_{m}^{-1}\right) \bD_{r} \Bigr] \text{ d} \vechop\left(\bS_{m}\right)
\end{split} \notag
\end{align}
so that
\begin{align}
D\bF_{\bmu}&(\bS_{m}) = \bF_{\bmu\bS} \notag \\
\begin{split}
=& - 2 \sum_{\xinX} \Bigl[\eta(t_{nm})\left(\bS_{m}^{-1} \bxt_{n} \bxt_{n}^{\top} \bS_{m}^{-1} \otimes \bxt_{n}^{\top} \bS_{m}^{-1}\right) 
\\&+ \psi(t_{nm}) \left( \bS_{m}^{-1} \otimes \bxt_{n}^{\top} \bS_{m}^{-1}\right) \Bigr]\bD_{r} .
\end{split}
\label{eqn:dmuS}
\end{align}
Evaluating $\bF_{\bmu\bS}$ at $\bShat_{m}$ and $\bmuhat_{m}$ from Appendix \ref{ch:ml} leads to
\begin{align}
\begin{split}
\bFhat_{\bmu\bS} =& -2 \sum_{\xinX} \Bigl[\eta(\that_{nm}) \left( \bShat_{m}^{-1} \bxhat_{n} \bxhat_{n}^{\top} \bShat_{m}^{-1} \otimes \bxhat_{n}^{\top}\bShat_{m}^{-1}  \right) \bD_{r}
\\&+ \psi(\that_{nm})\left( \bShat_{m}^{-1} \otimes \bxhat_{n}^{\top}\bShat_{m}^{-1} \right) \bD_{r}\Bigr] 
\end{split} \notag \\
\begin{split}
=& -2 \sum_{\xinX} \eta(\that_{nm}) \left( \bShat_{m}^{-1} \bxhat_{n} \bxhat_{n}^{\top} \bShat_{m}^{-1} \otimes \bxhat_{n}^{\top}\bShat_{m}^{-1}\right) \bD_{r}
\\&- 2 \left( \bShat_{m}^{-1} \otimes \left(\sum_{\xinX}\psi(\that_{nm})\bxhat_{n}^{\top} \right)\bShat_{m}^{-1} \right)\bD_{r}  \end{split} \notag \\
=& -2 \sum_{\xinX} \eta(\that_{nm})  \left( \bShat_{m}^{-1} \bxhat_{n} \bxhat_{n}^{\top} \bShat_{m}^{-1} \otimes \bxhat_{n}^{\top}\bShat_{m}^{-1} \right)\bD_{r}.
\end{align}
Here, we used that
\begin{align}
\begin{split}
\sum_{\xinX}&\psi(\that_{nm}) \bxhat_{n} 
= \sum_{\xinX}\psi(\that_{nm}) \xn 
\\&- \left(\sum_{\xinX}\psi(\that_{nm})\right) \frac{\sum_{\xinX} \psi(\that_{nm}) \xn}{\sum_{\xinX} \psi(\that_{nm})}=0
\end{split}
\label{eqn:omega_x_0}
\end{align}
Due to the symmetry of the FIM, it is evident that
\begin{equation}
\bF_{\bmu\bS} = \left(\bF_{\bS\bmu}\right)^{\top}, \qquad \bFhat_{\bmu\bS} = \left(\bFhat_{\bS\bmu}\right)^{\top}.
\label{eqn:FSmuSmuS}
\end{equation}

\subsection{First derivative with respect to the variance}
We define $\bF$ as a $1 \times 1$ scalar function of the $r \times r$ matrix $\bS_{m}$. Hence, the resulting Jacobian matrix should be of size $1 \times r^2$. Again, we only keep the unique elements, such that, $\bF_{\bS}$ is of size $r \times \frac{1}{2}r(r+1)$. Setting $\bF(\bS_{m})$ equal to the log-likelihood function we obtain
\begin{align}
\bF& (\bS_{m}) =\ln\left(\mathcal{L}(\btheta_{m}|\mathcal{X}_{m})\right) \notag \\
&= -\sum_{\xinX} \rho(t_{nm}) + N_{m} \ln \left(\frac{N_{m}}{N}\right) + \frac{N_{m}}{2} \ln\left(\left| \bS_{m}^{-1} \right|\right),
\end{align}
taking the differential yields
\begin{align}
\text{d}\bF&(\bS_{m}) = -\sum_{\xinX} \text{d}\rho(t_{nm}) - \frac{N_{m}}{2} \text{d}\ln\left(\left| \bS_{m} \right|\right)\notag\\
\begin{split}
=& \sum_{\xinX} \psi(t_{nm}) \bxt_{n}^{\top} \bS_{m}^{-1} \text{d}\bS_{m} \bS_{m}^{-1} \bxt_{n} 
- \frac{N_{m}}{2} \Tr\left(\bS_{m}^{-1} \text{d} \bS_{m}\right),
\end{split}
\end{align}
vectorization results in
\begin{align}
\text{d}\vecop &(\bF(\bS_{m})) \notag\\
\begin{split}
=& \sum_{\xinX} \psi(t_{nm}) \left(\left(\bS_{m}^{-1} \bxt_{n}\right)^{\top} \otimes\bxt_{n}^{\top} \bS_{m}^{-1}\right)\text{d}\vecop\left(\bS_{m}\right)
\\ &- \frac{N_{m}}{2} \Tr\left(\bS_{m}^{-1} \text{d} \bS_{m}\right)
\end{split}\notag\\
\begin{split}
=& \sum_{\xinX} \psi(t_{nm}) \left(\bxt_{n}^{\top} \bS_{m}^{-1} \otimes \bxt_{n}^{\top} \bS_{m}^{-1}\right) \bD_{r} \text{ d} \vechop\left(\bS_{m}\right)\\ &- \frac{N_{m}}{2} \vecop\left(\bS_{m}^{-1}\right)^{\top} \bD_{r} \text{ d} \vechop\left(\bS_{m}\right)
\end{split}
\end{align}
and the Jacobian matrix becomes
\begin{align}
\begin{split}
D\bF(\bS_{m})= \bF_{\bS} =& \sum_{\xinX} \psi(t_{nm}) \left(\bxt_{n}^{\top} \bS_{m}^{-1} \otimes \bxt_{n}^{\top} \bS_{m}^{-1}\right)\bD_{r} 
\\&- \frac{N_{m}}{2} \vecop\left(\bS_{m}^{-1}\right)^{\top}\bD_{r}.
\end{split}
\label{eqn:dS}
\end{align}

Defining $\bF_{\bS}$ as a $1 \times \frac{1}{2}r(r+1)$ scalar function of the $r \times r$ matrix $\bS_{m}$, the resulting Jacobian matrix should be of size $\frac{1}{2}r(r+1) \times r^2$. As before, only the unique elements are of interest. Hence, the final size is $\frac{1}{2}r(r+1) \times \frac{1}{2}r(r+1)$. Starting with the differential of \eqref{eqn:dS} yields
\begin{align}
\text{d} (\bF&_{\bS}(\bS_{m})) \notag \\
\begin{split}
=& \sum_{\xinX} \Bigl[\text{d}\psi(t_{nm}) \left(\bxt_{n}^{\top} \bS_{m}^{-1} \otimes \bxt_{n}^{\top} \bS_{m}^{-1}\right)\bD_{r} \\&+ \psi(t_{nm}) \text{d}\left(\bxt_{n}^{\top} \bS_{m}^{-1} \otimes \bxt_{n}^{\top} \bS_{m}^{-1}\right)\bD_{r}\Bigr] 
\\&- \frac{N_{m}}{2} \vecop\left(\text{d} \bS_{m}^{-1}\right)^{\top}\bD_{r}
\end{split}\notag\\
\begin{split}
=& -\sum_{\xinX} \Bigl[\eta(t_{nm}) \bxt_{n}^{\top} \bS_{m}^{-1} \text{d}\bS_{m} \bS_{m}^{-1} \bxt_{n} 
\\&\left(\bxt_{n}^{\top} \bS_{m}^{-1} \otimes \bxt_{n}^{\top} \bS_{m}^{-1}\right)\bD_{r} 
\\&- \psi(t_{nm}) \text{d}\left(\bxt_{n}^{\top} \bS_{m}^{-1} \otimes \bxt_{n}^{\top} \bS_{m}^{-1}\right)\bD_{r}\Bigr] 
\\&+ \frac{N_{m}}{2} \vecop\left(\bS_{m}^{-1} \text{d}\bS_{m} \bS_{m}^{-1}\right)^{\top}\bD_{r}
\end{split}
\end{align}
and applying the $\vecop$ operator leads to
\begin{align}
\begin{split}
\text{d} \vecop &(\bF_{\bS}(\bS_{m})) 
\\=& -\sum_{\xinX} \Bigl[\eta(t_{nm}) \vecop\Bigl(\bxt_{n}^{\top} \bS_{m}^{-1} \text{d}\bS_{m} \bS_{m}^{-1} \bxt_{n} 
\\&\times \left(\bxt_{n}^{\top} \bS_{m}^{-1} \otimes \bxt_{n}^{\top} \bS_{m}^{-1}\right)\bD_{r} \Bigr)
\\&- \psi(t_{nm})\vecop\left(\text{d}\left(\bxt_{n}^{\top} \bS_{m}^{-1} \otimes \bxt_{n}^{\top} \bS_{m}^{-1}\right)\bD_{r}\right)\Bigr] 
\\&+ \frac{N_{m}}{2} \vecop\left(\vecop\left(\bS_{m}^{-1} \text{d}\bS_{m} \bS_{m}^{-1}\right)^{\top}\bD_{r}\right)
\end{split}\notag\\
\begin{split}
=& -\sum_{\xinX} \Bigl[\eta(t_{nm}) \Bigl(\left(\bS_{m}^{-1} \bxt_{n} \left(\bxt_{n}^{\top} \bS_{m}^{-1} \otimes \bxt_{n}^{\top} \bS_{m}^{-1}\right)\bD_{r}\right)^{\top} 
\\&\otimes \bxt_{n}^{\top} \bS_{m}^{-1} \Bigr) \text{d}\vecop\left(\bS_{m}\right)- \psi(t_{nm}) \bD_{r}^{\top}\left(\bI_{r} \otimes \bK_{r,1} \otimes \bI_{1}\right)
\\&\times \Bigl[\left(\bI_{r} \otimes \vecop\left(\bxt_{n}^{\top} \bS_{m}^{-1}\right)\right) 
+ \left(\vecop\left(\bxt_{n}^{\top} \bS_{m}^{-1}\right)\otimes \bI_{r} \right) \Bigr] 
\\&\times \text{d}\vecop\left(\bxt_{n}^{\top} \bS_{m}^{-1}\right) \Bigr] + \frac{N_{m}}{2} \bD_{r}^{\top} \vecop\left(\bS_{m}^{-1} \text{d}\bS_{m} \bS_{m}^{-1}\right)
\end{split}\notag\\
\begin{split}
=& -\sum_{\xinX} \Bigl[\eta(t_{nm}) \bD_{r}^{\top} \left( \bS_{m}^{-1} \bxt_{n}\otimes  \bS_{m}^{-1} \bxt_{n}\right)
\\&\times \left( \bxt_{n}^{\top}\bS_{m}^{-1} \otimes \bxt_{n}^{\top} \bS_{m}^{-1}\right) \bD_{r} 
\\&+ \psi(t_{nm}) \bD_{r}^{\top} \left(\bI_{r} \otimes \bS_{m}^{-1}\bxt_{n} \right) \left(\bS_{m}^{-1} \otimes \bxt_{n}^{\top} \bS_{m}^{-1}\right)\bD_{r} 
\\&+ \psi(t_{nm}) \bD_{r}^{\top}\left(\bS_{m}^{-1}\bxt_{n}\otimes \bI_{r} \right) \left(\bxt_{n}^{\top} \bS_{m}^{-1} \otimes \bS_{m}^{-1}\right) \bD_{r} 
\\&+ \frac{1}{2} \bD_{r}^{\top} \left(\bS_{m}^{-1} \otimes \bS_{m}^{-1}\right)\bD_{r}\Bigr]\text{ d}\vechop\left(\bS_{m}\right)
\end{split}\notag\\
\begin{split}
=& -\sum_{\xinX} \Bigl[\eta(t_{nm}) \bD_{r}^{\top} \Bigl( \bS_{m}^{-1} \bxt_{n} \bxt_{n}^{\top}\bS_{m}^{-1} 
\\&\otimes \bS_{m}^{-1} \bxt_{n} \bxt_{n}^{\top}\bS_{m}^{-1}\Bigr) \bD_{r} 
\\&+ \psi(t_{nm}) \bD_{r}^{\top}\left(\bS_{m}^{-1} \otimes \bS_{m}^{-1}\bxt_{n}\bxt_{n}^{\top} \bS_{m}^{-1}\right)\bD_{r} 
\\&+ \psi(t_{nm}) \bD_{r}^{\top}\left(\bS_{m}^{-1}\bxt_{n}\bxt_{n}^{\top} \bS_{m}^{-1} \otimes \bS_{m}^{-1}\right) \bD_{r}
\\& + \frac{1}{2} \bD_{r}^{\top} \left(\bS_{m}^{-1} \otimes \bS_{m}^{-1}\right)\bD_{r} \Bigr] \text{ d}\vechop\left(\bS_{m}\right)
\end{split}\notag
\end{align}
with the commutation matrix $\bK_{r,1} = \bI_{r}$.

Now the Jacobian matrix is obtained as
\begin{align}
D\bF_{\bS}&(\bS_{m}) = \bF_{\bS\bS} \notag \\*
\begin{split}
=& -\sum_{\xinX} \Bigl[\eta(t_{nm}) \bD_{r}^{\top} \Bigl( \bS_{m}^{-1} \bxt_{n} \bxt_{n}^{\top}\bS_{m}^{-1} 
\\& \otimes \bS_{m}^{-1} \bxt_{n} \bxt_{n}^{\top}\bS_{m}^{-1}\Bigr) \bD_{r} 
\\&+ \psi(t_{nm}) \bD_{r}^{\top}\left(\bS_{m}^{-1} \otimes \bS_{m}^{-1}\bxt_{n}\bxt_{n}^{\top} \bS_{m}^{-1}\right)\bD_{r} 
\\&+ \psi(t_{nm}) \bD_{r}^{\top}\left(\bS_{m}^{-1}\bxt_{n}\bxt_{n}^{\top} \bS_{m}^{-1} \otimes \bS_{m}^{-1}\right) \bD_{r} \Bigr]  
\\&+ \frac{N_{m}}{2} \bD_{r}^{\top} \left(\bS_{m}^{-1} \otimes \bS_{m}^{-1}\right)\bD_{r}.
\end{split}
\label{eqn:dSS}
\end{align}
Evaluating $\bF_{\bS\bS}$ at $\bShat_{m}$ and $\bmuhat_{m}$ from Appendix \ref{ch:ml} leads to
\begin{align}
\bFhat&_{\bS\bS} \notag \\
\begin{split}
=& -\sum_{\xinX} \eta(\that_{nm}) \bD_{r}^{\top} \Bigl( \bShat_{m}^{-1} \bxhat_{n} \bxhat_{n}^{\top}\bShat_{m}^{-1} 
\\&\otimes   \bShat_{m}^{-1} \bxhat_{n} \bxhat_{n}^{\top}\bShat_{m}^{-1}\Bigr) \bD_{r}
\\&-\sum_{\xinX} \psi(t_{nm}) \bD_{r}^{\top}\left(\bShat_{m}^{-1} \otimes \bShat_{m}^{-1}\bxhat_{n}\bxt_{n}^{\top} \bShat_{m}^{-1}\right)\bD_{r}
\\&- \sum_{\xinX} \psi(t_{nm}) \bD_{r}^{\top}\left(\bShat_{m}^{-1}\bxhat_{n}\bxt_{n}^{\top} \bShat_{m}^{-1} \otimes \bShat_{m}^{-1}\right) \bD_{r} 
\\&+ \frac{N_{m}}{2} \bD_{r}^{\top} \left(\bShat_{m}^{-1} \otimes \bShat_{m}^{-1}\right)\bD_{r}
\end{split}\notag \\
\begin{split}
=& -\sum_{\xinX} \eta(\that_{nm}) \bD_{r}^{\top} \Bigl( \bShat_{m}^{-1} \bxhat_{n} \bxhat_{n}^{\top}\bShat_{m}^{-1} 
\\& \otimes   \bShat_{m}^{-1} \bxhat_{n} \bxhat_{n}^{\top}\bShat_{m}^{-1}\Bigr) \bD_{r}
\\&-  \bD_{r}^{\top}\left( \bShat_{m}^{-1} \otimes  \bShat_{m}^{-1} \left( \sum_{\xinX} \psi(\that_{nm}) \bxhat_{n} \bxhat_{n}^{\top}\right) \bShat_{m}^{-1} \right)\bD_{r} 
\\&-  \bD_{r}^{\top}\left( \bShat_{m}^{-1} \left( \sum_{\xinX} \psi(\that_{nm}) \bxhat_{n} \bxhat_{n}^{\top}\right) \bShat_{m}^{-1} \otimes \bShat_{m}^{-1}\right)\bD_{r} 
\\&+ \frac{N_{m}}{2} \bD_{r}^{\top} \left(\bShat_{m}^{-1} \otimes \bShat_{m}^{-1}\right)\bD_{r}
\end{split}\notag \\
& \text{with Eq.~\eqref{eqn:mlS}} \notag \\
\begin{split}
=& -\sum_{\xinX} \eta(\that_{nm}) \bD_{r}^{\top} \Bigl( \bShat_{m}^{-1} \bxhat_{n} \bxhat_{n}^{\top}\bShat_{m}^{-1} 
\\& \otimes   \bShat_{m}^{-1} \bxhat_{n} \bxhat_{n}^{\top}\bShat_{m}^{-1}\Bigr) \bD_{r} - \frac{N_{m}}{2}\bD_{r}^{\top}\left( \bShat_{m}^{-1} \otimes  \bShat_{m}^{-1} \right)\bD_{r} 
\\&- \frac{N_{m}}{2}\bD_{r}^{\top}\left( \bShat_{m}^{-1} \otimes  \bShat_{m}^{-1} \right)\bD_{r}
\\&+ \frac{N_{m}}{2} \bD_{r}^{\top} \left(\bShat_{m}^{-1} \otimes \bShat_{m}^{-1}\right)\bD_{r}
\end{split}\notag \\
\begin{split}
=& - \bD_{r}^{\top} \left( \bShat_{m}^{-1} \otimes \bShat_{m}^{-1} \right) \left(\sum_{\xinX} \eta(\that_{nm})\left( \bxhat_{n} \bxhat_{n}^{\top} \otimes \bxhat_{n} \bxhat_{n}^{\top}\right)\right) 
\\& \times\left( \bShat_{m}^{-1} \otimes \bShat_{m}^{-1} \right)\bD_{r} - \frac{N_{m}}{2}\bD_{r}^{\top}\left( \bShat_{m}^{-1} \otimes \bShat_{m}^{-1} \right)\bD_{r}.
\end{split}
\end{align}

%% file: Paper/ap_ML_RES_short.tex
\section{Maximum Likelihood Estimators for RES distributions}
\label{ch:ml}
\subsection{Maximum Likelihood Estimator for the mean}
Setting \eqref{eqn:dmu} equal to zero and solving for $\bmuhat_{m}$ leads to the ML estimator of $\bmu_{m}$ as
\begin{align}
 & \bmuhat_{m} = \frac{\sum_{\xinX} \psi(\that_{nm}) \xn}{\sum_{\xinX} \psi(\that_{nm})}
\label{eqn:mlmu}
\end{align}
with
\begin{equation}
	\that_{nm} =  \left(\xn - \bmuhat_{m}\right)^{\top} \bShat_{m}^{-1} \left(\xn - \bmuhat_{m}\right).
	\label{eqn:that}
\end{equation}

\subsection{Maximum Likelihood Estimator for the variance}
Setting the first derivative \eqref{eqn:dS} equal to zero yields
\begin{align}
\begin{split}
&\sum_{\xinX} \psi(t_{nm}) \left(\bxt_{n}^{\top} \bS_{m}^{-1} \otimes \bxt_{n}^{\top} \bS_{m}^{-1}\right) \bD_{r}\bD_{r}^{+}
 \\&= \frac{N_{m}}{2} \vecop\left(\bS_{m}^{-1}\right)^{\top}\bD_{r}\bD_{r}^{+}
\end{split}\notag\\
\begin{split}
\Rightarrow &\sum_{\xinX} \psi(t_{nm}) \left(\bxt_{n}^{\top} \otimes \bxt_{n}^{\top} \right) \left(\bS_{m}^{-1} \otimes \bS_{m}^{-1}\right)
\\&= \frac{N_{m}}{2} \vecop\left(\bS_{m}^{-1}\right)^{\top}
\end{split}\notag\\
\begin{split}
\Rightarrow &\sum_{\xinX}  \psi(t_{nm}) \left(\bxt_{n}^{\top} \otimes \bxt_{n}^{\top} \right) 
\\& = \frac{N_{m}}{2} \vecop\left(\bS_{m}^{-1}\right)^{\top} \left(\bS_{m} \otimes \bS_{m}\right)
\end{split}\notag\\
\Rightarrow &\sum_{\xinX} \psi(t_{nm}) \left(\bxt_{n} \otimes \bxt_{n} \right) = \frac{N_{m}}{2}  \vecop\left(\bS_{m}\bS_{m}^{-1}\bS_{m}\right) \notag\\
\Rightarrow &  \vecop\left(\bShat_{m}\right)  = \frac{2}{N_{m}}\sum_{\xinX} \psi(\that_{nm}) \left(\bxhat_{n} \otimes \bxhat_{n} \right)
\label{eqn:mlSvec}
\end{align}
leads to a vectorized form of the ML estimator with $\bxhat_{n} \triangleq \xn - \bmuhat_{m}$. To obtain the matrix form, we apply the inverse vec operator, which results in
\begin{align}
\bShat_{m}  =& \frac{2}{N_{m}}\sum_{\xinX}  \left(\vecop(\bI_r)^\top \otimes \bI_r\right)\left(\bI_r \otimes \bxhat_{n} \otimes \psi(\that_{nm})\bxhat_{n} \right)\notag\\
=& \frac{2}{N_{m}}\sum_{\xinX}  \left(\vecop(\bI_r)^\top \left(\bI_r \otimes \bxhat_{n}\right)\right) \otimes \psi(\that_{nm}) \bI_r \bxhat_{n}\notag\\
=& \frac{2}{N_{m}}\sum_{\xinX}  \bxhat_{n}^{\top} \otimes \psi(t_{nm})\bxhat_{n}\notag\\
=& \frac{2}{N_{m}}\sum_{\xinX}  \psi(\that_{nm})\bxhat_{n}\bxhat_{n}^{\top}.
\label{eqn:mlS}
\end{align}

\begin{figure}[!ht]
	\removelatexerror
	\begin{algorithm}[H]
		\KwIn{$\mathcal{X}$, $i_{\max}$, $l$, $g(t)$, $\psi(t)$}
		\KwOut{$\bmuhat_{m}$, $\bShat_{m}$, $\hat{\gamma}_{m}$}
		\For{$m = 1,\dots, l$}
		{
			Initialize $\bmuhat_{m}^{(0)}$ with K-medoids\;
			\[
			\bShat_{m}^{(0)} = \frac{1}{N_m}\sum_{\xinX} \left(\xn - \bmuhat_{m}^{(0)}\right) \left(\xn - \bmuhat_{m}^{(0)}\right)^{\top}
			\]
			\[
			\hat{\gamma}_{m}^{(0)} = N_{m}/N
			\]
		}
		\For{$i = 1,\dots, i_{\max}$}
		{%
			E-step:\;
			\For{$m = 1,\dots, l$}
			{
				\For{$n = 1,\dots, N$}
				{
					\[
					\hat{v}_{nm}^{(i)} = \frac{\hat{\gamma}_{m}^{(i-1)} \left| \bShat_{m}^{(i-1)} \right|^{-\frac{1}{2}} g\left( \that^{(i-1)}_{nm}\right)}{\sum_{j=1}^{l} \hat{\gamma}_{j}^{(i-1)} \left| \bShat_{j}^{(i-1)} \right|^{-\frac{1}{2}} g\left( \that^{(i-1)}_{nj}\right)}
					\]
					\[
					\hat{v}_{nm}^{\prime(i)} = \hat{v}_{nm}^{(i)} \psi\left(\that_{nm}^{(i-1)}\right)
					\]
			}}
			M-Step:\;
			\For{$m = 1,\dots, l$}
			{
				\[
				\bmuhat_{m}^{(i)} = \sum_{n = 1}^{N} \hat{v}_{nm}^{\prime(i)} \xn^{(i)} \Bigg/ \sum_{n = 1}^{N} \hat{v}_{nm}^{\prime(i)}
				\]
				\[
				\bShat_{m}^{(i)} = \frac{2 \sum_{n = 1}^{N} \hat{v}_{nm}^{\prime(i)} \left(\xn - \bmuhat_{m}^{(i)}\right) \left(\xn - \bmuhat_{m}^{(i)}\right)^{\top}}{\sum_{n = 1}^{N} \hat{v}_{nm}^{(i)}}
				\]
				\[
				\hat{\gamma}_{m}^{(i)} = \frac{1}{N} \sum_{n = 1}^{N} \hat{v}_{nm}^{(i)}
				\]
			}
			Calculate log-likelihood:
			\[
			\begin{split}
			\ln\Bigl(\mathcal{L}(&\bPhihat_{l}^{(i)}|\mathcal{X} )\Bigr) 
			\\&= \sum_{n = 1}^{N} \ln \left(\sum_{m=1}^{l} \hat{\gamma}_{m}^{(i)} \left| \bShat_{m}^{(i)} \right|^{-\frac{1}{2}} g\left( \that_{nm}^{(i)}\right)\right)
			\end{split}
			\]
			\If{$\left|\ln\left(\mathcal{L}\left(\bPhihat_{l}^{(i)}|\mathcal{X} \right)\right)-\ln\left(\mathcal{L}\left(\bPhihat_{l}^{(i-1)}|\mathcal{X} \right)\right)\right| < \delta$}
			{
				break loop\;
			}
		}
		\caption{EM algorithm for RES distributions}
		\label{alg:em}
	\end{algorithm}
\end{figure}

%% file: Appendix/ap_FIM_RES.tex
\section{Derivatives for the FIM of the RES Distribution}
\label{ch:appDerivative}
The FIM requires the calculation of the second derivative of the log-likelihood function. In this appendix this is done for the set of RES distributions. Since the differentiation of matrices is not straight forward, the derivation is shown in detail. A short introduction on matrix calculus can be found in \cite{Magnus.2010}, a more detailed explanation is provided in \cite{Magnus.2007} and a large number of examples are discussed in \cite{Abadir.2005}. Most of the used matrix calculus rules can be found in these references and are also noted in Appendix \ref{ch:matcalc}.

\subsection{First derivative with respect to the mean}
First, we define $\bF$ as a $1 \times 1$ scalar function of the $r \times 1$ vector $\bmu_{m}$. Hence, the resulting Jacobian matrix is of size $1 \times r$. Setting $\bF(\bmu_{m})$ equal to the log-likelihood function we get
\begin{equation}
\bF(\bmu_{m}) = \ln\left(\mathcal{L}(\btheta_{m}|\mathcal{X}_{m})\right) = -\sum_{\xinX} \rho(t_{nm}) + N_{m} \ln \left(\frac{N_{m}}{N}\right) + \frac{N_{m}}{2} \ln\left(\left| \bS_{m}^{-1} \right|\right)
\end{equation}
and afterwards apply the differential 
\begin{align}
\text{d}\bF(\bmu_{m}) =& -\sum_{\xinX} \text{d}\rho(t_{nm}) \notag\\
& \text{with } \psi(t_{nm}) = \frac{\partial \rho(t_{nm})}{\partial t_{nm}} \notag \\
=& -\sum_{\xinX} \psi(t_{nm}) \text{d}\left(\left(\xn - \bmu_{m}\right)^{\top} \bS_{m}^{-1} \left(\xn - \bmu_{m}\right)\right)\notag\\
=& -\sum_{\xinX} \psi(t_{nm}) \left( \left(- \text{d}\bmu_{m}\right)^{\top} \bS_{m}^{-1} \left(\xn - \bmu_{m}\right) +  \left(\xn - \bmu_{m}\right)^{\top} \bS_{m}^{-1} \left(- \text{d}\bmu_{m}\right)\right) \notag\\
& \text{with } \alpha = \alpha^{\top}, \alpha \text{ being a scalar} \notag \\
=& \sum_{\xinX} \psi(t_{nm}) \left( \left(\left(\text{d}\bmu_{m}\right)^{\top} \bS_{m}^{-1} \left(\xn - \bmu_{m}\right)\right)^{\top} +  \left(\xn - \bmu_{m}\right)^{\top} \bS_{m}^{-1} \left( \text{d}\bmu_{m}\right)\right)\notag\\
& \text{with } \left(\bA\bB\right)^{\top} = \bB^{\top}\bA^{\top} \notag \\
=& \sum_{\xinX} \psi(t_{nm}) \left( \left(\xn - \bmu_{m}\right)^{\top} \bS_{m}^{-1} \left( \text{d}\bmu_{m}\right) +  \left(\xn - \bmu_{m}\right)^{\top} \bS_{m}^{-1} \left( \text{d}\bmu_{m}\right)\right)\notag\\
=& \sum_{\xinX} 2 \psi(t_{nm}) \left(\xn - \bmu_{m}\right)^{\top} \bS_{m}^{-1} \text{d}\bmu_{m}.
\label{eqn:dt}
\end{align}
Finally, the Jacobian matrix of $\bF(\bmu_{m})$, which we will denote as $\bF_{\bmu}$, becomes
\begin{equation}
D\bF(\bmu_{m}) = \bF_{\bmu} = 2 \sum_{\xinX}  \psi(t_{nm}) \left(\xn - \bmu_{m}\right)^{\top} \bS_{m}^{-1}.
\label{eqn:dmu}
\end{equation}

For the second derivative, $\bF_{\bmu}$ is a $1 \times r$ vector function of the $r \times 1$ vector $\bmu_{m}$, hence the resulting Jacobian matrix is of size $r \times r$. Starting with the differential of \eqref{eqn:dmu}
\begin{align}
\text{d}\bF_{\bmu}(\bmu_{m}) =& 2 \sum_{\xinX} \left[\text{d}\psi(t_{nm}) \left(\xn - \bmu_{m}\right)^{\top} \bS_{m}^{-1} + \psi(t_{nm}) \text{d}\left(\left(\xn - \bmu_{m}\right)^{\top} \bS_{m}^{-1} \right) \right] \notag\\
& \text{with } \eta(t_{nm}) = \frac{\partial \psi(t_{nm})}{\partial t_{nm}} \notag \\
=& -2 \sum_{\xinX} \left[2 \eta(t_{nm}) \left(\xn - \bmu_{m}\right)^{\top} \bS_{m}^{-1} \text{d}\bmu_{m} \left(\xn - \bmu_{m}\right)^{\top} \bS_{m}^{-1} + \psi(t_{nm}) \left(\text{d}\bmu_{m}\right)^{\top} \bS_{m}^{-1}\right]
\end{align}
and applying the $\vecop$ operator
\begin{align}
\begin{split}
\text{d}\vecop (\bF_{\bmu}(\bmu_{m})) =& - \sum_{\xinX} \left[4\eta(t_{nm}) \vecop\left(\left(\xn - \bmu_{m}\right)^{\top} \bS_{m}^{-1} \text{d}\bmu_{m} \left(\xn - \bmu_{m}\right)^{\top} \bS_{m}^{-1}\right) \right.\\ &\left.+ 2\psi(t_{nm}) \vecop\left( \left(\text{d}\bmu_{m}\right)^{\top} \bS_{m}^{-1}\right)\right]
\end{split}\notag\\
& \text{with \eqref{eqnA:vecABC} and \eqref{eqnA:vecAB}} \notag \\
\begin{split}
=& - \sum_{\xinX} \left[4\eta(t_{nm}) \left(\bS_{m}^{-1}\left(\xn - \bmu_{m}\right) \otimes \left(\xn - \bmu_{m}\right)^{\top}\bS_{m}^{-1} \right) \text{d}\vecop(\bmu_{m}) \right.\\ &\left. +  2\psi(t_{nm})\left( \bS_{m}^{-1} \otimes \bI_{1}  \right)\text{d}\vecop(\bmu_{m}) \right]
\end{split}\notag\\
& \text{with \eqref{eqnA:veckron} and } \bI_{1} = 1 \notag \\
=& - \sum_{\xinX} \left[4\eta(t_{nm}) \left(\bS_{m}^{-1}\left(\xn - \bmu_{m}\right) \left(\xn - \bmu_{m}\right)^{\top} \bS_{m}^{-1}\right) +  2\psi(t_{nm}) \bS_{m}^{-1} \right]\text{d}\bmu_{m}.
\end{align}
Hence, we obtain the Jacobian matrix $\bF_{\bmu\bmu}$ as
\begin{equation}
D\bF_{\bmu}(\bmu_{m}) = \bF_{\bmu\bmu} = - \sum_{\xinX} \left[4\eta(t_{nm}) \bS_{m}^{-1}\left(\xn - \bmu_{m}\right) \left(\xn - \bmu_{m}\right)^{\top} \bS_{m}^{-1} +  2\psi(t_{nm}) \bS_{m}^{-1} \right].
\label{eqn:dmumu}
\end{equation}

Evaluating $\bF_{\bmu\bmu}$ at $\bShat_{m}$ and $\bmuhat_{m}$ from Appendix \ref{ch:ml} leads to
\begin{equation}
\bFhat_{\bmu\bmu} = - 4 \bShat_{m}^{-1} \left(\sum_{\xinX} \eta(\that_{nm})\left(\xn - \bmuhat_{m}\right) \left(\xn - \bmuhat_{m}\right)^{\top}\right) \bShat_{m}^{-1} - 2 \bShat_{m}^{-1} \sum_{\xinX}\psi(\that_{nm})
\label{eqn:dmumuhat}
\end{equation}

For the other second derivative, $\bF_{\bmu}$ is a $1 \times r$ vector function of the $r \times r$ matrix $\bS_{m}$, hence the resulting Jacobian matrix should be of size $r \times r^{2}$, but because $\bS_{m}$ is a symmetric matrix and only the unique elements are needed, we use the duplication matrix \eqref{eqnA:vech} to only keep the unique elements of $\bS_{m}$. Therefore the resulting matrix only has the size $r \times \frac{1}{2}r(r+1)$. Starting with the differential of \eqref{eqn:dmu}
\begin{align}
\text{d}\bF_{\bmu}(\bS_{m}) =& 2 \sum_{\xinX} \left[\text{d}\psi(t_{nm}) \left(\xn - \bmu_{m}\right)^{\top} \bS_{m}^{-1} + \psi(t_{nm}) \text{d}\left(\left(\xn - \bmu_{m}\right)^{\top} \bS_{m}^{-1} \right) \right] \notag\\
& \text{with \eqref{eqnA:dinv}} \notag \\
\begin{split}
=& -2 \sum_{\xinX} \left[\eta(t_{nm}) \left(\xn - \bmu_{m}\right)^{\top} \bS_{m}^{-1} \text{d}\bS_{m} \bS_{m}^{-1} \left(\xn - \bmu_{m}\right) \left(\xn - \bmu_{m}\right)^{\top} \bS_{m}^{-1} \right.\\ &\left. + \psi(t_{nm}) \left(\xn - \bmu_{m}\right)^{\top} \bS_{m}^{-1} \text{d}\bS_{m} \bS_{m}^{-1} \right]
\end{split}
\end{align}
For ease of notation, we introduce
\begin{equation}
	\bxt_{n} \triangleq \xn - \bmu_{m}
\end{equation}
and continue with the application of the $\vecop$ operator
\begin{align}
\begin{split}
\text{d}\vecop (\bF_{\bmu}(\bS_{m})) =& - 2 \sum_{\xinX} \left[\eta(t_{nm}) \vecop\left(\bxt_{n} ^{\top} \bS_{m}^{-1} \text{d}\bS_{m} \bS_{m}^{-1} \bxt_{n} \bxt_{n}^{\top} \bS_{m}^{-1}\right) \right.\\ &\left.+ \psi(t_{nm}) \vecop\left( \bxt_{n}^{\top} \bS_{m}^{-1} \text{d}\bS_{m} \bS_{m}^{-1}\right)\right] 
\end{split}\notag\\
& \text{with \eqref{eqnA:vech}} \notag \\
\begin{split}
=& - 2 \sum_{\xinX} \Bigl[\eta(t_{nm}) \left(\left(\bS_{m}^{-1} \bxt_{n} \bxt_{n}^{\top} \bS_{m}^{-1}\right)^{\top} \otimes \bxt_{n}^{\top} \bS_{m}^{-1}\right)\bD_{r} \text{ d}\vechop \left(\bS_{m}\right) \\ & + \psi(t_{nm}) \left( \bS_{m}^{-1} \otimes \bxt_{n}^{\top} \bS_{m}^{-1}\right) \bD_{r} \text{ d}\vechop\left(\bS_{m}\right)\Bigr]
\end{split}
\end{align}
so that
\begin{equation}
D\bF_{\bmu}(\bS_{m}) = \bF_{\bmu\bS} = - 2 \sum_{\xinX} \Bigl[\eta(t_{nm})\left(\bS_{m}^{-1} \bxt_{n} \bxt_{n}^{\top} \bS_{m}^{-1} \otimes \bxt_{n}^{\top} \bS_{m}^{-1}\right)\bD_{r} + \psi(t_{nm}) \left( \bS_{m}^{-1} \otimes \bxt_{n}^{\top} \bS_{m}^{-1}\right)\bD_{r} \Bigr].
\label{eqn:dmuS}
\end{equation}

Evaluating $\bF_{\bmu\bS}$ at $\bShat_{m}$ and $\bmuhat_{m}$ with $\bxhat_{n} \triangleq \xn - \bmuhat_{m}$ from Appendix \ref{ch:ml} leads to
\begin{align}
\bFhat_{\bmu\bS} =& -2 \sum_{\xinX} \left[\eta(\that_{nm}) \left( \bShat_{m}^{-1} \bxhat_{n} \bxhat_{n}^{\top} \bShat_{m}^{-1} \otimes \bxhat_{n}^{\top}\bShat_{m}^{-1}  \right) \bD_{r}
+ \psi(\that_{nm})\left( \bShat_{m}^{-1} \otimes \bxhat_{n}^{\top}\bShat_{m}^{-1} \right) \bD_{r}\right]  \notag \\
=& -2 \sum_{\xinX} \eta(\that_{nm})  \left( \bShat_{m}^{-1} \bxhat_{n} \bxhat_{n}^{\top} \bShat_{m}^{-1} \otimes \bxhat_{n}^{\top}\bShat_{m}^{-1}  \right)\bD_{r}
- 2 \sum_{\xinX}\psi(\that_{nm})\left( \bShat_{m}^{-1} \otimes \bxhat_{n}^{\top}\bShat_{m}^{-1} \right)\bD_{r} \notag \\
=& -2 \sum_{\xinX} \eta(\that_{nm}) \left( \bShat_{m}^{-1} \bxhat_{n} \bxhat_{n}^{\top} \bShat_{m}^{-1} \otimes \bxhat_{n}^{\top}\bShat_{m}^{-1}\right) \bD_{r}
- 2 \left( \bShat_{m}^{-1} \otimes \left(\sum_{\xinX}\psi(\that_{nm})\bxhat_{n}^{\top} \right)\bShat_{m}^{-1} \right)\bD_{r}  \notag \\
& \text{with \eqref{eqn:omega_x_0}} \notag \\
=& -2 \sum_{\xinX} \eta(\that_{nm})  \left( \bShat_{m}^{-1} \bxhat_{n} \bxhat_{n}^{\top} \bShat_{m}^{-1} \otimes \bxhat_{n}^{\top}\bShat_{m}^{-1} \right)\bD_{r}
\label{eqn:dmuShat}
\end{align}

\subsection{First derivative with respect to the variance}
We define $\bF$ as a $1 \times 1$ scalar function of the $r \times r$ matrix $\bS_{m}$. Hence, the resulting Jacobian matrix should be of size $1 \times r^2$. Again, we only keep the unique elements, such that, $\bF_{\bS}$ is of size $r \times \frac{1}{2}r(r+1)$. Setting $\bF(\bS_{m})$ equal to the log-likelihood function we get
\begin{equation}
\bF(\bS_{m}) = \ln\left(\mathcal{L}(\btheta_{m}|\mathcal{X}_{m})\right) = -\sum_{\xinX} \rho(t_{nm}) + N_{m} \ln \left(\frac{N_{m}}{N}\right) + \frac{N_{m}}{2} \ln\left(\left| \bS_{m}^{-1} \right|\right)
\end{equation}
and taking the differential yields
\begin{align}
\text{d}\bF(\bS_{m}) =& -\sum_{\xinX} \text{d}\rho(t_{nm}) - \frac{N_{m}}{2} \text{d}\ln\left(\left| \bS_{m} \right|\right)\notag\\
& \text{with \eqref{eqnA:dinv} and \eqref{eqnA:dlndet}} \notag \\
=& \sum_{\xinX} \psi(t_{nm}) \left(\xn - \bmu_{m}\right)^{\top} \bS_{m}^{-1} \text{d}\bS_{m} \bS_{m}^{-1} \left(\xn - \bmu_{m}\right) - \frac{N_{m}}{2} \Tr\left(\bS_{m}^{-1} \text{d} \bS_{m}\right)\notag\\
=& \sum_{\xinX} \psi(t_{nm}) \left(\xn - \bmu_{m}\right)^{\top} \bS_{m}^{-1} \text{d}\bS_{m} \bS_{m}^{-1} \left(\xn - \bmu_{m}\right) - \frac{N_{m}}{2} \Tr\left(\bS_{m}^{-1} \text{d} \bS_{m}\right)
\end{align}
with vectorization
\begin{align}
\begin{split}
\text{d}\vecop (\bF(\bS_{m})) =& \sum_{\xinX} \psi(t_{nm}) \vecop\left(\left(\xn - \bmu_{m}\right)^{\top} \bS_{m}^{-1} \text{d}\bS_{m} \bS_{m}^{-1} \left(\xn - \bmu_{m}\right)\right)\\ 
&- \frac{N_{m}}{2} \vecop\left(\Tr\left(\bS_{m}^{-1} \text{d} \bS_{m}\right)\right)
\end{split}\notag\\
\begin{split}
=& \sum_{\xinX} \psi(t_{nm}) \left(\left(\bS_{m}^{-1} \left(\xn - \bmu_{m}\right)\right)^{\top} \otimes \left(\xn - \bmu_{m}\right)^{\top} \bS_{m}^{-1}\right)\text{d}\vecop\left(\bS_{m}\right)
\\ &- \frac{N_{m}}{2} \Tr\left(\bS_{m}^{-1} \text{d} \bS_{m}\right)
\end{split}\notag\\
\begin{split}
& \text{with \eqref{eqnA:vecTr}} \notag \\
=& \sum_{\xinX} \psi(t_{nm}) \left(\left(\xn - \bmu_{m}\right)^{\top} \bS_{m}^{-1} \otimes \left(\xn - \bmu_{m}\right)^{\top} \bS_{m}^{-1}\right)\text{d}\vecop\left(\bS_{m}\right)\\ &- \frac{N_{m}}{2} \vecop\left(\bS_{m}^{-1}\right)^{\top} \text{d}\vecop\left(\bS_{m}\right)
\end{split}\notag\\
\begin{split}
=& \sum_{\xinX} \psi(t_{nm}) \left(\left(\xn - \bmu_{m}\right)^{\top} \bS_{m}^{-1} \otimes \left(\xn - \bmu_{m}\right)^{\top} \bS_{m}^{-1}\right) \bD_{r} \text{ d} \vechop\left(\bS_{m}\right)\\ &- \frac{N_{m}}{2} \vecop\left(\bS_{m}^{-1}\right)^{\top} \bD_{r} \text{ d} \vechop\left(\bS_{m}\right)
\end{split}
\end{align}
and the Jacobian matrix
\begin{align}
D\bF(\bS_{m})=& \bF_{\bS} \notag\\
=& \sum_{\xinX} \psi(t_{nm}) \left(\left(\xn - \bmu_{m}\right)^{\top} \bS_{m}^{-1} \otimes \left(\xn - \bmu_{m}\right)^{\top} \bS_{m}^{-1}\right)\bD_{r} - \frac{N_{m}}{2} \vecop\left(\bS_{m}^{-1}\right)^{\top}\bD_{r}.
\label{eqn:dS}
\end{align}

Defining $\bF_{\bS}$ as a $1 \times \frac{1}{2}r(r+1)$ scalar function of the $r \times 1$ vector $\bmu_{m}$, the resulting Jacobian matrix is of size $\frac{1}{2}r(r+1) \times r$. Starting with the differential of \eqref{eqn:dS}
\begin{align}
\begin{split}
\text{d}\bF_{\bS}&(\bmu_{m}) 
\\=& \sum_{\xinX} \text{d}\left(\psi(t_{nm})\left(\left(\xn - \bmu_{m}\right)^{\top} \bS_{m}^{-1} \otimes \left(\xn - \bmu_{m}\right)^{\top} \bS_{m}^{-1}\right)\bD_{r} \right)
\end{split}\notag\\
\begin{split}
=& \sum_{\xinX} \Biggl[\text{d}\psi(t_{nm})\left(\left(\xn - \bmu_{m}\right)^{\top} \bS_{m}^{-1} \otimes \left(\xn - \bmu_{m}\right)^{\top} \bS_{m}^{-1}\right)\bD_{r} \\ &+  \psi(t_{nm})\text{d}\left(\left(\xn - \bmu_{m}\right)^{\top} \bS_{m}^{-1} \otimes \left(\xn - \bmu_{m}\right)^{\top} \bS_{m}^{-1}\right)\bD_{r}\Biggr]
\end{split}\notag\\
& \text{with \eqref{eqnA:dkron}} \notag \\
\begin{split}
=& \sum_{\xinX} \Biggl[-2 \eta(t_{nm}) \left(\xn - \bmu_{m}\right)^{\top} \bS_{m}^{-1} \text{d}\bmu_{m}\left(\left(\xn - \bmu_{m}\right)^{\top} \bS_{m}^{-1} \otimes \left(\xn - \bmu_{m}\right)^{\top} \bS_{m}^{-1}\right)\bD_{r} \\ &+  \psi(t_{nm})\left((-\text{d}\bmu_{m})^{\top} \bS_{m}^{-1} \otimes \left(\xn - \bmu_{m}\right)^{\top} \bS_{m}^{-1} + \left(\xn - \bmu_{m}\right)^{\top} \bS_{m}^{-1} \otimes (-\text{d}\bmu_{m})^{\top} \bS_{m}^{-1}\right)\bD_{r}\Biggr]
\end{split}
\end{align}
and the vectorization
\begin{align}
\begin{split}
\text{d}\vecop &(\bF_{\bS}(\bmu_{m})) 
\\=& \sum_{\xinX} \Biggl[-2 \eta(t_{nm}) \vecop \left(\bxt_{n}^{\top} \bS_{m}^{-1} \text{d}\bmu_{m}\left(\bxt_{n}^{\top} \bS_{m}^{-1} \otimes \bxt_{n}^{\top} \bS_{m}^{-1}\right)\bD_{r} \right) 
\\ &+  \psi(t_{nm})\Bigl[\vecop \left(\left((-\text{d}\bmu_{m})^{\top} \bS_{m}^{-1} \otimes \bxt_{n}^{\top} \bS_{m}^{-1}\right)\bD_{r}\right) 
\\&+ \vecop \left(\left(\bxt_{n}^{\top} \bS_{m}^{-1} \otimes (-\text{d}\bmu_{m})^{\top} \bS_{m}^{-1}\right)\bD_{r}\right)\Bigr]\Biggr]
\end{split}\notag\\
\begin{split}
=& \sum_{\xinX} \Biggl[-2 \eta(t_{nm}) \left[\left(\left(\bxt_{n}^{\top} \bS_{m}^{-1} \otimes \bxt_{n}^{\top} \bS_{m}^{-1}\right)\bD_{r} \right)^{\top} \otimes \bxt_{n}^{\top} \bS_{m}^{-1}\right] \text{d}\vecop \left(\bmu_{m}\right)
\\ &+  \psi(t_{nm})\Bigl[\left(\bD_{r}^{\top} \otimes \bI_{1}\right)
\vecop \left((-\text{d}\bmu_{m})^{\top} \bS_{m}^{-1} \otimes \bxt_{n}^{\top} \bS_{m}^{-1}\right) \\ &+ \left(\bD_{r}^{\top} \otimes \bI_{1}\right)\vecop \left(\bxt_{n}^{\top} \bS_{m}^{-1} \otimes (-\text{d}\bmu_{m})^{\top} \bS_{m}^{-1}\right)\Bigr]\Biggr]
\end{split}\notag\\
& \text{with \eqref{eqnA:veckronvec}} \notag \\
\begin{split}
=& \sum_{\xinX} \Biggl[-2 \eta(t_{nm}) \left[\left(\bD_{r}^{\top} \left( \bS_{m}^{-1} \bxt_{n} \otimes \bS_{m}^{-1} \bxt_{n}\right)\right)  \otimes \bxt_{n}^{\top} \bS_{m}^{-1}\right] \text{d}\vecop \left(\bmu_{m}\right)
\\ &+  \psi(t_{nm})\bD_{r}^{\top}\Bigl[ \left(\bI_{r} \otimes \bK_{r1} \otimes \bI_{1}\right)\left(\vecop \left((-\text{d}\bmu_{m})^{\top} \bS_{m}^{-1}\right) \otimes \vecop \left(\bxt_{n}^{\top} \bS_{m}^{-1}\right)\right) \\ &+ \left(\bI_{r} \otimes \bK_{r1} \otimes \bI_{1}\right)\left(\vecop \left(\bxt_{n}^{\top} \bS_{m}^{-1}\right) \otimes \vecop \left((-\text{d}\bmu_{m})^{\top} \bS_{m}^{-1}\right)\right)\Bigr]\Biggr]
\end{split}\notag\\
& \text{with \eqref{eqnA:veckron} and \eqref{eqnA:Kn1I}} \notag \\
\begin{split}
=& \sum_{\xinX} \Biggl[-2 \eta(t_{nm}) \left[\bD_{r}^{\top} \left( \bS_{m}^{-1} \bxt_{n} \otimes \bS_{m}^{-1} \bxt_{n}\right)\bxt_{n}^{\top} \bS_{m}^{-1}\right] \text{d}\vecop \left(\bmu_{m}\right)
\\ &-  \psi(t_{nm})\bD_{r}^{\top}\Bigl[ \left(\bI_{r} \otimes \bI_{r}\right)\left( \left( \bS_{m}^{-1} \otimes \bI_{1}\right) \vecop\left((\text{d}\bmu_{m})^{\top}\right) \otimes \left( \bS_{m}^{-1} \otimes \bI_{1}\right) \vecop\left(\bxt_{n}^{\top}\right)\right) \\ &+ \left(\bI_{r} \otimes \bI_{r}\right)\left(\left( \bS_{m}^{-1} \otimes \bI_{1}\right) \vecop\left(\bxt_{n}^{\top}\right) \otimes \left( \bS_{m}^{-1} \otimes \bI_{1}\right) \vecop\left((\text{d}\bmu_{m})^{\top}\right) \right)\Bigr]\Biggr]
\end{split}\notag\\
& \text{with \eqref{eqnA:vec}} \notag \\
\begin{split}
=& \sum_{\xinX} \Biggl[-2 \eta(t_{nm}) \left[\bD_{r}^{\top} \left( \bS_{m}^{-1} \bxt_{n} \otimes \bS_{m}^{-1}\bxt_{n}\right) \bxt_{n}^{\top} \bS_{m}^{-1}\right] \text{d}\vecop \left(\bmu_{m}\right)
\\ &-  \psi(t_{nm})\bD_{r}^{\top}\Bigl[\bI_{r^{2}}\left( \bS_{m}^{-1} \text{d}\vecop\left(\bmu_{m}\right) \otimes \bS_{m}^{-1} \bxt_{n}\right) + \bI_{r^{2}} \left(\bS_{m}^{-1}\bxt_{n} \otimes  \bS_{m}^{-1} \text{d}\vecop\left(\bmu_{m}\right) \right)\Bigr]\Biggr]
\end{split}\notag\\
\begin{split}
=& \sum_{\xinX} \Biggl[-2 \eta(t_{nm}) \bD_{r}^{\top} \left( \bS_{m}^{-1}\bxt_{n} \otimes \bS_{m}^{-1} \bxt_{n}\right) \bxt_{n}^{\top} \bS_{m}^{-1} \text{d}\vecop \left(\bmu_{m}\right)
\\ &-  \psi(t_{nm})\bD_{r}^{\top}\Bigl[\left( \bS_{m}^{-1} \otimes \bS_{m}^{-1} \bxt_{n}\right) +  \left(\bS_{m}^{-1}\bxt_{n} \otimes  \bS_{m}^{-1} \right)\Bigr]\text{d}\vecop\left(\bmu_{m}\right)\Biggr]
\end{split}\notag\\
& \text{with \eqref{eqnA:commAB6} and \eqref{eqnA:KDK}} \notag \\
\begin{split}
=& -2 \sum_{\xinX} \Biggl[\eta(t_{nm}) \bD_{r}^{\top} \left( \bS_{m}^{-1} \bxt_{n} \otimes \bS_{m}^{-1} \bxt_{n}\right) \bxt_{n}^{\top} \bS_{m}^{-1} \text{d}\vecop \left(\bmu_{m}\right)
\\ &+ \psi(t_{nm})\bD_{r}^{\top}\left( \bS_{m}^{-1} \otimes \bS_{m}^{-1} \bxt_{n}\right)\text{d}\vecop\left(\bmu_{m}\right)\Biggr]
\end{split}
\end{align}
and the final Jacobian matrix
\begin{align}
D\bF_{\bS}(\bmu_{m}) =& \bF_{\bS\bmu} \notag\\
=& -2 \sum_{\xinX} \left[\eta(t_{nm}) \bD_{r}^{\top} \left( \bS_{m}^{-1} \bxt_{n} \otimes \bS_{m}^{-1} \bxt_{n} \bxt_{n}^{\top} \bS_{m}^{-1}\right)
+ \psi(t_{nm})\bD_{r}^{\top}\left( \bS_{m}^{-1} \otimes \bS_{m}^{-1} \bxt_{n}\right)\right].
\label{eqn:dSmu}
\end{align}
Comparing \eqref{eqn:dmuS} with \eqref{eqn:dSmu} it is evident that
\begin{equation}
\bF_{\bmu\bS} = \left(\bF_{\bS\bmu}\right)^{\top}.
\label{eqn:FSmuSmuS}
\end{equation}

Evaluating $\bF_{\bS\bmu}$ at $\bShat_{m}$ and $\bmuhat_{m}$ from Appendix \ref{ch:ml} leads to
\begin{align}
\bFhat_{\bS\bmu} =& -2 \sum_{\xinX} \left[\eta(\that_{nm}) \bD_{r}^{\top} \left( \bShat_{m}^{-1} \bxhat_{n} \otimes \bShat_{m}^{-1} \bxhat_{n} \bxhat_{n}^{\top} \bShat_{m}^{-1}\right)
+ \psi(\that_{nm})\bD_{r}^{\top}\left( \bShat_{m}^{-1} \otimes \bShat_{m}^{-1} \bxhat_{n}\right)\right]  \notag \\
=& -2 \sum_{\xinX} \eta(\that_{nm}) \bD_{r}^{\top} \left( \bShat_{m}^{-1} \bxhat_{n} \otimes \bShat_{m}^{-1} \bxhat_{n} \bxhat_{n}^{\top} \bShat_{m}^{-1}\right)
-2 \sum_{\xinX} \psi(\that_{nm})\bD_{r}^{\top}\left( \bShat_{m}^{-1} \otimes \bShat_{m}^{-1} \bxhat_{n}\right)  \notag \\
=& -2 \sum_{\xinX} \eta(\that_{nm}) \bD_{r}^{\top} \left( \bShat_{m}^{-1} \bxhat_{n} \otimes \bShat_{m}^{-1} \bxhat_{n} \bxhat_{n}^{\top} \bShat_{m}^{-1}\right)
-2 \bD_{r}^{\top}\left( \bShat_{m}^{-1} \otimes \bShat_{m}^{-1} \sum_{\xinX} \psi(\that_{nm})\bxhat_{n}\right)  \notag \\
& \text{with \eqref{eqn:omega_x_0}} \notag \\
=& -2 \sum_{\xinX} \eta(\that_{nm}) \bD_{r}^{\top} \left( \bShat_{m}^{-1} \bxhat_{n} \otimes \bShat_{m}^{-1} \bxhat_{n} \bxhat_{n}^{\top} \bShat_{m}^{-1}\right)
\label{eqn:dSmuhat}
\end{align}
and equivalently to \eqref{eqn:FSmuSmuS}
\begin{equation}
\bFhat_{\bmu\bS} = \left(\bFhat_{\bS\bmu}\right)^{\top}.
\end{equation}

Defining $\bF_{\bS}$ as a $1 \times \frac{1}{2}r(r+1)$ scalar function of the $r \times r$ matrix $\bS_{m}$, the resulting Jacobian matrix should be of size $\frac{1}{2}r(r+1) \times r^2$. As before, only the unique elements are of interest. Hence, the final size is $\frac{1}{2}r(r+1) \times \frac{1}{2}r(r+1)$. Starting with the differential of \eqref{eqn:dS}
\begin{align}
\text{d} (\bF_{\bS}(\bS_{m})) =& \sum_{\xinX} \text{d} \left(\psi(t_{nm}) \left(\bxt_{n}^{\top} \bS_{m}^{-1} \otimes \bxt_{n}^{\top} \bS_{m}^{-1}\right)\right)\bD_{r} - \frac{N_{m}}{2} \vecop\left(\text{d} \bS_{m}^{-1}\right)^{\top}\bD_{r}\notag\\
\begin{split}
=& \sum_{\xinX} \left[\text{d}\psi(t_{nm}) \left(\bxt_{n}^{\top} \bS_{m}^{-1} \otimes \bxt_{n}^{\top} \bS_{m}^{-1}\right)\bD_{r} + \psi(t_{nm}) \text{d}\left(\bxt_{n}^{\top} \bS_{m}^{-1} \otimes \bxt_{n}^{\top} \bS_{m}^{-1}\right)\bD_{r}\right] 
\\&- \frac{N_{m}}{2} \vecop\left(\text{d} \bS_{m}^{-1}\right)^{\top}\bD_{r}
\end{split}\notag\\
\begin{split}
=& -\sum_{\xinX} \Bigl[\eta(t_{nm}) \bxt_{n}^{\top} \bS_{m}^{-1} \text{d}\bS_{m} \bS_{m}^{-1} \bxt_{n} \left(\bxt_{n}^{\top} \bS_{m}^{-1} \otimes \bxt_{n}^{\top} \bS_{m}^{-1}\right)\bD_{r} 
\\&- \psi(t_{nm}) \text{d}\left(\bxt_{n}^{\top} \bS_{m}^{-1} \otimes \bxt_{n}^{\top} \bS_{m}^{-1}\right)\bD_{r}\Bigr] + \frac{N_{m}}{2} \vecop\left(\bS_{m}^{-1} \text{d}\bS_{m} \bS_{m}^{-1}\right)^{\top}\bD_{r}
\end{split}
\end{align}
and applying the $\vecop$ operator
\begin{align}
\begin{split}
\text{d} \vecop &(\bF_{\bS}(\bS_{m})) 
\\=& -\sum_{\xinX} \Bigl[\eta(t_{nm}) \vecop\left(\bxt_{n}^{\top} \bS_{m}^{-1} \text{d}\bS_{m} \bS_{m}^{-1} \bxt_{n} \left(\bxt_{n}^{\top} \bS_{m}^{-1} \otimes \bxt_{n}^{\top} \bS_{m}^{-1}\right)\bD_{r} \right)
\\&- \psi(t_{nm})\vecop\left(\text{d}\left(\bxt_{n}^{\top} \bS_{m}^{-1} \otimes \bxt_{n}^{\top} \bS_{m}^{-1}\right)\bD_{r}\right)\Bigr] + \frac{N_{m}}{2} \vecop\left(\vecop\left(\bS_{m}^{-1} \text{d}\bS_{m} \bS_{m}^{-1}\right)^{\top}\bD_{r}\right)
\end{split}\notag\\
& \text{ with Equations \eqref{eqnA:vecAB} and \eqref{eqnA:d_veckronvec}}\notag\\
\begin{split}
=& -\sum_{\xinX} \Bigl[\eta(t_{nm}) \left(\left(\bS_{m}^{-1} \bxt_{n} \left(\bxt_{n}^{\top} \bS_{m}^{-1} \otimes \bxt_{n}^{\top} \bS_{m}^{-1}\right)\bD_{r}\right)^{\top} \otimes \bxt_{n}^{\top} \bS_{m}^{-1} \right) \text{d}\vecop\left(\bS_{m}\right)
\\&- \psi(t_{nm}) \bD_{r}^{\top}\left(\bI_{r} \otimes \bK_{r1} \otimes \bI_{1}\right)\Bigl[\left(\bI_{r} \otimes \vecop\left(\bxt_{n}^{\top} \bS_{m}^{-1}\right)\right) 
\\&+ \left(\vecop\left(\bxt_{n}^{\top} \bS_{m}^{-1}\right)\otimes \bI_{r} \right) \Bigr] \text{d}\vecop\left(\bxt_{n}^{\top} \bS_{m}^{-1}\right) \Bigr] + \frac{N_{m}}{2} \bD_{r}^{\top} \vecop\left(\bS_{m}^{-1} \text{d}\bS_{m} \bS_{m}^{-1}\right)
\end{split}\notag\\
\begin{split}
=& -\sum_{\xinX} \Bigl[\eta(t_{nm}) \left(\bD_{r}^{\top} \left( \bS_{m}^{-1} \bxt_{n}\otimes  \bS_{m}^{-1} \bxt_{n}\right) \bxt_{n}^{\top}\bS_{m}^{-1} \otimes \bxt_{n}^{\top} \bS_{m}^{-1} \right) \bD_{r} \text{ d}\vechop\left(\bS_{m}\right)
\\&+ \psi(t_{nm}) \bD_{r}^{\top}\bI_{r^{2}} \Bigl[\left(\bI_{r} \otimes \bS_{m}^{-1}\bxt_{n} \right) + \left(\bS_{m}^{-1}\bxt_{n}\otimes \bI_{r} \right) \Bigr] \left(\bS_{m}^{-1} \otimes \bxt_{n}^{\top} \bS_{m}^{-1}\right) \bD_{r} \text{ d}\vechop\left(\bS_{m}\right)  \Bigr]
\\&+ \frac{N_{m}}{2} \bD_{r}^{\top} \left(\bS_{m}^{-1} \otimes \bS_{m}^{-1}\right) \bD_{r} \text{ d}\vechop\left(\bS_{m}\right)
\end{split}\notag\\
& \text{with \eqref{eqn:KT_K}, \eqref{eqnA:commAB5} and \eqref{eqnA:KDK}} \notag \\
\begin{split}
=& -\sum_{\xinX} \Bigl[\eta(t_{nm}) \bD_{r}^{\top} \left( \bS_{m}^{-1} \bxt_{n}\otimes  \bS_{m}^{-1} \bxt_{n}\right)\left( \bxt_{n}^{\top}\bS_{m}^{-1} \otimes \bxt_{n}^{\top} \bS_{m}^{-1}\right) \bD_{r} 
\\&+ \psi(t_{nm}) \bD_{r}^{\top} \left(\bI_{r} \otimes \bS_{m}^{-1}\bxt_{n} \right) \left(\bS_{m}^{-1} \otimes \bxt_{n}^{\top} \bS_{m}^{-1}\right)\bD_{r} 
\\&+ \psi(t_{nm}) \bD_{r}^{\top}\left(\bS_{m}^{-1}\bxt_{n}\otimes \bI_{r} \right) \left(\bxt_{n}^{\top} \bS_{m}^{-1} \otimes \bS_{m}^{-1}\right) \bD_{r} + \frac{1}{2} \bD_{r}^{\top} \left(\bS_{m}^{-1} \otimes \bS_{m}^{-1}\right)\bD_{r}\Bigr]\text{ d}\vechop\left(\bS_{m}\right)
\end{split}\notag\\
\begin{split}
=& -\sum_{\xinX} \Bigl[\eta(t_{nm}) \bD_{r}^{\top} \left( \bS_{m}^{-1} \bxt_{n} \bxt_{n}^{\top}\bS_{m}^{-1} \otimes \bS_{m}^{-1} \bxt_{n} \bxt_{n}^{\top}\bS_{m}^{-1}\right) \bD_{r} 
\\&+ \psi(t_{nm}) \bD_{r}^{\top}\left(\bS_{m}^{-1} \otimes \bS_{m}^{-1}\bxt_{n}\bxt_{n}^{\top} \bS_{m}^{-1}\right)\bD_{r} + \psi(t_{nm}) \bD_{r}^{\top}\left(\bS_{m}^{-1}\bxt_{n}\bxt_{n}^{\top} \bS_{m}^{-1} \otimes \bS_{m}^{-1}\right) \bD_{r}
\\& + \frac{1}{2} \bD_{r}^{\top} \left(\bS_{m}^{-1} \otimes \bS_{m}^{-1}\right)\bD_{r} \Bigr] \text{ d}\vechop\left(\bS_{m}\right)
\end{split}
\end{align}
we finally obtain the Jacobian matrix
\begin{equation}
\begin{split}
D\bF_{\bS}(\bS_{m}) = \bF_{\bS\bS} =& -\sum_{\xinX} \Bigl[\eta(t_{nm}) \bD_{r}^{\top} \left( \bS_{m}^{-1} \bxt_{n} \bxt_{n}^{\top}\bS_{m}^{-1} \otimes \bS_{m}^{-1} \bxt_{n} \bxt_{n}^{\top}\bS_{m}^{-1}\right) \bD_{r} 
\\&+ \psi(t_{nm}) \bD_{r}^{\top}\left(\bS_{m}^{-1} \otimes \bS_{m}^{-1}\bxt_{n}\bxt_{n}^{\top} \bS_{m}^{-1}\right)\bD_{r} 
\\&+ \psi(t_{nm}) \bD_{r}^{\top}\left(\bS_{m}^{-1}\bxt_{n}\bxt_{n}^{\top} \bS_{m}^{-1} \otimes \bS_{m}^{-1}\right) \bD_{r} \Bigr]  + \frac{N_{m}}{2} \bD_{r}^{\top} \left(\bS_{m}^{-1} \otimes \bS_{m}^{-1}\right)\bD_{r}.
\end{split}
\label{eqn:dSS}
\end{equation}

Evaluating $\bF_{\bS\bmu}$ at $\bShat_{m}$ and $\bmuhat_{m}$ from Appendix \ref{ch:ml} leads to
\begin{align}
\begin{split}
\bFhat_{\bS\bS} =& -\sum_{\xinX} \eta(\that_{nm}) \bD_{r}^{\top} \left( \bShat_{m}^{-1} \bxhat_{n} \bxhat_{n}^{\top}\bShat_{m}^{-1} \otimes   \bShat_{m}^{-1} \bxhat_{n} \bxhat_{n}^{\top}\bShat_{m}^{-1}\right) \bD_{r}
\\&-\sum_{\xinX} \psi(t_{nm}) \bD_{r}^{\top}\left(\bShat_{m}^{-1} \otimes \bShat_{m}^{-1}\bxhat_{n}\bxt_{n}^{\top} \bShat_{m}^{-1}\right)\bD_{r}
\\&- \sum_{\xinX} \psi(t_{nm}) \bD_{r}^{\top}\left(\bShat_{m}^{-1}\bxhat_{n}\bxt_{n}^{\top} \bShat_{m}^{-1} \otimes \bShat_{m}^{-1}\right) \bD_{r} + \frac{N_{m}}{2} \bD_{r}^{\top} \left(\bShat_{m}^{-1} \otimes \bShat_{m}^{-1}\right)\bD_{r}
\end{split}\notag \\
\begin{split}
=& -\sum_{\xinX} \eta(\that_{nm}) \bD_{r}^{\top} \left( \bShat_{m}^{-1} \bxhat_{n} \bxhat_{n}^{\top}\bShat_{m}^{-1} \otimes   \bShat_{m}^{-1} \bxhat_{n} \bxhat_{n}^{\top}\bShat_{m}^{-1}\right) \bD_{r}
\\&-  \bD_{r}^{\top}\left( \bShat_{m}^{-1} \otimes  \bShat_{m}^{-1} \left( \sum_{\xinX} \psi(\that_{nm}) \bxhat_{n} \bxhat_{n}^{\top}\right) \bShat_{m}^{-1} \right)\bD_{r} 
\\&-  \bD_{r}^{\top}\left( \bShat_{m}^{-1} \left( \sum_{\xinX} \psi(\that_{nm}) \bxhat_{n} \bxhat_{n}^{\top}\right) \bShat_{m}^{-1} \otimes \bShat_{m}^{-1}\right)\bD_{r} + \frac{N_{m}}{2} \bD_{r}^{\top} \left(\bShat_{m}^{-1} \otimes \bShat_{m}^{-1}\right)\bD_{r}
\end{split}\notag \\
& \text{with \eqref{eqn:mlS}} \notag \\
\begin{split}
=& -\sum_{\xinX} \eta(\that_{nm}) \bD_{r}^{\top} \left( \bShat_{m}^{-1} \bxhat_{n} \bxhat_{n}^{\top}\bShat_{m}^{-1} \otimes   \bShat_{m}^{-1} \bxhat_{n} \bxhat_{n}^{\top}\bShat_{m}^{-1}\right) \bD_{r}
\\&- \frac{N_{m}}{2}\bD_{r}^{\top}\left( \bShat_{m}^{-1} \otimes  \bShat_{m}^{-1} \right)\bD_{r} - \frac{N_{m}}{2}\bD_{r}^{\top}\left( \bShat_{m}^{-1} \otimes  \bShat_{m}^{-1} \right)\bD_{r}
+ \frac{N_{m}}{2} \bD_{r}^{\top} \left(\bShat_{m}^{-1} \otimes \bShat_{m}^{-1}\right)\bD_{r}
\end{split}\notag \\
\begin{split}
=& - \bD_{r}^{\top} \left( \bShat_{m}^{-1} \otimes \bShat_{m}^{-1} \right) \left(\sum_{\xinX} \eta(\that_{nm})\left( \bxhat_{n} \bxhat_{n}^{\top} \otimes \bxhat_{n} \bxhat_{n}^{\top}\right)\right) \left( \bShat_{m}^{-1} \otimes \bShat_{m}^{-1} \right)\bD_{r}
\\&- \frac{N_{m}}{2}\bD_{r}^{\top}\left( \bShat_{m}^{-1} \otimes \bShat_{m}^{-1} \right)\bD_{r}.
\end{split}
\label{eqn:dSShat}
\end{align}

%% file: Appendix/ap_ML_RES.tex
\section{Maximum Likelihood Estimators for RES distributions}
\label{ch:ml}
\subsection{Maximum Likelihood Estimator for the mean}
Setting \eqref{eqn:dmu} equal to zero leads to the ML estimator $\bmuhat_{m}$ of $\bmu_{m}$, which results in
\begin{align}
& 2 \sum_{\xinX}  \psi(t_{nm}) \left(\xn - \bmu_{m}\right)^{\top} \bS_{m}^{-1} \overset{!}{=} 0 \notag\\
\Rightarrow & \sum_{\xinX}  \psi(t_{nm}) \left(\xn - \bmu_{m}\right)^{\top} = 0 \notag\\
\Rightarrow & \bmuhat_{m} = \frac{\sum_{\xinX} \psi(\that_{nm}) \xn}{\sum_{\xinX} \psi(\that_{nm})}
\label{eqn:mlmu}
\end{align}
with
\begin{equation}
	\that_{nm} =  \left(\xn - \bmuhat_{m}\right)^{\top} \bShat_{m}^{-1} \left(\xn - \bmuhat_{m}\right).
	\label{eqn:that}
\end{equation}

\subsection{Maximum Likelihood Estimator for the variance}
Again setting the first derivative \eqref{eqn:dS} equal to zero
\begin{align}
&\sum_{\xinX} \psi(t_{nm}) \left(\bxt_{n}^{\top} \bS_{m}^{-1} \otimes \bxt_{n}^{\top} \bS_{m}^{-1}\right)\bD_{r} - \frac{N_{m}}{2} \vecop\left(\bS_{m}^{-1}\right)^{\top}\bD_{r} \overset{!}{=} 0\notag\\
\Rightarrow &\sum_{\xinX} \psi(t_{nm}) \left(\bxt_{n}^{\top} \bS_{m}^{-1} \otimes \bxt_{n}^{\top} \bS_{m}^{-1}\right) \bD_{r}\bD_{r}^{+}= \frac{N_{m}}{2} \vecop\left(\bS_{m}^{-1}\right)^{\top}\bD_{r}\bD_{r}^{+}\notag\\
& \text{with \eqref{eqnA:Kvec}, \eqref{eqn:KT_K}, \eqref{eqnA:commAB5} and \eqref{eqnA:dup3}}\notag\\
\Rightarrow &\sum_{\xinX} \psi(t_{nm}) \left(\bxt_{n}^{\top} \otimes \bxt_{n}^{\top} \right) \left(\bS_{m}^{-1} \otimes \bS_{m}^{-1}\right)= \frac{N_{m}}{2} \vecop\left(\bS_{m}^{-1}\right)^{\top}\notag\\
\Rightarrow &\sum_{\xinX}  \psi(t_{nm}) \left(\bxt_{n}^{\top} \otimes \bxt_{n}^{\top} \right) = \frac{N_{m}}{2} \vecop\left(\bS_{m}^{-1}\right)^{\top} \left(\bS_{m} \otimes \bS_{m}\right)\notag\\
\Rightarrow &\sum_{\xinX} \psi(t_{nm}) \left(\bxt_{n} \otimes \bxt_{n} \right) = \frac{N_{m}}{2} \left(\bS_{m} \otimes \bS_{m}\right) \vecop\left(\bS_{m}^{-1}\right) \notag\\
\Rightarrow &\sum_{\xinX} \psi(t_{nm}) \left(\bxt_{n} \otimes \bxt_{n} \right) = \frac{N_{m}}{2}  \vecop\left(\bS_{m}\bS_{m}^{-1}\bS_{m}\right) \notag\\
\Rightarrow &  \vecop\left(\bShat_{m}\right)  = \frac{2}{N_{m}}\sum_{\xinX} \psi(\that_{nm}) \left(\bxhat_{n} \otimes \bxhat_{n} \right)
\label{eqn:mlSvec}
\end{align}
leads to a vectorized form of the ML estimator with $\bxhat_{n} \triangleq \xn - \bmuhat_{m}$. To obtain the matrix form, we apply \eqref{eqn:vecinv}
\begin{align}
\bShat_{m}  =& \frac{2}{N_{m}}\sum_{\xinX}  \left(\vecop(\bI_r)^\top \otimes \bI_r\right)\left(\bI_r \otimes \bxhat_{n} \otimes \psi(\that_{nm})\bxhat_{n} \right)\notag\\
=& \frac{2}{N_{m}}\sum_{\xinX}  \left(\vecop(\bI_r)^\top \left(\bI_r \otimes \bxhat_{n}\right)\right) \otimes \psi(\that_{nm}) \bI_r \bxhat_{n}\notag\\
=& \frac{2}{N_{m}}\sum_{\xinX}  \left(\left(\bI_r \otimes \bxhat_{n}\right)\vecop(\bI_r)\right)^\top  \otimes \psi(\that_{nm}) \bxhat_{n}\notag\\
=& \frac{2}{N_{m}}\sum_{\xinX}  \left(\vecop\left(\bxhat_{n}^{\top}\bI_r\bI_r\right) \right)^\top  \otimes \psi(\that_{nm})\bxhat_{n}\notag\\
=& \frac{2}{N_{m}}\sum_{\xinX}  \bxhat_{n}^{\top} \otimes \psi(t_{nm})\bxhat_{n}\notag\\
=& \frac{2}{N_{m}}\sum_{\xinX}  \psi(\that_{nm})\bxhat_{n}\bxhat_{n}^{\top}
\label{eqn:mlS}
\end{align}

\subsection{Interesting Identities}
Using the ML estimators, some interesting identities can be shown, which can be used to further simplify the final results. Firstly in \cite{Kent.1994} we find
\begin{align}
\bShat_{m}  &= \frac{2}{N_{m}}\sum_{\xinX}  \psi(\that_{nm})\bxhat_{n}\bxhat_{n}^{\top} \notag\\
\Rightarrow \bI_{r} &= \frac{2}{N_{m}}\sum_{\xinX}  \psi(\that_{nm})\bxhat_{n}\bxhat_{n}^{\top} \bShat_{m}^{-1} \notag\\
\Rightarrow \Tr \left(\bI_{r}\right) &= \frac{2}{N_{m}}\sum_{\xinX}  \psi(\that_{nm}) \Tr \left(\bxhat_{n} \bxhat_{n}^{\top} \bShat_{m}^{-1}\right) \notag\\
\Rightarrow r &= \frac{2}{N_{m}}\sum_{\xinX}  \psi(\that_{nm}) \Tr \left( \bxhat_{n}^{\top} \bShat_{m}^{-1}\bxhat_{n}\right) \notag\\
\Rightarrow r &= \frac{2}{N_{m}}\sum_{\xinX}  \psi(\that_{nm}) \that_{nm}.
\label{eqn:r_psi_t}
\end{align}

Also, one can find
\begin{align}
\sum_{\xinX}\psi(\that_{nm}) \bxhat_{n} &= \sum_{\xinX}\psi(\that_{nm})(\xn - \bmuhat_{m}) \notag\\
&=  \sum_{\xinX}\psi(\that_{nm}) \xn- \left(\sum_{\xinX}\psi(\that_{nm})\right) \bmuhat_{m}\notag\\
&= \sum_{\xinX}\psi(\that_{nm}) \xn - \left(\sum_{\xinX}\psi(\that_{nm})\right) \frac{\sum_{\xinX} \psi(\that_{nm}) \xn}{\sum_{\xinX} \psi(\that_{nm})}\notag\\
&= 0
\label{eqn:omega_x_0}
\end{align}

%% file: Appendix/ap_matcalc.tex
\section{Matrix Calculus}
\label{ch:matcalc}

In this Appendix, a brief overview of the used matrix calculus is given. Most of the formulae can be found in \cite{Magnus.2010, Abadir.2005, Magnus.2007} with some additions from \cite{Harmeling.2013, Petersen.2012, invvec.2016}.

\subsection{vec-Operator and inverse vec-Operator}
$\ba$ is a $m \times 1$ column vector
\begin{equation}
	\vecop(\ba) = \vecop\left(\ba^{\top}\right) = \ba
	\label{eqnA:vec}
\end{equation}

\begin{equation}
	\vecop\left(\ba\bb^{\top}\right) = \bb \otimes \ba
\end{equation}
$\bA = \left[\ba_{1} \cdots \ba_{n} \right]$ is a $m \times n$ matrix
\begin{equation}
	\vecop\left(\bA\right) = \begin{bmatrix}
	\ba_{1} \\
	\vdots \\
	\ba_{n}
	\end{bmatrix}, \quad mn \times 1\text{ column vector}
\end{equation}
\begin{equation}
	\vecop_{m\times n}^{-1}(\vecop\left(\bA\right)) = \bA
\end{equation}
\begin{equation}
	\vecop_{m\times n}^{-1}(\ba) = \left(\vecop(\bI_n)^\top \otimes \bI_m\right)\left(\bI_n \otimes \ba\right)
	\label{eqn:vecinv}
\end{equation}
\begin{equation}
	\vecop(\bA\bB\bC) = \left(\bC^{\top} \otimes \bA\right) \vecop(\bB)
	\label{eqnA:vecABC}
\end{equation}
$\bB$ is a $n \times q$ matrix
\begin{align}
	\vecop(\bA\bB) =& \left(\bB^{\top} \otimes \bI_{m}\right) \vecop(\bA) \notag \\
	=&  \left(\bI_{q} \otimes \bA\right) \vecop(\bB)
	\label{eqnA:vecAB}
\end{align}
$\bX$ is a $n \times q$ and $\bY$ is a $p \times r$ matrix
\begin{equation}
	\vecop\left(\bX \otimes \bY\right) = \left(\bI_{q} \otimes \bK_{r,n} \otimes \bI_{p}\right)\left(\vecop(\bX) \otimes \vecop(\bY)\right)
	\label{eqnA:veckronvec}
\end{equation}

\subsection{Trace}

\begin{equation}
	\Tr\left(\bA^{\top}\bB\right) = \vecop\left(\bA\right)^{\top} \vecop\left(\bB\right) 
	\label{eqnA:vecTr}
\end{equation}
\begin{equation}
	\Tr\left(\bA + \bB\right) = \Tr\left(\bA\right) + \Tr\left(\bB\right)
	\label{eqnA:TrA+B}
\end{equation}
\begin{equation}
\Tr\left(\alpha \bA\right) = \alpha \Tr\left( \bA\right) 
\label{eqnA:alphaTrA}
\end{equation}

\subsection{Kronecker Product}
\begin{equation}
	\ba^{\top} \otimes \bb  = \bb\otimes \ba^{\top} = \bb \ba^{\top}
	\label{eqnA:veckron}
\end{equation}
\begin{equation}
	\bA \otimes \bB \otimes \bC  = (\bA \otimes \bB) \otimes \bC = \bA \otimes (\bB \otimes \bC)
\end{equation}
\begin{equation}
	(\bA + \bB) \otimes (\bC + \bD) = \bA \otimes \bC + \bA \otimes \bD + \bB \otimes \bC + \bB \otimes \bD
\end{equation}
\begin{equation}
	\sum_{n = 1}^{N} (\bA \otimes \bB_{n}) = (\bA \otimes \bB_{1}) + \dots + (\bA \otimes \bB_{N}) =\bA \otimes \sum_{n = 1}^{N}\bB_{n}
\end{equation}
\begin{equation}
	(\bA \otimes \bB) (\bC \otimes \bD) = \bA\bC \otimes \bB\bD
\end{equation}
\begin{equation}
	\alpha \otimes \bA = \alpha \bA = \bA\alpha = \bA \otimes \alpha
\end{equation}
\begin{equation}
	\alpha (\bA \otimes \bB) =  (\alpha\bA) \otimes \bB = \bA \otimes (\alpha\bB)
\end{equation}
\begin{equation}
	(\bA \otimes \bB)^{\top}  = \bA^{\top} \otimes \bB^{\top}
\end{equation}
\begin{equation}
	(\bA \otimes \bB)^{-1}  = \bA^{-1} \otimes \bB^{-1}
\end{equation}

\subsection{Definition of the Matrix Derivative}
$\bF$ is a differentiable $m \times p$ matrix function of a $n \times q$ matrix $\bX$. Then, the Jacobian matrix of $\bF$ at $\bX$ is a $mp \times nq$ matrix
\begin{equation}
D\bF(\bX) = \frac{\partial \vecop(\bF(\bX))}{\partial (\vecop(\bX))^{\top}}.
\end{equation}

\subsection{Differentials}
\begin{equation}
	\text{d}\left(\bX^{\top}\right) = \left(\text{d}\bX\right)^{\top}
\end{equation}
\begin{equation}
	\text{d}\vecop\left(\bX\right) = \vecop\left(\text{d}\bX\right)
\end{equation}
\begin{equation}
	\text{d}\Tr\left(\bX\right) = \Tr\left(\text{d}\bX\right)
\end{equation}
$\phi$ is a scalar function
\begin{equation}
	\text{d}\left(\phi^{\alpha}\right) = \alpha\phi^{\alpha-1}\text{d}\phi
	\label{eqnA:dscalar_alpha}
\end{equation}
\begin{equation}
	\text{d}\bX^{-1} = - \bX^{-1}\text{d}\bX\bX^{-1}
	\label{eqnA:dinv}
\end{equation}
\begin{equation}
	\text{d}\left| \bX \right| = \left| \bX \right|\Tr\left(\bX^{-1} \text{d} \bX\right)
	\label{eqnA:ddet}
\end{equation}
\begin{equation}
	\text{d}\ln\left(\left| \bX \right|\right) = \text{Tr}\left(\bX^{-1} \text{d} \bX\right)
	\label{eqnA:dlndet}
\end{equation}
\begin{equation}
	\text{d}\left(\bX \otimes \bY \right) = \text{d}\bX \otimes \bY + \bX \otimes \text{d}\bY
	\label{eqnA:dkron}
\end{equation}
$\bx$ is a $n \times 1$ vector
\begin{equation}
	\text{d}\vecop\left( \bx\bx^{\top} \right) =  \left(\left(\bx \otimes \bI_{n}\right)+ \left(\bI_{n}\otimes \bx \right)\right)\text{ d}\vecop\left( \bx\right) 
\end{equation}
$\bA$ is symmetric
\begin{equation}
	\text{d}\vecop\left( \bx^{\top} \bA\bx\right) =  2 \bx^{\top} \bA \text{ d}\vecop\left( \bx\right) 
	\label{eqnA:d_xAx}
\end{equation}
$\bX$ is a $n \times q$ and $\bY$ is a $p \times r$ matrix
\begin{equation}
\text{d}\vecop\left(\bX \otimes \bY\right) = \left(\bI_{q} \otimes \bK_{r,n} \otimes \bI_{p}\right)\left[\left(\bI_{nq} \otimes \vecop(\bY)\right)\text{d}\vecop\left(\bX\right) + \left(\vecop(\bX) \otimes \bI_{pr}\right)\text{d}\vecop\left(\bY\right)\right]
\label{eqnA:d_veckronvec}
\end{equation}

\subsection{Commutation Matrix}
$\bA$ is a $m \times n$ matrix, $\bK_{m,n}$ is a $mn \times mn$ matrix such that
\begin{equation}
	\bK_{m,n}\vecop\left(\bA\right) =   \vecop\left(\bA^{\top}\right)
	\label{eqnA:Kvec}
\end{equation}
with the properties
\begin{equation}
	\bK_{m,n}^{\top} = \bK_{m,n}^{-1} = \bK_{n,m}
	\label{eqn:KT_K}
\end{equation}
\begin{equation}
	\bK_{n,n} = \bK_{n}
\end{equation}
\begin{equation}
	\bK_{n,m}\bK_{m,n} = \bI_{n}
\end{equation}
\begin{equation}
	\bK_{n,1} = \bK_{1,n} = \bI_{n}
	\label{eqnA:Kn1I}
\end{equation}
$\bB$ is a $p \times q$ matrix, $\bb$ is a $p \times 1$ vector
\begin{equation}
	\bK_{p,m}(\bA \otimes \bB) = (\bB \otimes \bA)\bK_{q,n}
	\label{eqnA:commAB1}
\end{equation}
\begin{equation}
	\bK_{p,m}(\bA \otimes \bB)\bK_{n,q} = (\bB \otimes \bA)
	\label{eqnA:commAB2}
\end{equation}
\begin{equation}
	\bK_{p,m}(\bA \otimes \bb) = (\bb \otimes \bA)
	\label{eqnA:commAB3}
\end{equation}
\begin{equation}
	\bK_{m,p}(\bb \otimes \bA) = (\bA \otimes \bb)
	\label{eqnA:commAB4}
\end{equation}
\begin{equation}
	(\bA \otimes \bb^{\top})\bK_{n,p} =  (\bb^{\top} \otimes \bA)
	\label{eqnA:commAB5}
\end{equation}
\begin{equation}
	(\bb^{\top} \otimes \bA)\bK_{p,n} = (\bA \otimes \bb^{\top})
	\label{eqnA:commAB6}
\end{equation}

\subsection{Duplication Matrix}
$\bA$ is a symmetric $n \times n$ matrix with $\frac{1}{2}n(n+1)$ unique elements, $\bD_{n}$ is a $n^{2} \times \frac{1}{2}n(n+1)$ matrix, such that
\begin{equation}
	\vecop\left(\bA\right) =  \bD_{n} \vechop\left(\bA\right), \quad \bA = \bA^{\top}
	\label{eqnA:vech}
\end{equation}
\begin{equation}
	\bK_{n}\bD_{n} = \bD_{n}
	\label{eqnA:KDK}
\end{equation}
\begin{equation}
	\bD_{n}^{+} = \left(\bD_{n}^{\top}\bD_{n}\right)^{-1} \bD_{n}^{\top}
	\label{eqnA:dup1}
\end{equation}
\begin{equation}
	\bD_{n}^{+}\bD_{n} = \bI_{\frac{1}{2}n(n+1)}
	\label{eqnA:dup2}
\end{equation}
\begin{equation}
	\bD_{n}\bD_{n}^{+} = \frac{1}{2}\left(\bI_{n^{2}} + \bK_{n}\right)
	\label{eqnA:dup3}
\end{equation}
$\bb$ is a $n \times 1$ vector
\begin{equation}
	\bD_{n}\bD_{n}^{+}\left(\bb \otimes \bA\right) = \frac{1}{2}\left(\bb \otimes \bA + \bA \otimes \bb\right)
	\label{eqnA:dup4}
\end{equation}

Why are we using the duplication matrix for derivatives with respect to symmetric matrices?
\begin{remark}
	Since $\bA$ is symmetric, say of order $n$, its $n^{2}$ elements cannot move independently. The symmetry imposes $n(n-1)/2$ restrictions. The “free” elements are precisely the $n(n+1)/2$ elements in $\vechop(\bA)$, and the derivative is therefore defined by considering $\bF$ as a function of $\vechop(\bA)$ and not as a function of $\vecop(\bA)$. \mbox{({\cite[p. 367]{Abadir.2005}})}
	\label{re:dup_mat}
\end{remark}